\documentclass[11pt,reqno]{amsart}

\usepackage[colorlinks,citecolor=blue,urlcolor=blue,linkcolor=blue]{hyperref}
\usepackage{amsmath}
\usepackage{amssymb}
\usepackage{bm}
\usepackage{graphicx}
\usepackage{xcolor}
\usepackage[latin1]{inputenc}
\usepackage{amsfonts}
\usepackage{amsthm}
\usepackage{newlfont}
\usepackage{natbib}
\usepackage{latexsym}
\usepackage{float}
\usepackage{capt-of}\usepackage{stackrel}
\usepackage{mathrsfs}
\usepackage{xr}
\usepackage{enumerate}
\usepackage{natbib}
\usepackage{url} 
\usepackage{tikz}
\usetikzlibrary{trees}

\newlength{\defbaselineskip}
\newcommand{\setlinespacing}[1]{\setlength{\baselineskip}{#1\defbaselineskip}}

\numberwithin{equation}{section}
\newtheorem{theorem}{Theorem}[section]
\newtheorem{corollary}[theorem]{Corollary}
\newtheorem{lemma}[theorem]{Lemma}

\newtheorem{ass}[theorem]{Assumption}

\newtheorem{defn}[theorem]{Definition}

\theoremstyle{remark}
\newtheorem{remark}[theorem]{Remark}
\theoremstyle{remark}

\newcommand{\mtI}{\mathcal{I}}
\newcommand{\ml}{\lambda}
\newcommand{\argmax}{\operatorname{argmax}}

\newcommand{\diag}{\operatorname{diag}}
\newcommand{\rank}{\operatorname{rank}}
\newcommand{\col}{\operatorname{col}}
\newcommand{\miff}{\Leftrightarrow}
\newcommand{\cp}{\clearpage}
\newcommand{\BC}{\mathbb{C}}
\newcommand{\BR}{\mathbb{R}}
\newcommand{\BN}{\mathbb{N}}
\newcommand{\BZ}{\mathbb{Z}}
\newcommand{\wt}[1]{\widetilde{#1}}
\newcommand{\wh}[1]{\widehat{#1}}
\newcommand{\conv}{\rightarrow}
\newcommand{\toinf}{\conv\infty}
\newcommand{\tozero}{\conv 0}
\newcommand{\toone}{\conv 1}
\newcommand{\convp}{\stackrel{p}{\conv}}

\newcommand{\convw}{\stackrel{w}{\conv}}
\newcommand{\convas}{\stackrel{a.s.}\conv }
\newcommand{\asyp}{\stackrel{p}\asymp}
\newcommand{\aspos}{\stackrel{a.s.}{>}0}
\newcommand{\asgeq}{\stackrel{a.s.}{\geq}}
\newcommand{\asleq}{\stackrel{a.s.}{\leq}}
\newcommand{\veca}{\operatorname{vec}}
\newcommand\mfl{{\lfloor Tu\rfloor}}
\newcommand{\myV}{\wh U}
\newcommand{\V}{\mathbb{V}}
\def\E{\mathbb{E}}
\def\myPr{\mathbb{P}}
\newcommand{\tr}{\operatorname{tr}}
\newcommand{\rd}{\ensuremath{\mathrm{d}}}
\newcommand{\eig}{\operatorname{eig}}
\newcommand{\tref}[1]{$\ref{#1}$}
\newcommand{\cca}{\textsc{cca}}

\newcommand{\plim}{\operatorname{plim}}

\def\cadlag{c\`adl\`ag}
\newcommand{\KL}{Karhunen-Lo\`eve}
\newcommand{\KLB}{KL basis} 
\newcommand{\CCA}{\textsc{cca}}
\newcommand{\SCC}{\textsc{scc}}
\newcommand{\myiid}{\text{i.i.d.}}
\newcommand{\cs}{cointegrating space}
\newcommand{\as}{attractor space}

\usepackage[normalem]{ulem}

\newcommand{{\Ntext}}{$N=10^4$}
\newcommand{{\N}}{10000}

\begin{document}

\setlinespacing{1.5}

\title[]{Canonical correlation analysis of stochastic trends via functional approximation}

\author[]{M. Franchi, I. Georgiev, P. Paruolo\\\\\today}
\thanks{\\ \emph{Acknowledgments}: The views expressed in the paper are those of the authors and do not necessarily reflect the opinion of the institutions of affiliation. The authors thankfully acknowledge useful comments from:
 the Associate Editor, two anonymous Referees,
 S\o ren Johansen, Ye Lu, Anders Rahbek, and participants to paper presentations at
 University of Copenhagen,
 University of Sydney,
 Workshop in memory of Stefano Fachin at Sapienza University of Rome,
 University of Oxford,
 University of Aarhus,
 Workshop on Methodological and Computational Issues in Large-Scale Time Series Models for Economics and Finance at University of Messina,
 Econometric Workshop in memory of Carlo Giannini at University of Pavia,
 Villa Mondragone Time Series Symposium in honor of Marco Lippi at Tor Vergata University of Rome,
 Eleventh Italian Congress of Econometrics and Empirical Economics at University of Palermo. M. Franchi gratefully acknowledges partial financial support from Ministero dell'Universit\`a e della Ricerca for the project PRIN 20223725WE ``Methodological and computational issues in large-scale time series models for economics and finance''.\\
 \\\indent
 M. Franchi, Sapienza University of Rome,
\href{https://orcid.org/0000-0002-3745-2233}{ORCID: 0000-0002-3745-2233}
 E-mail: massimo.franchi@uniroma1.it,
 Address: P.le A. Moro 5, 00185 Rome, Italy.
 \\\indent
 I. Georgiev,
 University of Bologna,
\href{https://orcid.org/0000-0003-3392-806X}{ORCID: 0000-0003-3392-806X}.
 E-mail: i.georgiev@unibo.it,
 Address:
 Piazza Scaravilli 2, 40126 Bologna, Italy.
 \\\indent
 P. Paruolo.
 European Commission, Joint Research Centre (JRC)
\href{https://orcid.org/0000-0002-3982-4889}{ORCID: 0000-0002-3982-4889}, Email: paolo.paruolo@ec.europa.eu,
 Address: Via Enrico Fermi 2749, 21027 Ispra (VA), Italy.
 }
 
 \begin{abstract}
This paper proposes a novel approach for semiparametric inference on the number $s$ of common trends and their loading matrix $\psi$ in $I(1)/I(0)$ systems. It combines functional approximation of limits of random walks and canonical correlations analysis, performed between the $p$ observed time series of length $T$ and the first $K$ discretized elements of an $L^2$ basis. Tests and selection criteria on $s$, and estimators and tests on $\psi$ are proposed; their properties are discussed as $T$ and $K$ diverge sequentially for fixed $p$ and $s$. It is found that tests on $s$ are asymptotically pivotal, selection criteria of $s$ are consistent, estimators of $\psi$ are $T$-consistent, mixed-Gaussian and efficient, so that Wald tests on $\psi$ are asymptotically Normal or $\chi^2$. The paper also discusses asymptotically pivotal misspecification tests for checking model assumptions. The approach can be coherently applied to subsets or aggregations of variables in a given panel. Monte Carlo simulations show that these tools have reasonable performance for $T\geq 10 p$ and $p\leq 300$. An empirical analysis of 20 exchange rates illustrates the methods.
\end{abstract}

\begingroup
\def\uppercasenonmath#1{} 
\maketitle
\endgroup


\section{Introduction}\label{sec_intro}

The analysis of multiple time series with nonstationarity and comovements has been steadily developing since the introduction of the notion of cointegration in \citet{EG:87}, and the seminal contributions of \citet{SW:88}, \citet{Joh:88, Joh:91}, \citet{AR:90}, \citet{Phillips1990}; see \citet{Wat:94} for an early survey. These contributions discussed asymptotic results for diverging sample size $T$ and fixed cross-sectional dimension $p$. Associated applications considered a cross-sectional dimension $p$ in the low single digits.

A central parameter in this context is the $p\times r$ cointegration matrix $\beta$ and its rank $r$, the cointegration rank; inference on these quantities can be approached either parametrically, as in the context of Vector Autoregressive (VAR) processes, see \citet{Joh:91, Joh:96}, or semiparametrically, as in \citet{Phillips1990}, and \citet{Bie:97}.

Data with $p$ in the double digits have recently become pervasive. The label `moderate dimensional' has been used when such data are analyzed assuming a fixed $p$ in the asymptoptics for diverging $T$, see \citet{CS:24}; the present paper falls into this framework. Contributions for the moderate dimensional case include modifications of parametric procedures, such as Bartlett corrections, see \citet{Joh:00,Joh:02b,Joh:02}, bootstrap implementations, see \citet{Swe:06}, \citet{CRT:12} and \citet{CNR:15}, and Lasso-type procedures, see \citet{CS:24}.

One alternative stream of literature, called `high dimensional', considers $p$ and $T$ diverging proportionally; leading contributions are \citet{OW:18,OW:19}, \citet{BG:22,BG:24}, \citet{LS:19}. Finally, the cross sectional dimension $p$ can also be infinite from the start, as in functional time series, see \citet{PZK22}, \citet{NSS:22}. Combinations of the moderate and high dimensional cases have been studied in \citet{ZRY:19}, \citet{BCT:22}, \citet{FGP:23}.


\subsection{Contributions and relation with the literature}

The present paper provides novel inferential tools for semiparametric inference on $I(1)/I(0)$ $p$-dimensional vectors $X_t$ that admit the Common Trends (CT) representation
$$
X_t=\gamma+\psi\kappa'\sum_{i=1}^t\varepsilon_i +C_1(L)\varepsilon_t,\qquad t=1,2,\dots,
$$
where $\gamma$ is a vector of initial values, $\kappa'\sum_{i=1}^t\varepsilon_i$ are $0\leq s\leq p$ stochastic trends, $\psi $ is a $p\times s$ full column rank loading matrix, $L$ is the lag operator and $C_1(L)\varepsilon_t$ is an $I(0)$ linear process.

The analysis can be coherently applied to subsets or aggregations of variables. This fact rests on the property, called \emph{dimensional coherence}, that linear combinations of the observables $X_t$ also admit a common trends representation with a number of stochastic trends that is at most equal to the one of the original system.

The proposed approach is based on the empirical canonical correlations between the $p$ observed variables $\{X_t\}_{t=1}^T$ and $K$ deterministic variables $\{d_t\}_{t=1}^T$, constructed as the first $K$ elements of an orthonormal $L^2[0,1]$ basis discretized over the equispaced grid $1/T,2/T,\dots,1$. In the asymptotic analysis, the cross-sectional dimension $p$ is kept fixed while $T$ and $K$ diverge sequentially, $K$ after $T$, denoted by $(T,K)_{seq}\toinf$.

The techniques in the present paper build on the results in \citet{Phi:98,Phi:05,Phi:14}, where functional approximations with an increasing number $K$ of $L^2[0,1]$ basis elements are used, respectively, for understanding regressions between $I(1)$ and deterministic variables, for the heteroskedasticity and autocorrelation-consistent estimation of long-run variance (LRV) matrices and for efficient instrumental-variables (IV) estimation of cointegrating vectors.

Related approaches based on a fixed number $K$ of cosine functions are presented in \citet{Bie:97} and in \citet{MW:08,MW:17,MW:18}, who respectively estimate the number of stochastic trends and the cointegrating relations using an IV approach, and analyze long-run variability and covariability of filtered time series.

The present study combines canonical correlation analysis (\CCA) and the functional approximation of limits of random walks. It is shown that the $s$ largest squared canonical correlations (\SCC) tend to one while the $p-s$ smallest \SCC\ tend to zero for any cross-sectional dimension $p$, for any number of stochastic trends $0\leq s\leq p$ and for any choice of $L^2[0,1]$ basis, all fixed, as $(T,K)_{seq}\toinf$. This implies that a criterion based on the maximal gap between \SCC\ selects the correct number of common trends consistently. The same result applies to criteria designed following \citet{Bie:97} and \citet{AH:13}.

The cause of these results is that the stochastic trends in $X_t$ converge to Brownian motions as $T$ diverges while the deterministic variables $d_t$ used in the \CCA\ converge to the first $K$ elements of an $L^2[0,1]$ basis; Brownian motions have an $L^2$-representation (unlike $I(0)$ components), and in the $K$-limit they are completely recovered. This implies that the largest \SCC\ are associated to the nonstationary directions in $X_t$ and the smallest ones to the stationary linear combinations.

This is the opposite of what happens in the cases analyzed in \citet{OW:18,OW:19} and in \citet{BG:22,BG:24}, which involves the \CCA\ between differences $\Delta X_t$ and lagged levels $X_{t-1}$ (possibly correcting for lagged differences and deterministic terms) derived from the Gaussian Quasi Maximum Likelihood (QML) estimator in cointegrated VARs, see \citet{Joh:88,Joh:91} and \citet{AR:90}.

The present approach also differs from the one in \cite{FGP:23}, who discuss the \CCA\ between levels $X_t$ and their cumulations $\sum_{i=1}^{t}X_i$, and consider an asymptotic regime in which $T$ and $p$ diverge sequentially. The advantage of the proposal here is that it leads to efficient asymptotic inference on the loading matrix $\psi$, as well as to tests on $s$ and misspecification tests, as explained next.

The canonical variates in the \CCA\ are used to define estimators of the space spanned by the loading matrix $\psi$ and of the {\cs} spanned by the cointegrating matrix $\beta$; these estimators are shown to be $T$-consistent for any cross-sectional dimension $p$, for any number of stochastic trends $0\leq s\leq p$ and for any basis of $L^2[0,1]$, all fixed. When the \CCA\ is iterated once in a specific way, for $(T,K)_{seq}\toinf$ the associated estimators of $\psi$ and $\beta$ are asymptotically mixed-Gaussian and efficient, as in \citet{Joh:96} and in \citet{Phi:05,Phi:14}. The mixed-Gaussian asymptotic distribution is then used to construct asymptotically Normal and $\chi^2$ Wald test statistics for hypotheses on $\psi$ and $\beta$. Identification of $\psi$ and $\beta$ is discussed and an inferential criterion for the validity of the chosen identification condition is proposed.


Further asymptotic results are derived for the specific basis in the \KL\ (KL) representation of a Brownian Motion, see \citet{LoII:78}. Using the \KLB, canonical correlations are used (i) to estimate $s$ by a top-down sequences of asymptotically pivotal tests for $H_0:s=j$ versus $H_1:s<j$, for $j=p,p-1,\dots,1$ and (ii) to perform misspecification tests for model assumptions. The sequential tests are shown to select the correct number of stochastic trends with limit probability equal to 1 minus the size of each test in the sequence, similarly to what happens for sequences of likelihood ratio (LR) tests in the VAR case, see \citet{Joh:96}.

\subsection{Organization and notation}

The remainder of the paper is organized as follows. Section \ref{sec_setup} discusses assumptions on the Data Generating Process (DGP) and Section \ref{sec_CCA} defines the main estimators. Section \ref{sec_asympt_s} discusses the asymptotic properties of estimators of the number of stochastic trends and Section \ref{sec_asympt_psi} those of the loading and the cointegrating matrices. Section \ref{sec_M} contains the Monte Carlo study, Section \ref{sec_E} presents an empirical application to exchange rates and Section \ref{sec_conc} concludes. The Supplementary material contains lemmas, proofs and additional material on simulations and on the empirical application.

Concerning notation, for any $a \in \BR$, $\lfloor a \rfloor$ and $\lceil a \rceil$ denote the floor and ceiling functions; for any condition $c$, $1_{c}$ is the indicator function of $c$, taking value 1 if $c$ is true and 0 otherwise; $\iota_{n}$ indicates the $n \times 1$ vectors of ones. The space of right continuous functions $f:[0,1]\mapsto\BR^n$ having finite left limits, endowed with the Skorokhod topology, is denoted by $D_n[0,1]$, see \citet[][Ch. VI]{JS:03} for details, also abbreviated to $D[0,1]$ if $n=1$. A similar notation is employed for the space of square integrable functions $L_n^2[0,1]$, abbreviated to $L^2[0,1]$ if $n=1$. Riemann integrals $\int_0^1 a(u)b(u)'\rd u$, Ito stochastic integrals $\int_0^1\rd a(u)b(u)'$ and Stratonovich stochastic integrals $\int_0^1\partial a(u)b(u)'$ are abbreviated respectively as $\int ab'$, $\int\rd a b'$ and $\int\partial a b'$.

\section{Generating process, identification and $L^2$-representation}\label{sec_setup}

This section describes the class of processes studied in the paper and their limiting $L^2$-representation; it also discusses identification of the parameters of interest.

\subsection{Data Generating Process}

Let $\{X_t\}_{t\in \BZ}$ be a $p\times 1$ linear process generated by
\begin{equation}\label{eq_DX}
\Delta X_t=C(L)\varepsilon_t,
\end{equation}
where $L$ is the lag operator, $\Delta:=1-L$, $C(z)=\sum_{n=0}^\infty C_n z^n$, $z\in \BC$, satisfies Assumption \tref{ass_DGP} below, and the innovations $\{\varepsilon_t\}_{t\in \BZ}$ are independently and identically distributed (i.i.d.) with expectation $\E(\varepsilon_t)=0$, finite moments of order $2+\epsilon$, $\epsilon>0$, and positive definite variance-covariance matrix $\V(\varepsilon_t)=\Omega_\varepsilon$, indicated as $\varepsilon_t\:\myiid\: (0,\Omega_\varepsilon)$, $\Omega_\varepsilon>0$.

Assumption \tref{ass_DGP} below concerns the first-order expansion of $C(z)$, $C(z)=C+C_1(z)(1-z)$, and the rank $s:=\rank C(1)=\rank C$. For $s>0$, let $\breve\psi$ denote a matrix whose columns form a basis of $\col C$; then $\breve\kappa$ is the unique matrix such that $C=\breve\psi\breve\kappa'$.\footnote{For any matrix $a$, $\col a$ indicates the linear space spanned by its columns; when $(\col a)^\bot\neq\{0\}$, $a_\bot$ is used to indicate any basis of $(\col a)^\bot$.} Moreover, for $s<p$, let $\breve\beta:=\breve\psi_\bot$ and $\breve\kappa_\bot$ be matrices whose columns form bases of $(\col C)^{\bot}$ and $(\col C')^{\bot}$ respectively; thus, for $0<s<p$, it holds that $(\col \breve \psi)^{\bot}=\col(\breve\psi_\bot)$ and likewise for $\breve\kappa$. The space $\col C$ is called the {\as} and its orthogonal complement $(\col C)^\bot$ is the {\cs} of $X_t$, of dimension $0\leq r:=p-s\leq p$ (the cointegrating rank).

\begin{ass}[Assumptions on $C(z)$]\label{ass_DGP}
$C(z)$ in \eqref{eq_DX} is of dimension $p\times n_\varepsilon$, $n_\varepsilon \geq p$, satisfying
\begin{enumerate}
\item[(i)] $C(z)=\sum_{n=0}^\infty C_n z^n$ converges for all $|z|<1+\delta$, $\delta >0$;
\item[(ii)] $\rank C(z)<p$ may only hold at isolated points $z$, where either $z=1$ or $|z|>1$;
\item[(iii)] if $s<p$, then $\breve\psi_\bot'C_1\breve\kappa_\bot$ has full row rank $r:=p-s$, with $C_1:=C_1(1)$.
\end{enumerate}
\end{ass}

Assumption \tref{ass_DGP} includes possibly nonstationary VARMA processes (for $n_\varepsilon = p$) in line with e.g. \citet{SW:88} and Dynamic Factor Models (for $n_\varepsilon > p$) as in e.g. \citet{Bai:04}.

Cumulating \eqref{eq_DX}, one finds the Common Trends (CT) representation
\begin{equation}\label{eq_X}
X_t=\gamma+\breve\psi\breve\kappa'\sum_{i=1}^t\varepsilon_i +C_1(L)\varepsilon_t,\qquad t=1,2,\dots,
\end{equation}
which gives a decomposition of $X_t$ into initial conditions $\gamma:=X_0-C_1(L)\varepsilon_0$, the $s$-dimensional random walk component
$\breve\kappa'\sum_{i=1}^t\varepsilon_i$ with loading matrix $\breve\psi$ and the stationary component $C_1(L)\varepsilon_t$.
Observe that $X_t$ is $I(0)$ if $s=0$, i.e. $r=p$; it is $I(1)$ and cointegrated if $0<r,s<p$ and it is $I(1)$ and non-cointegrated if $s=p$, i.e. $r=0$.

The aim of the paper is to make inferences about $s$ and $\breve\psi$ (and on their complements $r$ and $\breve\beta$) from a sample $\{X_t\}_{t = 0,1,\dots,T}$ observed from \eqref{eq_X}.

\begin{remark}[Initial conditions]\label{rem_init_cond}
Definition 3.4 in \citet{Joh:96} stipulates that a vector $v\neq 0$ is a cointegrating vector if $v' X_t$ can be made stationary by a suitable choice of its initial distribution; in the present case, this corresponds to setting $\breve\beta'\gamma=0$, i.e. $\breve\beta'X_0=\breve\beta'C_1(L)\varepsilon_0$, so that $\gamma:=X_0-C_1(L)\varepsilon_0=\breve\psi\gamma_0$ and hence
\begin{equation}\label{eq_X_SJ}
X_t=\breve\psi\left(\gamma_0+\breve\kappa'\sum_{i=1}^t\varepsilon_i\right)+C_1(L)\varepsilon_t,\qquad t=1,2,\dots\:.
\end{equation}
Alternatively, as in \citet{Ell:99}, see also \citet{MW:08} and \citet{OW:18}, one can consider
\begin{equation}\label{eq_X_MW}
X_t-X_0=\breve\psi\breve\kappa'\sum_{i=1}^t\varepsilon_i +C_1(L)(\varepsilon_t-\varepsilon_0),\qquad t=1,2,\dots\:.
\end{equation}
\end{remark}
The statistical analysis is performed on $x_t:=X_t$ or $x_t:=X_t-X_0$ according to whether specification \eqref{eq_X_SJ} or \eqref{eq_X_MW} is chosen.

\subsection{Identification}\label{sec_ident}

The parameters $\breve\psi$, $\breve\kappa$ and $\breve\beta$ are not identified; in fact $C=\breve\psi\breve\kappa'= (\breve\psi a) (a^{-1}\breve\kappa')$ for any nonsingular $a\in \BR^{s\times s}$, so that $\breve\psi$ and $\breve\psi a$ are bases of the {\as} $\col C$, and $\breve\kappa$ and $\breve\kappa (a^{-1})'$ of $\col (C')$; similarly, $\breve\beta$ and $\breve\beta a$ with $a\in \BR^{r\times r}$ nonsingular are bases of the {\cs} $(\col C)^\bot$.


As done in \cite{Joh:96} for $\breve\beta$, just-identification of the loading matrix $\breve\psi$ is achieved by means of linear restrictions, employing a known and full column rank matrix $b \in \BR^{p\times s}$ such that $b'\breve\psi\in \BR^{s\times s}$ is nonsingular. Similarly, just-identification of the cointegrating vectors is obtained using a known and full column rank matrix $c\in \BR^{p\times r}$ such that $c'\breve\beta\in \BR^{r\times r}$ is nonsingular, where $c$ is a basis of $(\col b)^\bot$. The following Lemma discusses the relationship between the identified parameters obtained in this way; here $\bar a:=a(a'a)^{-1}$ for a full column rank matrix $a$ and $\breve \psi$ and $\psi$ (respectively $\breve \beta$ and $\beta$) are 
bases of $\col C$ (respectively $\col (C)^\perp$).

\begin{lemma}[Duality]\label{lem_ident}
Let $C=\breve\psi\breve\kappa'$ be a rank factorization of $C$, let $b \in \BR^{p\times s}$ be such that $b'\breve\psi\in \BR^{s\times s}$ is nonsingular and let $c \in \BR^{p\times r}$ be a basis of $(\col b)^\bot$; then
\renewcommand{\theenumi}{(\roman{enumi})}
\renewcommand{\labelenumi}{\theenumi}
\begin{enumerate}
\item for $s>0$, $\psi:=\breve\psi( b'\breve\psi)^{-1}$ and $\kappa:=\breve\kappa(\breve\psi' b)=C' b$ are identified as the unique solution of the system of equations $C=\psi\kappa'$ and $b'\psi=I_s$;
\item for $s<p$, $c'\breve\beta\in \BR^{r\times r}$ is nonsingular and $\beta:=\breve\beta( c'\breve\beta)^{-1}$ is identified as the unique solution of the system of equations $\beta'C=0$ and
$c'\beta=I_r$;
\item for $0<s<p$, the unrestricted coefficients $\psi_*:=\bar c'\psi$ and $\beta_*:=\bar b'\beta$ satisfy $\psi_*=-\beta_*'$;
\item for any estimator $\wt \psi$ of $\psi$ normalized as $ b'\wt \psi = b'\psi = I_{s}$ and any estimator $\wt \beta$ of $\beta$ normalized as $c'\wt \beta = c'\beta = I_{r}$, one has $\bar c'(\wt \psi -\psi) = \beta'(\wt \psi -\psi)$ and similarly $\bar b'(\wt \beta -\beta) = \psi'(\wt \beta -\beta)$.
\end{enumerate}
\end{lemma}

Point (iii) shows that the unrestricted coefficients in the identified parameters are linearly dependent, $\psi_*=-\beta_*'$; this implies a duality between hypotheses on $\psi$ and $\beta$, see \eqref{eq_dual_H0} below.

Importantly, one does not know a priori which restrictions are just-identifying, i.e., which $b$ delivers a nonsingular $b'\breve\psi$. Take for instance $b=(I_s,0)'$: because $\breve\psi$ has rank $s$, there is always at least one permutation of the order of variables for which these restrictions are just-identifying; however, one does not know which permutation should be used. A consistent decision rule on the nonsingularity of $b'\breve\psi$ is proposed in Section \ref{sec_iden_psi} below.

The choice of a specific $b$ implies a unique space $\col (b)^\bot = \col c$, but leaves the choice of the specific basis $c$ open, so that $c$ can be chosen to best fit the application. Observe also that the choice of $b$ and then of $c$ can be reversed, by interchanging the role of $(\breve\psi, b)$ and $(\breve\beta, c)$ in Lemma \tref{lem_ident}. In the remainder of the paper it is assumed that $\psi$, $\kappa$ and $\beta$ in \eqref{eq_X} are identified by the choice of $b$ and $c=b_\bot$ such that $b' \psi = I_s$, $c' \beta = I_r$ and $\kappa= C' b$.

\subsection{Dimensional coherence}\label{sec_dim_coh}

This section shows that whenever $X_t$ has a CT representation \eqref{eq_X}, also linear combinations of $X_t$ admit a CT representation, with a number of stochastic trends that is at most equal to that in $X_t$. This property, called dimensional coherence, is formally stated in the next theorem.

\begin{theorem}[Dimensional coherence]\label{theorem_dim_coh}
Let $H$ be a $p\times m$ full column rank matrix; if $X_t$ satisfies \eqref{eq_DX} with $C(z)$ fulfilling Assumption \tref{ass_DGP}, then the same holds for $H'X_t$, i.e. $\Delta H'X_t=G(L)\varepsilon_t$, where $G(z):=H'C(z)$ satisfies Assumption \tref{ass_DGP} with $G(1)=H'C$ of rank $q\leq s$.
\end{theorem}

Special cases of $H$ are the selection matrix $H=(I_m,0)'$ of the first $m$ variables, and the group-aggregation matrix $H=(I_{q}\otimes\iota_{n})$, where $p=qn$, $m=q$ and $\iota_{n}$ is an $n \times 1$ vector of ones. One consequence of Theorem \ref{theorem_dim_coh} is that the semiparametric approach described in this paper applies to subsets or aggregations of the original variables $X_t$; this property is exploited in the empirical application in Section \tref{sec_E}.

\subsection{Functional central limit theorem and long-run variance}

By the functional central limit theorem, see \citet{PS:92}, the partial sums $T^{-\frac12}\sum_{i=1}^t\varepsilon_i$ converge weakly to an $n_{\varepsilon}$-dimensional Brownian Motion $W_\varepsilon(u)$, $u\in[0,1]$, with variance $\Omega_\varepsilon>0$; by the CT representation \eqref{eq_X},
one hence has that
\begin{equation}\label{eq_x}
T^{-\frac12}
\left(\begin{array}c
\bar\psi' x_t\\
\beta'\sum_{i=1}^t x_i
\end{array}
\right)=
T^{-\frac12}\left(\begin{array}c
\kappa'\\
\beta'C_1
\end{array}
\right)\sum_{i=1}^{\lfloor Tu\rfloor}\varepsilon_i +o_p(1)
\underset{T\toinf}{\convw}
\left(\begin{array}c
 W_1(u)\\
 W_2(u)
\end{array}
\right):=
\left(\begin{array}c
\kappa'\\
\beta'C_1
\end{array}
\right)W_\varepsilon(u),
\end{equation}
where $t =\lfloor Tu\rfloor\in \BN$ and the $o_p(1)$-term is infinitesimal uniformly in $u\in[0,1]$. Here $\convw$ indicates weak convergence of probability measures on $D_p[0,1]$ and $W(u):=( W_1(u)', W_2(u)')'$ is a $p\times 1$ Brownian motion with variance $\Omega:=D\Omega_\varepsilon D'$, $D:=(\kappa,C_1'\beta)'$ and $\beta=\psi_{\bot}$.

Remark that long-run variance (LRV) $\Omega$ is positive definite because $D$ is nonsingular by Assumption \tref{ass_DGP}.(iii); in fact,
$$
\rank D=
\rank\left(\begin{array}c
\kappa'\\
\beta'C_1
\end{array}
\right) (\bar\kappa,\kappa_\bot)
 =\rank\left(\begin{array}{cc}
I_s & 0\\
\beta'C_1\bar\kappa &\beta'C_1\kappa_\bot
\end{array}
\right)=s+\rank\beta'C_1\kappa_\bot.
$$
Partition $\Omega$ conformably to $W(u)=( W_1(u)', W_2(u)')'$, namely
\begin{equation}\label{eq_Xi}
\Omega:=
\left(
\begin{array}{cc}
\Omega_{11}&\Omega_{12}\\
\Omega_{21} &\Omega_{22}\end{array}
\right)=
\left(
\begin{array}{cc}
\kappa'\Omega_\varepsilon \kappa&\kappa'\Omega_\varepsilon C_1'\beta\\
\beta'C_1\Omega_\varepsilon \kappa &\beta'C_1\Omega_\varepsilon C_1'\beta\end{array}%
\right),
\end{equation}
and define the independent Brownian motions
\begin{equation}\label{eq_def_b_q}
	W_{2.1}(u):= W_2(u)-\Omega_{21}\Omega_{11}^{-1} W_1(u),\qquad B_1(u):=\Omega_{11}^{-\frac12} W_1(u)
\end{equation}
with nonsingular variance matrices $\Omega_{22.1}:=\Omega_{22}-\Omega_{21}\Omega_{11}^{-1}\Omega_{12}$ and $I_s$ respectively.

\subsection{$L^2$-representation}

Recall, see e.g. \cite{LoII:78} section 37.5B, that the $p$-dimensional Brownian motion $W(u)$ admits the representation
\begin{equation}\label{eq_d_L2}
	W(u)\simeq\sum_{k=1}^\infty c_k\phi_k(u),\quad c_k:=\int_0^1 W(u)\phi_k(u)\rd u=\nu_k\xi_k,\quad \nu_k\in \BR_+, \qquad \xi_k\sim N(0,\Omega),
\end{equation}
where $\{\phi_k(u)\}_{k=1}^\infty$, $\int_0^1\phi_j(u)\phi_k(u)\rd u=1_{j=k}$, is an orthonormal basis of $L^2[0,1]$ and $\simeq$ indicates that the series in \eqref{eq_d_L2} is a.s. convergent in the $L^2$ sense to the l.h.s..

In the special case where $(\nu_k^2,\phi_k(u))$, $k=1,2,\dots$, is an eigenvalue-eigenvector pair of the covariance kernel of the standard Brownian motion, i.e. $\nu_k^2\phi_k(u)=\int_0^1\min(u,v)\phi_k(v)\rd v$, \eqref{eq_d_L2} is the {\KL} (KL) representation of $W(u)$, for which one has
\begin{equation}\label{eq_KLB}
\nu_k=\frac1{(k-\frac12)\pi},\qquad\phi_k(u)=\sqrt 2\sin \left(u/\nu_k\right),\qquad\xi_k\sim \myiid\:N(0,\Omega).
\end{equation}
In this case the series in \eqref{eq_d_L2} is a.s. uniformly convergent in $u$ and $\simeq$ is replaced by $=$. In the following, a basis $\{\phi_k(u)\}_{k=1}^\infty$ is indicated as the {\KLB} when $\phi_k(u)$ is chosen as in \eqref{eq_KLB}. Expansion \eqref{eq_d_L2} underlies the large-$K$ asymptotic theory in this paper; see Appendix \ref{sec_app_lemmas} for details and references.

\section{Definition of the estimators}\label{sec_CCA}

This section presents the proposed estimators of the number $s$ of stochastic trends and the cointegrating rank $r=p-s$, as well as of $\psi$ (the identified basis of the {\as}) and $\beta$ (the identified basis of the {\cs}).

Let $\varphi_K(u):=(\phi_1(u),\dots,\phi_K(u))'$ be a $K \times 1$ vector function of $u \in[0,1]$, where $\phi_1(u),\dots,\phi_K(u)$ are the first $K$ elements of some fixed orthonormal \cadlag\ basis of $L^2[0,1]$, see \eqref{eq_d_L2}. Let $d_t$ be the $K\times 1$ vector constructed by evaluating $\varphi_K(\cdot)$ at the discrete sample points $1/T,\dots, (T-1)/T, 1$, i.e.
\begin{equation}\label{eq_d}
d_t:=\varphi_K(t/T):=(\phi_1(t/T),\dots,\phi_K(t/T))',\qquad K\geq p,\qquad t=1,\dots,T.
\end{equation}

For a generic $p$-dimensional variable $f_t$ observed for $t=1,\dots,T$, the sample canonical correlation analysis of $f_t$ and $d_t$ in \eqref{eq_d}, denoted as $\cca(f_t,d_t)$, consists in solving the following generalized eigenvalue problem, see e.g. \citet{Joh:96} and references therein,
\begin{equation}\label{eq_BEP}
\cca(f_t,d_t):\qquad |\ml M_{ff}-M_{fd}M_{dd}^{-1}M_{df}|=0,\qquad M_{ij}:=T^{-1}\sum_{t=1}^Ti_tj_t'.
\end{equation}
This delivers eigenvalues $1\geq \ml_1\geq\ml_2\geq\dots\geq\ml_p \geq 0$, collected in $\Lambda=\diag(\ml_1,\dots,\ml_p)$, and corresponding eigenvectors $V=(v_1,\dots,v_p)$, organized as
\begin{equation}\label{eq_LV}
\Lambda_1:=\diag(\ml_1,\dots,\ml_s),\quad
\Lambda_0:=\diag(\ml_{s+1},\dots,\ml_p), \quad
V_1:=(v_1,\dots,v_s), \quad
V_0:=(v_{s+1},\dots,v_p),
\end{equation}
with the convention that $(\Lambda_1,V_1)=(\Lambda,V)$ when $s=p$ and $(\Lambda_0,V_0)=(\Lambda,V)$ when $s=0$.

\begin{defn}[Max-gap estimators of $s$ and $r$]\label{def_s_hat}
Define the maximal gap $($max-gap$)$ estimator of $s$ as
\begin{equation}\label{eq_def_shat}
\wh s :=\underset{i\in \{0,\dots, p\}}{\argmax}(\ml_i-\ml_{i+1}),\qquad\ml_0:=1,\qquad\ml_{p+1}:=0,
\end{equation}
where $1\geq \ml_1\geq\ml_2\geq\dots\geq\ml_p \geq 0$ are the eigenvalues of $\cca(x_t,d_t)$, $x_t$ is defined after \eqref{eq_X_MW} in Remark \tref{rem_init_cond}; further define the estimator of $r$ as $\wh r:=p-\wh s$.
\end{defn}

The choice $\ml_0:=1$ and $\ml_{p+1}:=0$  permits the calculation of $\ml_i-\ml_{i+1}$ also for $i=0$ and $i=p$, and it is motivated by the results in Theorem \tref{theorem_lim_eig} below.

First stage estimators of $\psi$ and $\beta$ are defined in terms of the eigenvectors in $\cca(x_t,d_t)$, and they are indicated as $\wh\psi^{(1)}$ and $\wh\beta^{(1)}$. The Iterated Canonical Correlation (ICC) estimators are denoted as $\wh\psi$ and $\wh\beta$, and they are defined in terms of the eigenvectors in $\cca(e_t,d_t)$, where $e_t$ denotes the residual of $x_t$ regressed on the fitted values from the regression of $\wh\psi^{(1)'}\Delta x_t$ on $d_t$. The motivation for the ICC estimation is to obtain mixed-Gaussian limit distributions, see Theorem \tref{eq_asy_iter} below.

\begin{defn}[Estimators of $\psi$ and $\beta$]\label{def_phi_hat}
Let $0<s<p$.\footnote{For $s=0$ (resp. $s=p$), $\psi$ (resp. $\beta$) is undefined and $\beta=(c')^{-1}$ (resp. $\psi=(b')^{-1}$) is known.} Let $b \in \BR^{p\times s}$ and $c= b_\bot \in \BR^{p\times r}$ 
be a pair of matrices for the just-identification of $\breve\psi$ and $\breve\beta$, see Section \tref{sec_ident}, with
$\psi:=\breve\psi( b'\breve\psi)^{-1}$,
$\beta:=\breve\beta( c'\breve\beta)^{-1}$; for any $V$ as in \eqref{eq_LV},
the first stage estimators of $\psi$ and $\beta$ are defined as
$$
\wh\psi^{(1)} := M_{xx} V_1( b'M_{xx} V_1)^{-1},\qquad\wh\beta^{(1)} := V_0( c' V_0)^{-1},
$$
where the eigenvectors $V$ are those of $\cca(x_t,d_t)$. The ICC estimators are defined as
$$
\wh\psi := M_{ee} V_1( b'M_{ee} V_1)^{-1},\qquad\wh\beta := V_0( c' V_0)^{-1},
$$
where the eigenvectors $V$ are those of $\cca(e_t,d_t)$, with
\begin{equation}\label{eq_gt_def}
e_t:=x_t-M_{xg}M_{gg}^{-1} g_t,\qquad g_t:=\wh\psi^{(1)'} M_{\Delta x d}M_{dd}^{-1}d_t.
\end{equation}
\end{defn}

The asymptotic analysis implies that the inverse matrices in Definition \tref{def_phi_hat} are well-defined with probability approaching one. Note moreover that, (i), the estimators in Definition \tref{def_phi_hat} are invariant to the normalization of the eigenvectors $V$ and, (ii), the ICC estimators $\wh\psi $ and $\wh\beta $ depend on $\wh\psi^{(1)}$ only through $\col(\wh\psi^{(1)})$.

Remark that the estimators in Definition \tref{def_phi_hat} satisfy the identifying constraints of $\psi$ and $\beta$, i.e.
\begin{equation}\label{eq_dual_psi_foo}
 b'\psi= b'\wh\psi^{(1)}= b'\wh\psi=I_s,\qquad c'\beta= c'\wh\beta^{(1)}= c'\wh\beta=I_r
\end{equation}
and that the duality in Lemma \tref{lem_ident}(iii) equally applies to them, i.e.
\begin{equation}\label{eq_dual_psi}
\psi_*=-\beta_*',\quad \wh\psi^{(1)}_*=-\wh\beta^{(1)'}_*,\quad \wh\psi_*=-\wh\beta_*',
\end{equation}
where $a_*:=\bar c'a$ and $h_*:=\bar b'h$, for $a=\psi,\wh\psi^{(1)},\wh\psi$ and $h=\beta,\wh\beta^{(1)},\wh\beta$.

\section{Inference on the number of common trends}\label{sec_asympt_s}

This section discusses the limits of the eigenvalues of $\cca(x_t,d_t)$ and their implications for estimating the number $s$ of stochastic trends (and the number $r$ of cointegrating relations).

First, results are provided for the max-gap estimator and for alternative argmax estimators based on \SCC. Second, by employing the \KLB, several statistics are introduced (i) to perform misspecification tests for model assumptions and (ii) to define top-down sequences of asymptotically pivotal tests of $H_0:s=j$ versus $H_1:s<j$ for $j=p,p-1,\dots,1$, which provide alternative estimators of $s$.

\subsection{Consistency}\label{sec_cons_s}

The asymptotic behavior of \SCC\ is given in the following theorem.

\begin{theorem}[Limits of the eigenvalues of $\cca(x_t,d_t)$]\label{theorem_lim_eig}
Let $W_2(u)$, $ B_1(u)$ be defined as in \eqref{eq_x}, \eqref{eq_def_b_q}, let $d_t$ in \eqref{eq_d} be constructed using any orthonormal \cadlag\ basis of $L^2[0,1]$ and let $1\geq \ml_1\geq\ml_2\geq\dots\geq\ml_p \geq 0$ be the eigenvalues of \textnormal{$\cca(x_t,d_t)$}, see \eqref{eq_BEP} and \eqref{eq_LV}; then for fixed $K\geq p$ and $T\toinf$,
\begin{enumerate}
\item[(i)] the $s$ largest eigenvalues $\ml_1\geq\ml_2\geq\dots\geq\ml_s$ converge weakly to the ordered eigenvalues of
\begin{equation}\label{eq_large_eigs}
\Upsilon_K:=\left(\int B_1 B_1'\right)^{-1}\int B_1\varphi_K'\int\varphi_K B_1'\aspos;
\end{equation}
\item[(ii)] the $r$ smallest eigenvalues $\ml_{s+1}\geq\ml_{s+2}\geq\dots\geq\ml_p$ multiplied by $T$ converge weakly to the ordered eigenvalues of
\begin{equation}\label{eq_small_eigs}
\Psi_K:=Q_{00}^{-1}\int\rd W_2\varphi_K'Q_{\varphi\varphi}\int\varphi_K\rd W_2'\aspos,
\end{equation}
where $Q_{00}:=\E(\beta'x_t x_t'\beta)>0$ and $Q_{\varphi\varphi}:=I_K-\int\varphi_K B_1'\left(\int B_1\varphi_K'\int\varphi_K B_1'\right)^{-1}\int B_1\varphi_K'$.
\end{enumerate}
Moreover,
\begin{enumerate}
\item[(iii)] for $K\toinf$, $\Upsilon_K\convp I_s$ and $K^{-1}\Psi_K\convp Q_{00}^{-1}\Omega_{22}$;
\item[(iv)] for $(T,K)_{seq}\toinf$,
all the $s$ largest eigenvalues $\ml_1\geq\ml_2\geq\dots\geq\ml_s$ converge in probability to $1$, while all the $r$ smallest eigenvalues $\ml_{s+1}\geq\ml_{s+2}\geq\dots\geq\ml_p$ converge in probability to $0$.
\end{enumerate}
\end{theorem}

Let $a \asyp g(T,K)$ for some function $g$ indicate that $a /g(T,K)$ is bounded and bounded away from 0 in probability as $(T,K)_{seq}\toinf$; points (i), (ii) and (iii) in Theorem \tref{theorem_lim_eig} imply that
\begin{equation}\label{eq_order_eigs2}
\ml_i\asyp\left\{
\begin{array}{ll}
1 & i=1,\dots,s\\
K/T & i=s+1,\dots,p
\end{array}\right.
\end{equation}
for $(T,K)_{seq}\toinf$. Because $K/T\conv 0$ as $(T,K)_{seq}\toinf$, this implies $\ml_i \convp 1 $ for $i=1,\dots,s$ while  $\ml_i \convp 0 $ for $i=s+1,\dots,p$, see Theorem \tref{theorem_lim_eig}.(iv).

\begin{corollary}[Consistency of the max-gap estimator $\wh s$]\label{coro_cons_s}
Let $\wh s$ be the max-gap estimator in \eqref{eq_def_shat} computed on \textnormal{$\cca(x_t,d_t)$} and let the assumptions of Theorem \tref{theorem_lim_eig} hold;
then $\myPr(\wh s = s)\conv 1$ as $(T,K)_{seq}\toinf$ for any $0\leq s \leq p$.
\end{corollary}

This result holds because Theorem \tref{theorem_lim_eig}.(iv) implies that $f_0(i):=\ml_i-\ml_{i+1}\convp 1_{i=s}$ as $(T,K)_{seq}\toinf$; note that the function $1_{i=s}$ provides the best discrepancy that the \SCC\ $\{\lambda_i\}_{i=1}^{p}$ can give between the $I(1)$ and $I(0)$ components.

In the context of eigenvalue problems different from the ones considered here, \citet{Bie:97} and \citet{AH:13} propose to estimate $s$ selecting the integer $i$ that maximizes some other function $f_j(i)$ of the eigenvalues, possibly using the rates of convergence of $\lambda_{i}$ in the definition of $f_j(i)$. These approaches, applied to the $\{\lambda_i\}_{i=1}^{p}$ eigenvalues from $\cca(x_t,d_t)$, suggest the following additional estimators of $s$, called here \emph{alternative argmax estimators}:
\begin{align}\label{eq_criteria}
&\wt s^{(j)} :=\underset{i\in \mtI_j}{\argmax}\:f_j(i),\qquad j=1,2,3,\\\nonumber
f_1(i):=\frac{\prod_{h=1}^i\ml_h}{\prod_{h=i+1}^p\left(\frac{T}{K}\ml_h\right)},\qquad & f_2(i):=\frac{\ml_i}{\ml_{i+1}},\qquad f_3(i):=\frac{\log\left(1+\ml_i/\sum_{h=i+1}^p\ml_h\right)}{\log\left(1+\ml_{i+1}/\sum_{h=i+2}^p\ml_h\right)},
\end{align}
where $\mtI_1:=\{0,1,\dots, p\}$, $\mtI_2:=\{1,\dots, p-1\}$, $\mtI_3:=\{1,\dots, p-2\}$, empty sums (products) are equal to 0 (1), and $f_1(i)$ uses the rates in \eqref{eq_order_eigs2} for $(T,K)_{seq}\toinf$.

The estimators $\wt s^{(j)}$, $j=1,2,3$, optimize over the set of integers $\mtI_j$; note that $f_1$ is defined for $0\leq i \leq p$, $f_2$ for $1\leq i \leq p-1$ and $f_3$ for $1\leq i \leq p-2$. One may extend $f_2$ and $f_3$ to cover $i=0$ by defining $\lambda_0:=1$ as in \eqref{eq_def_shat}; this leads to replacing $\mtI_j$ in \eqref{eq_criteria} with the sets $\mtI_j^0:=\{0\}\cup\mtI_j$, $j=2,3$. Note however that adding $\lambda_{p+1}:=0$ is not an option, as $\lambda_{p+1}$ would appear in the denominators of $f_2$ and $f_3$. This implies that $f_2$ (respectively $f_3$) cannot be defined for $i=p$ (respectively $i=p-1,p$).

\begin{corollary}[Consistency of the alternative argmax estimators of $s$]\label{coro_cons_other}
Let $\wt s^{(j)}$, $j=1,2,3$, be the alternative argmax estimators of $s$ in \eqref{eq_criteria} computed on \textnormal{$\cca(x_t,d_t)$} and let the assumptions of Theorem \tref{theorem_lim_eig} hold; then $\myPr(\wt s^{(j)}=s)\toone$ for $(T,K)_{seq}\toinf$ for any $s\in \mtI_j$, $j=1,2,3$. The same holds replacing $\mtI_j$ with $\mtI_j^0$ both in the optimization and in the class of DGPs.
\end{corollary}

These results are in line with the ones reported in \citet{Bie:97} and \citet{AH:13} in their respective setups.

\subsection{Testing identification}\label{sec_iden_psi}

Recall that the joint identification of $\psi$ and $\beta$ via $b$ and $c$ given in Lemma \ref{lem_ident} rests on the validity of the rank restriction $\rank( b'\breve\psi)=s$. Here a decision rule on the correctness of $\rank( b'\breve\psi)=s$ is proposed, using the property of dimensional coherence; let 
\begin{equation}\label{eq_test_on_c}
H_0:\rank ( b'\breve\psi ) =s\qquad\text{versus}\qquad H_1:\rank ( b'\breve\psi )<s.
\end{equation}
The proposed decision rule consists of estimating the number of stochastic trends of $b'x_t$, and rejecting $H_0$ if the estimated number of stochastic trends is lower than the one estimated for $x_t$. The rationale for the criterion is the following: from \eqref{eq_X} one has $b'x_t= b'\breve\psi\breve\kappa'\sum_{i=1}^t\varepsilon_i + b'C_1(L)\varepsilon_t$, so that the number of stochastic trends in $b'x_t$ coincides with the one in $x_t$ if and only if $b'\breve\psi$ is nonsingular. Formally,
\begin{equation}\label{eq_criterion_on_c}
\text{reject } H_0 \quad \text{ if } \quad \wh s ( b'x_t) <\wh s (x_t),
\end{equation}
where $\wh s (x_t)$ (respectively $\wh s ( b' x_t)$) is the max-gap estimator calculated for $x_t$ (respectively $b'x_t$), see \eqref{eq_def_shat}.

\begin{theorem}[Consistency of decision rule \eqref{eq_criterion_on_c}]\label{theorem_dec_rule}If $H_0$ in \eqref{eq_test_on_c} holds, one has $\myPr(\wh s ( b'x_t) <\wh s (x_t))\tozero$ and $\myPr(\wh s ( b'x_t) =\wh s (x_t))\toone$ as $(T,K)_{seq}\toinf$. If $H_1$ in \eqref{eq_test_on_c} holds, one has $\myPr(\wh s ( b'x_t) <\wh s (x_t))\toone$ and $\myPr(\wh s ( b'x_t) =\wh s (x_t))\tozero$ as $(T,K)_{seq}\toinf$. The same statements are true substituting $\wh s$ with $\wt s^{(j)}$ provided $s\in \mtI_j$ (respectively $s\in \mtI_j^0$) both in the DGP and in the optimization.
\end{theorem}

\subsection{Asymptotic distributions}\label{sec_asy_distr_s}

This subsection discusses misspecification tests and test sequences for the estimation of $s$; these are derived using the specific {\KLB} in \eqref{eq_KLB} in the definition of $d_t$ in \eqref{eq_d}.

\begin{theorem}[Asymptotic distribution of the eigenvalues with the {\KLB}]\label{theorem_asy_distr_s}
Let $ B_1(u)$ be defined as in \eqref{eq_def_b_q}, let $d_t$ in \eqref{eq_d} be constructed using the {\KLB} in \eqref{eq_KLB} and let $1\geq \ml_1\geq\ml_2\geq\dots\geq\ml_p \geq 0$ be the eigenvalues of\textnormal{$\cca(x_t,d_t)$}; then for $(T,K)_{seq}\toinf$,
\begin{enumerate}
\item[(i)] the $s$ largest eigenvalues $\ml_1\geq\ml_2\geq\dots\geq\ml_s$ satisfy
\begin{equation}\label{eq_K_IG}
K\pi^2\tau^{(s)}\convw\zeta^{(s)},
\end{equation}
where
$\tau^{(s)}:=(1-\ml_s,1-\ml_{s-1},\dots,1-\ml_1)'$, $\zeta^{(s)}:=(\zeta_1,\zeta_2,\dots,\zeta_s)'$, and
 $\zeta_1\geq\zeta_2\geq\dots\geq\zeta_s$ are the eigenvalues of
$
\omega:=\left(\int B_1 B_1'\right)^{-1};
$
\item[(ii)] the $r$ smallest eigenvalues $\ml_{s+1}\geq\ml_{s+2}\geq\dots\geq\ml_p$ satisfy $K\pi^2 (1-\ml_i)\stackrel{p}{\conv}\infty$, $i=s+1,\dots,p$.
\end{enumerate}
\end{theorem}

Observe that $\omega$ does not depend on nuisance parameters, and that tests based on $K\pi^2\tau^{(s)}$ are asymptotically pivotal.
Further, the limit distribution in Theorem \tref{theorem_asy_distr_s}(i) depends on the tail $\sum_{k=K+1}^\infty \nu_k^2$, see \eqref{eq_KLB},
and is therefore not invariant to the choice of an $L_2[0,1]$ basis.

The rest of this section discusses misspecification tests and sequences of tests on $s$ based on the results of Theorem \tref{theorem_asy_distr_s}.

\begin{remark}[Confidence stripe for misspecification analysis]\label{rem_stripe}
The results in Theorem \ref{theorem_asy_distr_s} can be used to construct confidence sets for misspecification analysis.
For example, consider $f(a)=\log(a)-\E(\log\zeta^{(s)})$, $a \in \BR^s$, where the $\log$ function is applied component-wise, and let $\|\cdot\|_n$ be the $n$-norm for vectors.
Define the confidence misspecification stripe $\mathcal{B}:=\{a \in \BR^s:\|f(a)\|_{\infty}<\delta \}$ with $\delta$,
such that $\myPr (\zeta^{(s)}\in \mathcal{B})=1-\eta$ for some pre-assigned confidence level $1-\eta$. The continuous mapping theorem and Theorem \ref{theorem_asy_distr_s} then imply that $\myPr(K\pi^2\tau^{(s)}\in \mathcal{B})\conv 1-\eta$ as $(T,K)_{seq}\conv\infty$. The set $\mathcal{B}$ is a stripe around $\E(\log\zeta^{(s)})$ with equal deviations above or below $\E(\log(\zeta_i))$ across $i=1,\dots,s$, see Figure \tref{fig_all_XR} below for an illustration.

In practice, one can compute $\mathcal{B}$ and check if $K\pi^2\tau^{(\wh s)}\in \mathcal{B}$. If so, the model assumptions appear valid, including the selection of $s$; otherwise it is necessary to question which assumptions need to be modified. The same approach can be used for continuous transformations other than $\log\zeta_i$ and any well-defined location indicator, such as the median.
\end{remark}

\begin{remark}[Sequential tests on $s$]\label{rem_seq_tests}
The results in Theorem \ref{theorem_asy_distr_s} can be used to test hypotheses of the following form
\begin{equation}\label{eq_hypo_s}
H_{0j}:s=j\qquad\text{versus}\qquad H_{1j} :s<j,\qquad j=p,p-1,\dots,1.
\end{equation}
Define the test statistics $F_{j,n}:=\|K\pi^2\tau^{(j)}\|_n$. Under $H_{0s}$, Theorem \ref{theorem_asy_distr_s} shows that $F_{s,n}\convw\|\zeta^{(s)}\|_n$, while under $H_{1s}$, $F_{s,n}\toinf$. For $n=1,\infty$, one finds
\begin{equation}\label{eq_def_F1_Finf}
F_{j,1}=\|K\pi^2\tau^{(j)}\|_1=K\pi^2\sum_{i=1}^j(1-\ml_i),\qquad F_{j,\infty}=\|K\pi^2\tau^{(j)}\|_\infty=K\pi^2(1-\ml_j),
\end{equation}
which are similar to the trace and $\ml$-max test statistics in \cite{Joh:91}, and satisfy
$ F_{s,1}\convw\tr (\omega)=\sum_{i=1}^s\zeta_i^{(s)}$, and $F_{s,\infty}\convw\max\eig (\omega)=\zeta_1^{(s)} $ as $(T,K)_{seq}\conv\infty$.

The $F_{j,n}$ test rejects for large values of the statistics, and critical values $c_{n,\eta}$ are defined as
$ \Pr(\|\zeta^{(s) }\|_{n} \leq c_{n,\eta})= 1- \eta $, see \eqref{eq_K_IG}; they can be estimated by simulation of the limit distribution.
The following test sequence is defined using these quantities: choose an asymptotic test size $\eta \in(0,1)$ and test $H_{0j}$ versus $H_{1j}$ in \eqref{eq_hypo_s}
for $j=p, p-1,\dots, 1$ using test statistic $F_{j,n}$ with significance level $\eta$ until a non-rejection is found. The resulting estimated value for $s$ equals the first non-rejected value of $j$; this test sequence is indicated as
$\{F_{j,n}\}_{j=p,\dots,1}$.
\end{remark}

\begin{corollary}[Asymptotic properties of test sequences]\label{theorem_sequence}
Under the assumptions of Theorem \tref{theorem_asy_distr_s}, as $(T,K)_{seq}\toinf$,
the test sequence $\{F_{j,n}\}_{j=p,\dots,1}$, has asymptotic probability to select the true $s$ equal to $1-\eta$ for $s>0$ and equal to $1$ if $s=0$, $n=1,\infty$.
\end{corollary}

Results for test sequences are similar to the ones for LR tests in a VAR context for cointegration rank determination, see \cite{Joh:96}.

\section{Inference on the attractor space}\label{sec_asympt_psi}

This section discusses the limit distributions of the first stage and ICC estimators of the identified parameters $\psi$ and $\beta$ as well as Wald tests of hypotheses on their unrestricted parameters $\psi_*$ and $\beta_*$, see \eqref{eq_dual_psi}, valid for any choice of basis of $L^2[0,1]$. Because $b'(\wh\psi-\psi)=0$ and $c'(\wh\beta-\beta)=0$, the unrestricted parameters are in $\bar c'(\wh\psi-\psi)=\wh\psi_*-\psi_*$ and $\bar b'(\wh\beta-\beta)=\wh\beta_*-\beta_*$, which are linearly related by the duality in Lemma \tref{lem_ident} and \eqref{eq_dual_psi}; see also Theorem \tref{theorem_asy_distr_psi} below.

Results can be summarized as follows. First stage and ICC estimators are $T$-consistent and the ICC estimators are asymptotically mixed-Gaussian, unlike the first stage estimators. Thus, asymptotically Normal or $\chi^2$ Wald test statistics for hypotheses on $\psi$ and $\beta$ can be constructed using the ICC estimators.

\begin{theorem}[Asymptotic distributions of the first stage and ICC estimators]\label{theorem_asy_distr_psi}
Let $d_t$ in \eqref{eq_d} be constructed using any orthonormal \cadlag\ basis of $L^2[0,1]$ and let the unrestricted coefficients be as in \eqref{eq_dual_psi_foo} and \eqref{eq_dual_psi}; then for $(T,K)_{seq}\toinf$, the first stage estimators satisfy
\begin{equation}\label{eq_asy_first}
T(\wh\psi^{(1)}_*-\psi_*)=- T(\wh\beta^{(1)}_*-\beta_*)' \convw \int \partial W_2 W_1'\left(\int W_1 W_1'\right)^{-1}=:Z^{(1)},
\end{equation}
where $\int \partial W_2 W_1'$ denotes a Stratonovich stochastic integral and $Z^{(1)}$ is not mixed-Gaussian in general. On the contrary, for the ICC estimators one has
\begin{equation}\label{eq_asy_iter}
T(\wh\psi_*-\psi_*)=- T(\wh\beta_*-\beta_*)' \convw \int \rd W_{2.1} W_1'\left(\int W_1 W_1'\right)^{-1}=:Z,
\end{equation}
where $\int \rd W_{2.1} W_1'$ denotes an Ito stochastic integral and $Z$ is mixed-Gaussian because $W_{2.1}$ and $W_1$ are independent.
\end{theorem}

\begin{remark}[Mixed Gaussianity]\label{rem_mixed_gauss}
Note that $W_2$ and $W_1$ in \eqref{eq_asy_first} are dependent because $\Omega_{21}$ in \eqref{eq_Xi} is in general different from 0; this implies that $Z^{(1)}$ is not mixed-Gaussian in general. On the contrary, $W_{2.1}$ and $W_1$ in \eqref{eq_asy_iter} are independent and thus $Z$ is mixed-Gaussian; specifically, conditionally on $W_1$, one has $\veca(Z)\sim N(0,\left(\int W_1 W_1'\right)^{-1}\otimes\Omega_{22.1})$, which is a variance mixture of Normals. This shows that the dependence between $W_2 $ and $W_1 $ is cleansed by correcting $x_t$ for the fitted values of $\psi'\Delta x_t$ on $d_t$, similarly to the heteroskedasticity and autocorrelation-consistent and LRV estimators in \citet{Phi:05} and the IV estimator in \citet{Phi:14}.
\end{remark}

The limiting mixed-Gaussianity of the ICC estimators in Theorem \ref{theorem_asy_distr_psi} can be exploited for testing hypothesis on $\psi_*$ and $\beta_*$ with Wald-type statistics, using consistent estimators of the LRV $\Omega_{22.1}$, see \eqref{eq_Xi}. Among others, in line with \cite{Phi:05,Phi:14}, the estimator $\wh\Omega_{22.1}:=\wh\Omega_{22}-\wh\Omega_{21}\wh\Omega_{11}^{-1}\wh\Omega_{12}$ could be used, with
\begin{equation}\label{eq_Xihat}
\wh\Omega:=
\left(
\begin{array}{cc}
\wh\Omega_{11}&\wh\Omega_{12}\\
\wh\Omega_{21} &\wh\Omega_{22}\end{array}%
\right)
:=\frac TK\left(\begin{array}c
\bar{a}'M_{\Delta x d}\\
h'M_{xd}
\end{array}
\right)M_{dd}^{-1}\left(\begin{array}c
\bar{a}'M_{\Delta x d}\\
h'M_{xd}
\end{array}
\right)' ,\quad
a=\wh\psi,\,\,\, h=\wh\beta.
\end{equation}

\begin{theorem}[Consistency of $\wh\Omega$]\label{theorem_Xihat_cons}
Let $d_t$ in \eqref{eq_d} be constructed using any orthonormal \cadlag\ basis of $L^2[0,1]$ and let $\wh\Omega$ be as in \eqref{eq_Xihat}; then $\wh\Omega\convp\Omega$ as $(T,K)_{seq}\toinf$, and as a consequence also $\wh\Omega_{22.1}\convp\Omega_{22.1}$.
\end{theorem}

Consider the following hypothesis on the unrestricted coefficients of $\psi$,
\begin{equation}\label{eq_H0_psi}
H_0:R'\veca(\psi_*) =h,
\end{equation}
where $R$ is a given $sr\times m$ matrix of full column rank and $h$ is a given $m \times 1$ vector.

\begin{remark}[Hypotheses on $\psi_*$ and $\psi$]\label{rem_fullrank}
Recall that $\psi_* := \bar c' \psi$ so that \eqref{eq_H0_psi} corresponds to a linear hypotheses on $\psi$ of the form $R_0'\veca(\psi)=h$ with $R_0':=R' (I_s\otimes\bar c')$; $R_0$ is also of full column rank $m$ because $R$ and $c$ are of full column rank. Symmetrically, one can write $-h=R'\veca(\beta_*') = R_1'\veca(\beta')$, where $R_1':=R' (\bar b'\otimes I_r)$ is of full row rank $m$.
\end{remark}

\begin{remark}[Duality between hypothesis on $\psi$ and $\beta$]\label{rem_dual}
Note that by
Lemma \ref{lem_ident}.(iii) and (iv), or \eqref{eq_dual_psi}, one has $R'\veca(\psi_*) - h = R'\veca(\beta_*') + h$. This shows that 
\begin{equation}\label{eq_dual_H0}
H_0:R'\veca(\psi_*) = h\qquad\miff\qquad H_0:R'\veca(\beta_*') = -h.
\end{equation}
\end{remark}

The next theorem discusses the Wald test of \eqref{eq_dual_H0}.

\begin{theorem}[Wald tests based on the ICC estimators]\label{theorem_inf_psi} Let the assumptions of Theorem \tref{theorem_asy_distr_psi} and the $m$ linear restrictions on $\psi_*$ and $\beta_*$ in \eqref{eq_dual_H0} hold. Let also $\wh\Omega_{22.1}$ be a consistent estimator of $\Omega_{22.1}$; then as $(T,K)_{seq}\toinf$ the Wald test statistic $Q$ satisfies
\begin{equation}\label{eq_Q}
\begin{array}l
Q:=T^2(R'\veca(\wh\psi_*)-h)'(R'\myV R)^{-1}(R'\veca(\wh\psi_*)-h)\\
\quad\:=T^2
(R'\veca(\wh\beta_*') +h)'
(R'\myV R)^{-1}
(R'\veca(\wh\beta_*') +h)\convw\chi^2_m,
\end{array}
\end{equation}
where $\myV:=(T^{-1}\bar a'M_{xx}\bar a)^{-1}\otimes\wh\Omega_{22.1}\convw\left(\int W_1W_1'\right)^{-1}\otimes\Omega_{22.1}$ with $a=\wh\psi$. Similarly when $m=1$, the $t$-ratio statistic satisfies
\begin{equation}\label{eq_t_ratio}
\frac{T(R'\veca(\wh\psi_*)-h)}{\sqrt{R'\myV R}}=\frac{T(R'\veca(\wh\beta_*')+h)}{\sqrt{R'\myV R}}\convw N(0,1).
\end{equation}
\end{theorem}

\subsection{Efficiency}\label{sec_eff}

This section compares $Z$ in \eqref{eq_asy_iter} with the distribution in \citet{Joh:96} of the ML estimator for VAR processes and with the one in \citet{Phi:14} of the optimal IV estimator in the semiparametric case; it is shown that $Z$ coincides with both of them, thus proving that the ICC estimators are asymptotically efficient.

\begin{theorem}[ML in \citet{Joh:96}]\label{theorem_MLE}
Assume that the DGP is a VAR satisfying the $I(1)$ condition in \textnormal{\citet{Joh:96}}; then the DGP has representation \eqref{eq_DX} with $C(z)$ satisfying Assumption \tref{ass_DGP} and the asymptotic distribution $Z$ in \eqref{eq_asy_iter} coincides with the asymptotic distribution of the corresponding ML estimator of $-\beta_\ast'$ in \textnormal{\citet[][Chapter 13]{Joh:96}} and of $\psi_\ast$ in \textnormal{\citet{Par:97}}.
\end{theorem}

Consider next the semiparametric triangular form in \citet{Phi:14}
\begin{equation}\label{eq_PCB}
\begin{array}{l}
y_t=Az_t+u_{0t}\\
\Delta z_t=u_{1t}
\end{array},
\end{equation}
where $u_t:=(u_{0t}',u_{1t}')'=Q(L)\varepsilon_t\sim I(0)$ with $Q(1)$ nonsingular,
$y_t$ is $r \times 1$ and $z_t$ is $s \times 1$. This corresponds to a partition of
$x_t$ into $(y_t',z_t')':=x_t$.

\begin{theorem}[IV in \citet{Phi:14}]\label{theorem_IV}
Assume \eqref{eq_DX} holds with $C(z)$ satisfying Assumption \tref{ass_DGP}; then $Q(z)$ satisfies the summability condition $(\textbf{L})$ in \textnormal{\citet{Phi:14}} and the asymptotic distribution $Z$ in \eqref{eq_asy_iter} coincides with the asymptotic distribution of the corresponding IV estimator of $A:=-\beta_\ast'$ in \textnormal{\citet{Phi:14}}.
\end{theorem}

\section{Monte Carlo simulations}\label{sec_M}

This section reports a set of Monte Carlo (MC) simulations. The DGP is taken from \citet{OW:18} and \citet{BG:22} and data is generated from
\begin{equation}\label{eq_DGP_OW}
\Delta X_t = \alpha \beta' X_{t-1} + \varepsilon_t, \qquad \varepsilon_t\sim \myiid N(0,I_p),\qquad t=1,\dots,T,\qquad \Delta X_0=X_0=0,
\end{equation}
with $\beta=(I_{p-s},0)'$ and $\alpha=-a\beta$. The DGP depends on the values of $(p,T,s,a)$ and identification is achieved with $b=(0,I_s)'$ and $c=(I_r,0)'$ except for $s=p$, where $\alpha$ and $\beta$ are undefined and $a$ plays no role.

When $s<p$, the first $p-s$ coordinate processes of $X_t$ in \eqref{eq_DGP_OW} are univariate AR(1)'s with coefficient $1-a$ and the remaining $s$ are pure random walks. In terms of \eqref{eq_X}, this implies that $\psi = (0,I_s)'$, $ \kappa' = (0,I_s)$, $C_1(L)=\diag(0,(1-(1-a)L)^{-1}I_r)$, so that the relevant LRV is $\Omega_{22.1}= a^{-2}I_r$. One expects small-sample inference on $s$ and $\psi$ to be more difficult when $a$ is smaller, because one needs to distinguish unit roots from autoregressive roots equal to $1-a$.

The properties of the max-gap and sequential estimators of $s$, see \eqref{eq_def_shat} and \eqref{eq_def_F1_Finf}, are simulated with $K=\lceil T^{3/4} \rceil$ for the following values of $(p,T,s,a)$: $p=10,20,50,100,300$, $T=(10,20,30)p$, $s=\lceil jp / 4 \rceil$, $j=0,1,2,3,4$, $a=0.25,0.5,0.75,1$. Tables \tref{table_max_gap}, \tref{table_mx} and \tref{table_tr} in the supplement report full MC results; Figure \tref{fig_freq_meadow} reports a representative subset for $p=20$.

\begin{figure}[htbp]
\includegraphics[width=\textwidth]{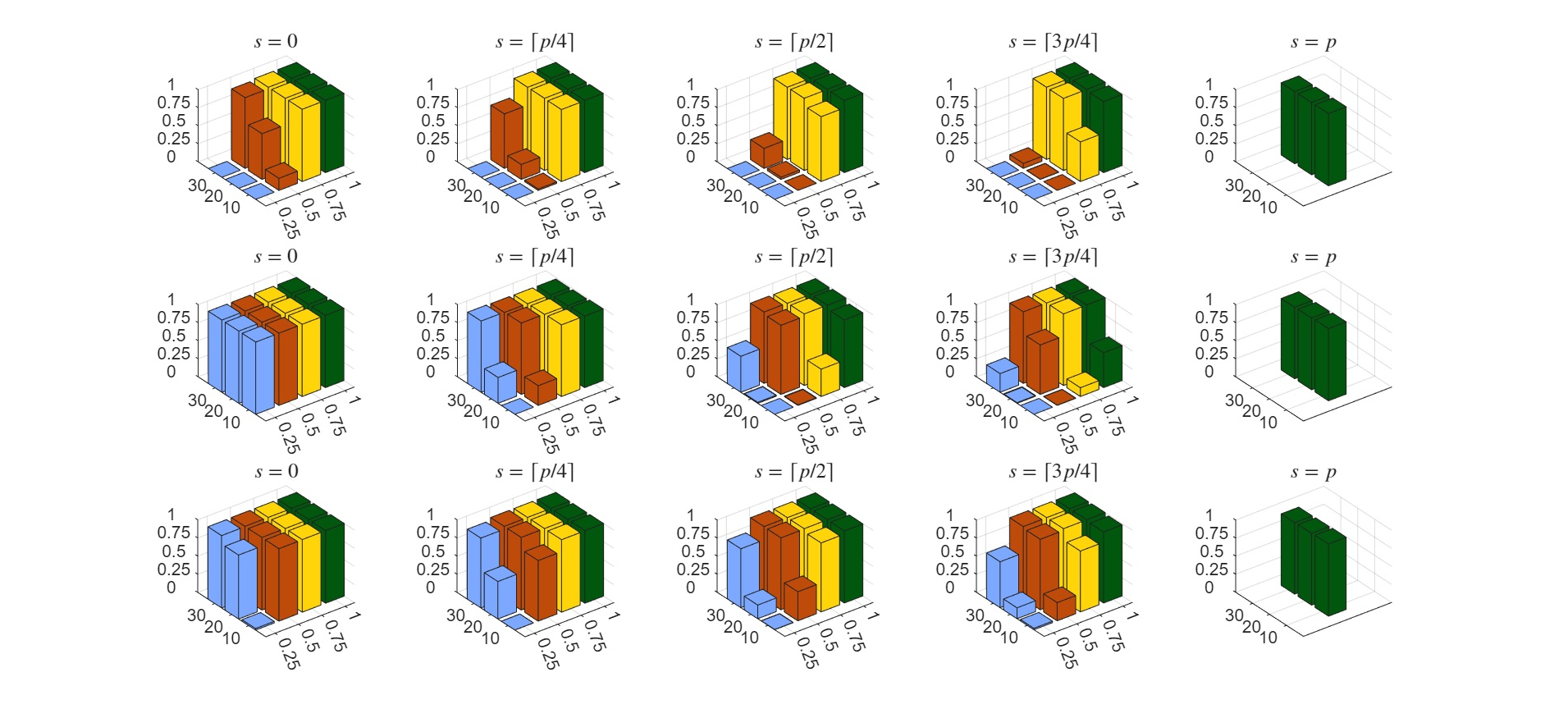}
\vspace{-5em}
\caption{\footnotesize MC relative frequency of correct selection (vertical axis) of $s=\lceil jp / 4 \rceil$, $j=0,1,2,3,4$, in a $p=20$ dimensional DGP \eqref{eq_DGP_OW} for $T=(10,20,30)p$, $a=0.25,0.5,0.75,1$, and {\Ntext} replications. Top row: max-gap estimator. Middle row: estimator based on test sequence $\{F_{j,1}\}_{j=p,\dots,1}$. Bottom row: hybrid estimator based on the test $F_{p,1}$ and the restricted max-gap estimator over $j\in\{0,\dots,p-1\}$. Significance level: $5\%$.}
\label{fig_freq_meadow}
\end{figure}

The top row in Figure \tref{fig_freq_meadow} reports results for the max-gap estimator. In absence of stationary AR components, i.e. when $s=p$ or $a=1$, the max-gap estimator selects the correct number of common trends $s$ with frequency 1 from $T=10p$ onward. When $a=0.75$ and $a=0.5$ the stationary AR components are increasingly autocorrelated, and the correct selection of $s$ with frequency 1 is observed for higher values of $T$. Finally, when $a=0.25$ and there are strong stationary AR components, the max-gap estimator never select the correct $s$ up to $T=30p$.

Overall, for fixed $(p,T,s)$ the performance of the max-gap worsens for lower values of $a$ because convergence to 0 of the smaller \SCC\ $\lambda_{j}$, $s<j\leq p$, appears to be slower for lower values of $a$, see Fig. \ref{fig_OW_eigs}.

The middle row in Fig. \tref{fig_freq_meadow} shows that when the max-gap estimator does not perform well, the estimator based on the test sequence $\{F_{j,n}\}_{j=p,\dots,1}$, $n=1$, has more reliable behavior (similar results hold for $n=\infty$, see Table \ref{table_mx}). Table \ref{table_mx} and \ref{table_tr} in the Supplement further document that the tests $F_{s,1}$ and $F_{s,\infty}$ are undersized in finite samples.

One can combine inference based on the max-gap estimator and on the test $F_{p,n}$, defining a hybrid estimator $\wh s _{n}$ that selects $p$ when $F_{p,n}$ does not reject, and uses the restricted max-gap estimator over $\{0,\dots,p-1\}$. The bottom row of Fig. \tref{fig_freq_meadow} shows that the hybrid estimator with $n=1$ performs well (similar results hold for $n=\infty$, see Tables \ref{table_hyb_mx} and \ref{table_hyb_tr}).

Next consider the Wald $Q$ and $t$-ratio statistics for linear hypotheses on $\psi$, see Theorem \tref{theorem_inf_psi}. Two hypotheses of type \eqref{eq_H0_psi} are considered, namely (i) the element 1,1 of $\psi_*$ equals 0, tested with the $t$-statistic in \eqref{eq_t_ratio}, and (ii) the first column of $\psi_*$ equals 0, tested using the Wald $Q$ statistic in \eqref{eq_Q} with a $\chi^2_r$ weak limit; both hypotheses are true under the DGP.

The main challenge in the implementation of the tests lies in the precise estimation of the LRV $\Omega_{22.1}$, which is notoriously difficult by nonparametric means, see e.g. \cite{Haug:02}. Three different estimators of $\Omega_{22.1}$ are considered in the computations of the test statistics: the first one is the true value in the DGP for $\Omega_{22.1}=a^{-2}I_r$, called LRV-U (unfeasible); the second option employs $\wh\Omega_{22.1}$ in \eqref{eq_Xihat} following \citet{Phi:14}, called LRV-P, and the third one (LRV-A) computes $\wh\Omega_{22.1}$ as in \cite{An:91} and \cite{AM:92} using Parzen's kernel.

\begin{figure}[htbp]
\includegraphics[width=\textwidth]{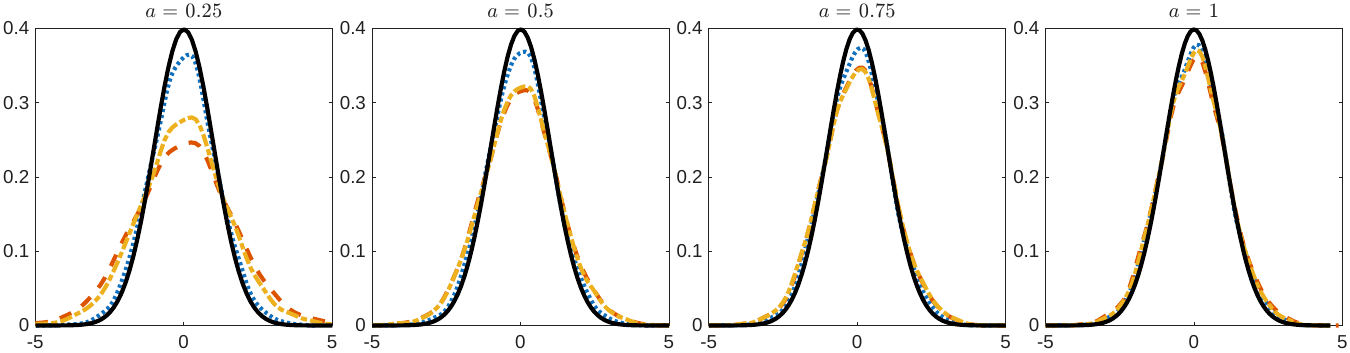}
\vspace{-3em}
\caption{\footnotesize Finite sample distribution of the $t$-ratio for the hypothesis that element 1,1 of $\psi_*$ equals 0 in DGP \eqref{eq_DGP_OW} for $p=20$, $T=30p$, $s=\lceil p / 4 \rceil$, $a=0.25,0.5,0.75,1$ and {\Ntext} replications. Solid black line: $N(0,1)$, dotted blue line: $t$-ratio with LRV-U, dashed red line: $t$-ratio with LRV-P, dot-dashed yellow line: $t$-ratio with LRV-A.}
\label{fig_distr_p_20_T_600_s_5}
\end{figure}

The properties of the tests are simulated with $K=\lceil T^{3/4} \rceil$ for the following values of $(p,T,s,a)$: $p=10,20$, $T=(30,60,90)p$, $s=\lceil jp /4 \rceil$, $j=1,2,3$, $a=0.25,0.5,0.75,1$. Table \tref{table_W1_Wr} in the supplement reports full Monte Carlo results, and Figure \tref{fig_distr_p_20_T_600_s_5} reports the subset for $T=30p$ and varying $a$. It can be seen that the tests are largely oversized in finite samples, in line with the literature \citep{Haug:02}. The unfeasible $t$-ratio LRV-U performs considerably better, while the behavior of LRV-P and LRV-A is comparable, thus suggesting that a major part of the size distortions are due to the imprecise LRV estimation. Finally, the $N(0,1)$ approximation improves with increasing $a$ (i.e. with decreasing level of autocorrelation of the stationary AR components.)

\section{Empirical application}\label{sec_E}

This section provides an illustration on a panel of daily exchange rates between Jan 4, 2022 and Aug 30, 2024 of the US dollar against 20 World Markets (WM) currencies, see \citet{OW:19} for a similar dataset and a review of the literature on the topic. The sample size is $T=667$ and $K=\lceil T^{3/4}\rceil = 132$. The illustration of the present methods is compared with the high dimensional VAR analysis of \citet{OW:18} and \citet{BG:24} for inference on $s$, and with the likelihood analysis of \citet{Joh:96} for inference on $s$ and $\psi$. Details are reported in the supplement.

\begin{figure}[htbp]
\includegraphics[width=\textwidth]{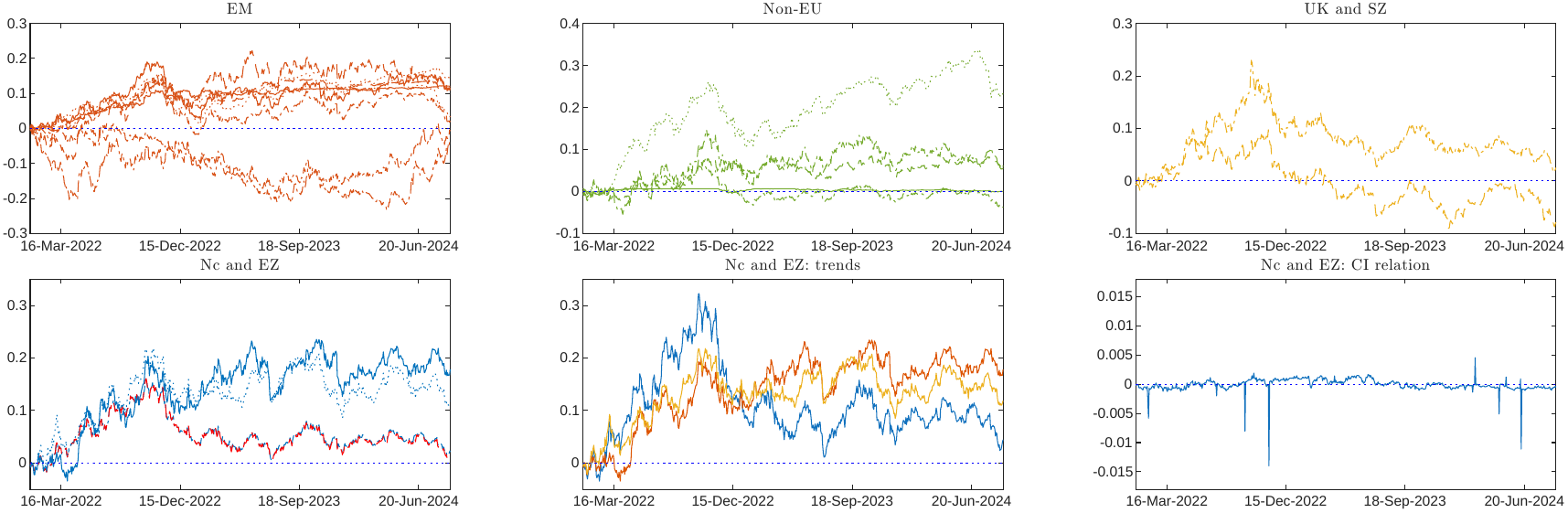}\vspace{-3em}
\caption{\footnotesize Top panels: EM, Non-EU, UK and SZ time series. Bottom left: Nc and EZ time series, Euro in red. Center bottom: Nc and EZ common stochastic trends $\wh \psi'X_t$. Bottom right: Nc and EZ cointegrating relation $\wh \beta'X_t$.}
\label{fig_data}
\end{figure}

Data (in logs and normalized to start at 0) are plotted in the first four panels of Figure \tref{fig_data}. Countries are grouped as follows: Emerging Markets (EM: Brazil, China, India, Malaysia, Mexico, South Africa, South Korea, Taiwan, Thailand) and Developed Markets (DM). DM are broken down into non European (Non-EU: Australia, Canada, Hong Kong, Japan, Singapore) and European (EU). EU is partitioned into UK and SZ (United Kindom and Switzerland), and Nordic and Eurozone (Nc and EZ: Denmark, Eurozone, Norway, Sweden).

\begin{figure}[htbp]
 \includegraphics[width=\textwidth]{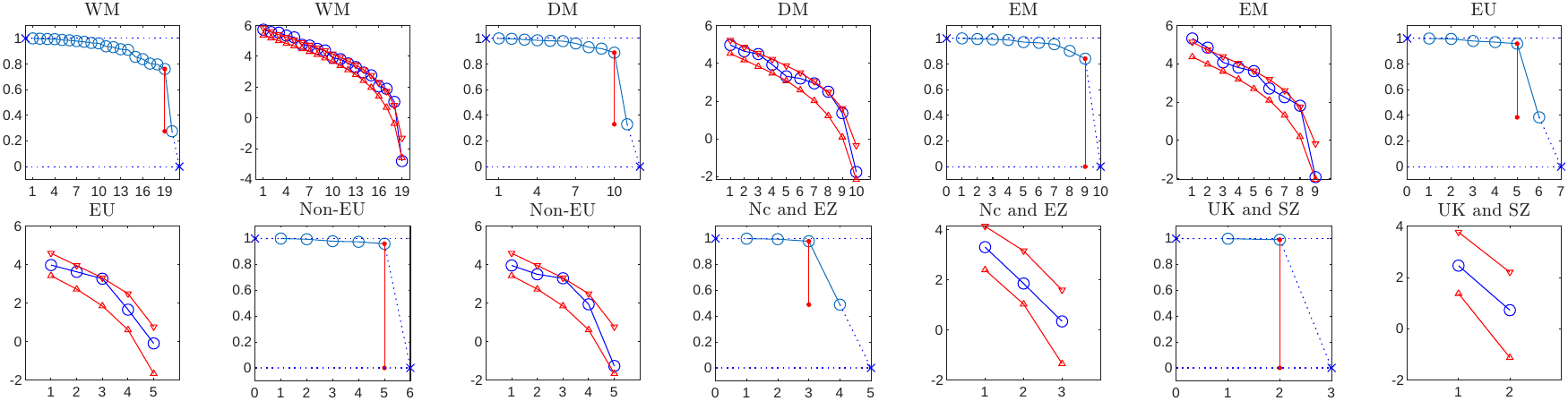}
 \vspace{-3em}
\caption{\footnotesize Pair of graphs for each \CCA; \SCC\ profile (first graph) with largest drop in red, misspecification stripe at 95\% confidence level (second graph). Numbering in the text is left-to-right, top-to-bottom.}
\label{fig_all_XR}
\end{figure}

The \CCA\ analysis applied to all the $p=20$ WM time series is summarized in the first two subgraphs in Figure \ref{fig_all_XR}, which report the associated \SCC\ profile and the misspecification stripe of Remark \tref{rem_stripe} at 95\% confidence level. The largest drop in the \SCC\ profile is between eigenvalue 19 and 20 (red vertical line), so that the max-gap estimator selects $\wh s = 19$, indicating $\wh r = 1$ cointegrating relation (CI).

The result on the presence of cointegration among the 20 WM currencies is further investigated using Wachter plots as in \citet{OW:18} and \citet{BG:24}, and the test of no cointegration ($s=p$) of \citet{BG:24} in a VAR$(k)$, $k=1,2,3,4$. For all values of $k$, Figure \tref{fig_BG} in the supplement displays eigenvalues to the right of the support of the Wachter distribution, which is an indication of the presence of cointegration ($s<20$). This is further confirmed by the results of their test at 0.01 significance level, see Table \tref{table_BG} in the supplement.

Thanks to the dimensional coherence of the present approach, the present analysis can be performed on subgroups separately, as indicated in Figure \tref{fig_tree}.

\begin{figure}[htbp]
	\begin{center}
		\footnotesize
		\begin{tikzpicture}
			[level distance=3.5cm]
			\node[align=center,rectangle,draw] {WM \\[-1em]$p=20,\wh s = 19$} 
			[edge from parent fork right,grow=right]
			child {node[align=center,rectangle,draw] {EM\\[-1em]$p=9,\wh s = 9$}}
			child {node[align=center,rectangle,draw] {DM\\[-1em]$p=11,\wh s = 10$}
				child {node[align=center,rectangle,draw] {Non-EU\\[-1em]$p=5,\wh s = 5$}}
				child {node[align=center,rectangle,draw] {EU\\[-1em]$p=6,\wh s = 5$}
					child {node[align=center,rectangle,draw] {UK and SZ\\[-1em]$p=2,\wh s = 2$}}
					child {node[align=center,rectangle,draw] {Nc and EZ\\[-1em]$p=4,\wh s = 3$}}}
			};
		\end{tikzpicture}
	\end{center}
	\vspace{-3em}
	\caption{\footnotesize Groups partitioning with dimension $p$ and max-gap estimator $\wh s$.}
	\label{fig_tree}
\end{figure}

The max-gap estimator finds no CI relations in EM and 1 CI relation in DM, see subgraphs 3-4 and 5-6 in Figure \ref{fig_all_XR}. In the next split into EU and Non-EU, see Figure \tref{fig_tree}, the max-gap estimator indicates no CI in Non-EU and 1 CI relation in Euro, see subgraphs subgraphs 7-8 and 9-10 in Figure \ref{fig_all_XR}. Finally the max-gap estimator indicates no CI in the UK and SZ group and 1 CI relation in the Nc and EZ group, see subgraphs 11-12 and 13-14 in Figure \ref{fig_all_XR}.

Next the Nc and EZ subgroup is used to estimate $\psi$ and $\beta$, with identification restrictions $c=e_2$ and $b=(e_1,e_3,e_4)$, where $e_i$ indicates the $i$-th unit vector; this identification restriction is tested using the criterion in \eqref{eq_test_on_c} and it is not rejected. The resulting $\wh\psi$ and $\wh \beta$ are\footnote{Danish Krone (DK), Eurozone (Euro), Norwegian Krone (NK), Swedish Krone (SK).}
\begin{equation}\label{eq_psi_hat}
\footnotesize
\wh\psi=\left(
\begin{array}{ccc}
	1 & 0 & 0\\
	\underset{(< 10^{-4})}{1.0147} & \underset{(< 10^{-4})}{-0.0153} & \underset{(0.7256)}{-0.0012}\\
	0 & 1 & 0\\
	0 & 0 & 1
\end{array}\right),\qquad
\wh \beta=\left(
\begin{array}{c}
	\underset{(< 10^{-4})}{-1.0147}\\
	1\\
	\underset{(< 10^{-4})}{0.0153} \\
	\underset{(0.7256)}{0.0012}
\end{array}\right),\qquad
X_t=\left(\begin{array}{c}
	\text{DK}_t \\
	\text{Euro}_t \\
	\text{NK}_t\\
	\text{SK}_t
\end{array}\right),
\end{equation}
with $p$-values reported in parenthesis for the unrestricted coefficients $\psi_*=e_2'\wh\psi=-\beta_*'$, which correspond to the second row of $\wh\psi$. The hypothesis $H_0:\psi_{2,1}=0$ and $H_0:\psi_{2,2}=0$ are strongly rejected while $H_0:\psi_{2,3}=0$ is not, i.e. the Euro loads the first and second stochastic trends but not the third trend, which only affects the Swedish Krone.

For the Nc and EZ subgroup, Table \ref{table_VAR} in the supplement reports the likelihood ratio test on $s$ and the QML estimate of $\beta$ in \citet{Joh:96}. The estimate of $s$ is 3 as in \eqref{eq_psi_hat}, and the estimates of $\beta$ are essentially indistinguishable.

The time series of the Nc and EZ common stochastic trends $\wh \psi'X_t$ and 
cointegrating relation $\wh \beta'X_t$ are reported in Figure \ref{fig_data}; the plot of $\wh \beta'X_t$ includes some large infrequent shocks, whose analysis, however, goes beyond the scope of the present illustration.

\section{Conclusions}\label{sec_conc}

This paper proposes a unified semiparametric framework for inference on the number and the loadings of stochastic trends (and on their duals, the number and the coefficients of the cointegrating relations) for $I(1)/I(0)$ processes, which includes VARIMA and Dynamic Factor Model processes. The setup directly applies also to subsets of variables or to their aggregation, i.e. it is dimensionally coherent.

The approach employs a novel canonical correlation analysis between the observed multiple time series and a finite subset of an $L^2[0,1]$ basis. Canonical correlations deliver test sequences for the number of stochastic trends, as well as consistent estimators of it. Canonical variates provide estimators of the loading matrix of stochastic trends and of their duals, the matrix of cointegrating coefficients. The ICC estimators are $T$-consistent, mixed-Gaussian and efficient. Wald tests on the parameters are developed, as well as misspecification tests for checking model assumptions.

The finite sample properties of the estimators and tests investigated via a Monte Carlo simulation study display reasonable performance.
An empirical analysis illustrates the range of available inferences in the proposed framework.

\bibliographystyle{Chicago}

\cp

\appendix
\setcounter{table}{0}
\renewcommand{\thetable}{A.\arabic{table}}
\setcounter{figure}{0}
\renewcommand{\thefigure}{A.\arabic{figure}}

\section{Lemmas}\label{sec_app_lemmas}

The following lemma collects well known facts on linear processes, see \citet{PS:92} or Theorem B.13 in \citet{Joh:96} (Thm SJ.B.13 henceforth).

\begin{lemma}[Limits for fixed $K\geq p$ and $T\toinf$]\label{lem_asympt_T}
Let $x_{1t}:=\bar\psi' x_t$ and $x_{0t}:=\beta' x_t$, and consider $\varphi_K(u):=(\phi_1(u),\dots,\phi_K(u))'$ for any orthonormal \cadlag\ basis
$\{\phi_i\}_{i=1}^{\infty}$ of $L^2[0,1]$; for fixed $K\geq p$ and $T\toinf$, one has that, jointly:
\begin{itemize}

\item[(i)] $T^{-\frac 12}\sum_{t=1}^{\lfloor Tu\rfloor}x_{0 t}\convw W_2(u):=\beta'C_1W_\varepsilon(u)$ in $D_r[0,1]$;

\item[(ii)] $T^{-\frac 12}x_{1\mfl}\convw W_1(u):=\kappa'W_\varepsilon(u)$ in $D_s[0,1]$;

\item[(iii)] $\varphi_K(\mfl/T)\conv\varphi_K(u)$ in $D_K[0,1]$ and $L_K^2[0,1]$;

\item[(iv)] $T^{-1}\bar\psi'M_{xx}\bar\psi=T^{-1}\sum_{t=1}^T(T^{-\frac 12}x_{1 t})(T^{-\frac 12}x_{1 t})'\convw\ell_{11}:=\int W_1 W_1'
$;

\item[(v)] $\bar\psi'M_{xx}\beta=T^{-\frac 12}\sum_{t=1}^T(T^{-\frac 12}x_{1 t})x_{0 t}'=O_p(1)$;

\item[(vi)] $\beta'M_{xx}\beta=T^{-1}\sum_{t=1}^Tx_{0 t}x_{0 t}'\convp Q_{00}:=\E(x_{0 t}x_{0 t}')>0$;

\item[(vii)] $T^{-\frac 12}\bar\psi'M_{xd}=T^{-1}\sum_{t=1}^T(T^{-\frac 12}x_{1 t})\varphi_K(t/T)'\convw\ell_{1\varphi}:=\int W_1 \varphi_K'
$;

\item[(viii)] $T^{1/2}\beta'M_{xd}=T^{-\frac 12}\sum_{t=1}^T x_{0 t}\varphi_K(t/T)'\convw\ell^\circ_{\rd2\varphi}:=\int\rd W_2 \varphi_K'$;

\item[(ix)] $T^{1/2}\bar\psi'M_{\Delta x d}=T^{-\frac 12}\sum_{t=1}^T\Delta x_{1 t}\varphi_K(t/T)'\convw\ell^\circ_{\rd1\varphi}:=\int\rd W_1 \varphi_K'$;

\item[(x)] $T^{1/2}\beta'M_{\Delta xd}=T^{-\frac 12}\sum_{t=1}^T\Delta x_{0 t}\varphi_K(t/T)'=o_p(1)$.

\end{itemize}
\end{lemma}
\begin{mproof}{\textit{Proof.}}
(i) From \eqref{eq_x} and $T^{-\frac 12}\sum_{t=1}^{\lfloor Tu\rfloor}x_{0 t}=T^{-\frac 12}\sum_{t=1}^{\lfloor Tu\rfloor}\beta'C_1\varepsilon_t+o_p(1)$ uniformly in $u\in[0,1]$. (ii) From \eqref{eq_x} and $T^{-\frac 12}x_{1\mfl}=T^{-\frac 12}\kappa'\sum_{t=1}^{\lfloor Tu\rfloor}\varepsilon_t+o_p(1)$ uniformly in $u\in[0,1]$. (iii) By the dominated convergence theorem for $L_K^2[0,1]$. (iv) By (B.19) in Thm SJ.B.13. (v) By (B.20) in Thm SJ.B.13. (vi) By the law of large numbers for linear processes. (vii) By (B.22) in Thm SJ.B.13. (viii)-(x) By (B.23) in Thm SJ.B.13.
\end{mproof}\smallskip

Let
\begin{equation}\label{eq_gt_def_sup}
g_t(a):=a' M_{\Delta x d}M_{dd}^{-1}d_t, \qquad e_t(a):=x_t-M_{xg}M_{gg}^{-1}g_t(a ).
\end{equation}
Note that substituting $a=\wh\psi^{(1)}$ in \eqref{eq_gt_def_sup} one finds \eqref{eq_gt_def}. The next lemma collects the limits of the sample moments $M_{ij}$ that appear in \eqref{eq_BEP} for $\CCA(f_t,d_t)$, $f=x,e$.

\begin{lemma}[$T\toinf$ limits for fixed $K\geq p$]\label{lem_asympt_T_BEP}
The following $T\toinf$ limits hold jointly and jointly with those in Lemma \tref{lem_asympt_T}, for fixed $K\geq p$. Consider $d_t$ in \eqref{eq_d} constructed using any orthonormal \cadlag\ basis of $L^2[0,1]$ and define $A_T=(T^{-\frac 12}\bar\psi,\beta)$; one has that:
\begin{itemize}

\item[(i)] $A_T'M_{xx}A_T\convw\diag(\ell_{11},Q_{00})$;

\item[(ii)] $M_{dd}=T^{-1}\sum_{t=1}^T\varphi_K(t/T)\varphi_K(t/T)'\conv\int\varphi_K\varphi_K'=I_K$;

\item[(iii)] $A_T'M_{xd}\convw(\ell_{1\varphi}',0')'$.
\end{itemize}
Let $g_t$, $e_t$ be $g_t(\psi)$ and $e_t(\psi)$ in \eqref{eq_gt_def_sup} using the true parameter value of $\psi$;\footnote{Points (iv) and (xii) involve $\beta$, which is a function of $\psi$, $\beta= (I-b \psi')\bar c$ in the present notation, see \citet{Par:97} eq. (B.1).} then one has that:
\begin{itemize}

\item[(iv)] $a_4(\psi):=T\beta'M_{xg}\convw\ell^\circ_{\rd2\varphi}\ell^\circ_{\varphi\rd1}$;

\item[(v)] $a_5(\psi):=TM_{gg}\convw\ell^\circ_{\rd1\varphi}\ell^\circ_{\varphi\rd1}$;

\item[(vi)] $a_6(\psi):=T^{1/2}M_{gd}\convw\ell^\circ_{\rd1\varphi}$;

\item[(vii)] $a_7(\psi):=A_T'M_{xg}M_{gg}^{-1}M_{gx}A_T\convw\diag(\ell_{1\varphi}U_\circ\ell_{\varphi1},0)$, where $U_\circ:=\ell^\circ_{\varphi\rd1}(\ell^\circ_{\rd1\varphi}\ell^\circ_{\varphi\rd1})^{-1}\ell^\circ_{\rd1\varphi}$;

\item[(viii)] $a_8(\psi):=A_T'M_{xg}M_{gg}^{-1}M_{gd}\convw ((\ell_{1\varphi}U_\circ)',0')'$;

\item[(ix)] $a_9(\psi):=M_{dg}M_{gg}^{-1}M_{gd}\convw U_\circ$;

\item[(x)] $a_{10}(\psi):=A_T'M_{ee}A_T\convw\diag(\ell_{11}-\ell_{1\varphi}U_\circ\ell_{\varphi1},Q_{00})$;

\item[(xi)] $a_{11}(\psi):=A_T'M_{ed}\convw ((\ell_{1\varphi}P_\circ)',0')'$, where $P_\circ:=I_K-U_\circ$;

\item[(xii)] $a_{12}(\psi):=T^{1/2}\beta'M_{ed}\convw\ell^\circ_{\rd2\varphi}P_\circ$.
\end{itemize}
Finally, local tightness results hold for the sample moments in items \textnormal{(vii)-(xii)}, in the following sense.
Let $\wt\psi$ be any estimator of $\psi$ such that for an arbitrary $L>0$
\begin{equation}\label{eq_L}
\wt\psi=\psi (h_1+ h_2 o_p(1))+\beta h_3 o_p(T^{-\frac 12}),\quad
\eig_{\min}(h_1'h_1)\asgeq L^{-1}, \quad \| h_i\| \asleq L, \quad i=1,2,3,
\end{equation}
with $h_1,h_2\in \mathbb{R}^{s\times s}$ and $h_3\in \mathbb{R}^{r\times r}$, and where $\eig_{\min}$ is the minimal eigenvalue of the argument matrix. Let $\tilde g_t:=g_t(\wt\psi)$, $\tilde e_t:=e_t(\wt\psi)$ be the quantities $g_t, e_t$ in \eqref{eq_gt_def_sup} calculated for this estimator $\wt\psi$; then the replacement of $e,g$ by $\tilde{e},\tilde{g}$ in items \textnormal{(vii)-(xii)} produces $o_p(1)$ perturbations uniformly over $h_1,h_2,h_3$, i.e. $\| a_i(\wt \psi)-a_i(\psi)\|=o_p(1)$, $i=4,\dots,12$.
\end{lemma}
\begin{mproof}{\textit{Proof.}}
(i) By Lemma \tref{lem_asympt_T}(iv),(v),(vi), one finds
$$
A_T'M_{xx}A_T=
T^{-1}\sum_{t=1}^T\left(\begin{array}{cc}
(T^{-\frac 12}x_{1 t})(T^{-\frac 12}x_{1 t})' & (T^{-\frac 12}x_{1 t})x_{0 t}'\\
x_{0 t}(T^{-\frac 12}x_{1 t})' & x_{0 t}x_{0 t}'
\end{array}\right)\convw\left(\begin{array}{cc}
\ell_{11} & 0\\
0 & Q_{00}
\end{array}\right).
$$
(ii) By the continuous mapping theorem and orthonormality of the basis in \eqref{eq_d_L2}. (iii) By Lemma \tref{lem_asympt_T}(vii),(viii), one finds
$$
A_T'M_{xd}=
T^{-1}\sum_{t=1}^T\left(\begin{array}c
(T^{-\frac 12}x_{1 t})d_t'\\
x_{0 t}d_t'
\end{array}\right)\convw\left(\begin{array}c
\ell_{1\varphi}\\
0
\end{array}\right).
$$
(iv) By (ii) and Lemma \tref{lem_asympt_T}(viii), (ix), one finds
$$
T\beta'M_{xg}=\left(T^{-\frac 12}\sum_{t=1}^Tx_{0 t}d_t'\right)M_{dd}^{-1}T^{1/2}M_{d\Delta x_1}\convw\ell^\circ_{\rd2\varphi}\ell^\circ_{\varphi\rd1}.
$$
(v) By (ii) and Lemma \tref{lem_asympt_T}(ix), one finds $TM_{gg}=(T^{1/2}M_{\Delta x_1 d})M_{dd}^{-1}(T^{1/2}M_{d\Delta x_1})\convw\ell^\circ_{\rd1\varphi}\ell^\circ_{\varphi\rd1}$. (vi) By Lemma \tref{lem_asympt_T}(ix), one finds $T^{1/2}M_{gd}=T^{-1}\sum_{t=1}^T(T^{1/2}g_t)d_t'=T^{1/2}M_{\Delta x_1 d}\convw\ell^\circ_{\rd1\varphi}$. (vii) Consider $A_T'M_{xg}M_{gg}^{-1}M_{gx}A_T$, where
\begin{align}
T^{1/2}A_T'M_{xg}&=
T^{-1}\sum_{t=1}^T\left(\begin{array}c
(T^{-\frac 12}x_{1 t})(T^{1/2}g_t)'\\
x_{0 t}(T^{1/2}g_t)'
\end{array}\right)=T^{-1}\sum_{t=1}^T\left(\begin{array}c
(T^{-\frac 12}x_{1 t})d_t'\\
x_{0 t}d_t'
\end{array}\right)M_{dd}^{-1}(T^{1/2}M_{d\Delta x_1}),\nonumber\\
&\convw\left(\begin{array}c
\ell_{1\varphi}\\
0\end{array}\right)\ell^\circ_{\varphi\rd1}\label{eq_ioo}
\end{align}
by (ii) and Lemma \tref{lem_asympt_T}(vii),(viii),(ix); hence, by (v) one has
$$
A_T'M_{xg}M_{gg}^{-1}M_{gx}A_T=(T^{1/2}A_T'M_{xg})(TM_{gg})^{-1}(T^{1/2}M_{gx}A_T)\convw\left(\begin{array}{cc}
\ell_{1\varphi}\ell^\circ_{\varphi\rd1}(\ell^\circ_{\rd1\varphi}\ell^\circ_{\varphi\rd1})^{-1}\ell^\circ_{\rd1\varphi}\ell_{\varphi1} & 0\\
0 & 0\end{array}\right).
$$
(viii) By \eqref{eq_ioo}, (v) and (vi), one has
$$
A_T'M_{xg}M_{gg}^{-1}M_{gd}=(T^{1/2}A_T'M_{xg})(TM_{gg})^{-1}(T^{1/2}M_{gd})\convw\left(\begin{array}c
\ell_{1\varphi}\ell^\circ_{\varphi\rd1}(\ell^\circ_{\rd1\varphi}\ell^\circ_{\varphi\rd1})^{-1}\ell^\circ_{\rd1\varphi}\\
0
\end{array}\right).
$$
(ix) By (v) and (vi), one has $M_{dg}M_{gg}^{-1}M_{gd}=(T^{1/2}M_{dg})(TM_{gg})^{-1}(T^{1/2}M_{gd})\convw\ell^\circ_{\varphi\rd1}(\ell^\circ_{\rd1\varphi}\ell^\circ_{\varphi\rd1})^{-1}\ell^\circ_{\rd1\varphi}$. (x) By (ii) and (ix), using $M_{ee}=M_{xx}-M_{xg}M_{gg}^{-1}M_{gx}$. (xi) By (iii) and (viii), using $M_{ed}=M_{xd}-M_{xg}M_{gg}^{-1}M_{gd}$. (xii) By Lemma \tref{lem_asympt_T}(vii),(iv),(v),(vi), one has
\begin{align*}
T^{1/2}\beta'M_{ed}&=T^{1/2}\beta'M_{xd}-(T\beta'M_{xg})(TM_{gg})^{-1}(T^{1/2}M_{gd})\\
&\convw\ell^\circ_{\rd2\varphi}-\ell^\circ_{\rd2\varphi}\ell^\circ_{\varphi\rd1}(\ell^\circ_{\rd1\varphi}\ell^\circ_{\varphi\rd1})^{-1}\ell^\circ_{\rd1\varphi}.
\end{align*}
The tight versions of items (vii)-(xii) follow from the tight versions
of items (iv)-(vi). As to the latter,
$$
T\|\beta'(M_{xg}h_1-M_{x\tilde{g}})\|\leq o_p(1)\| T\beta'M_{xg}\|\| h_2\| + o_p(T^{-\frac12})\|\beta'M_{xd}\|\| M_{dd}^{-1}\|\|M_{d\Delta x} \beta\|\| h_3\|=o_p(1)
$$
by Lemma \ref{lem_asympt_T}(viii),(x) and Lemma \ref{lem_asympt_T_BEP}(ii),(iv); similarly, uniformly over $h_1,h_2,h_3$ satisfying \eqref{eq_L}, one has
\begin{align*}
&T\| (h_1'M_{gg}h_1)-M_{\tilde{g}\tilde{g}}\|=o_p(1),\\
&T^{1/2}\| h_1'M_{gd}-M_{\tilde{g}d}\|=o_p(1),\\
&T^{-1}\| (h_1'M_{gg}h_1)^{-1}-M_{\tilde{g}\tilde{g}}^{-1}\|\leq\| (Th_1'M_{gg}h_1)^{-1}\|\|(Th_1'M_{gg}h_1)-TM_{\tilde{g}\tilde{g}}\|\| (TM_{\tilde{g}\tilde{g}})^{-1}\| =o_p(1).
\end{align*}
Finally note that $a_7, a_8, a_9$ are invariant with respect to the transformation from $g_t$ to $h_1' g_t $ with $h_1$ square and nonsingular. This completes the proof.
\end{mproof}\smallskip

The tightness results in Lemma \ref{lem_asympt_T_BEP} imply that the limits in items (x)-(xii) therein are invariant to the
replacement of $e_t(\psi)$ with $e_t(\wt\psi)$ for any estimator $\wt\psi\in \mathbb{R}^{p\times s}$ of $\psi$ such that
$\bar\psi' \wt\psi$ converges in distribution to an a.s. nonsingular matrix $h_1$ and $\beta' \wt\psi = o_p(T^{-\frac12})$,
as $T\conv\infty$. This is the case of $\wt\psi=\wh\psi^{(1)}$ with $h_1=I_s$, see Theorem \tref{lem_cons_psi}.

The following lemma discusses the limit of $S(\ml):=\ml M_{ff}-M_{fd}M_{dd}^{-1}M_{df}$, $f=x,e$, that appears in \eqref{eq_BEP} for $\CCA(f_t,d_t)$,
$f=x,e$, where $e_t:=e_t(\psi)$ (or asymptotically equivalently, $e_t(\wt\psi)$ with $\wt\psi$ as in the previous paragraph).

\begin{lemma}[Limits for fixed $K\geq p$ and $T\toinf$]\label{lem_asympt_T_BEP_S}
Consider $A_T=(T^{-\frac 12}\bar\psi,\beta)$ and let $S(\ml):=\ml M_{ff}-M_{fd}M_{dd}^{-1}M_{df}$, $f=x,e$; for fixed $K\geq p$ and $T\toinf$, one has that
$$
A_T'S(\ml)A_T\convw\ml \left(\begin{array}{cc}
\varsigma_{11} & 0\\
0 & Q_{00}
\end{array}\right)-\left(\begin{array}{cc}
\varsigma_{1\varphi}\varsigma_{\varphi1} & 0\\
0 & 0
\end{array}\right),
$$
where
\begin{equation}\label{eq_def_varsigma}
\varsigma_{11}=
\left\{
\begin{array}{cl}
\ell_{11} &\mbox{ for } f=x\\
\ell_{11}-\ell_{1\varphi}U_\circ\ell_{\varphi1} &\mbox{ for } f=e \\
\end{array}\right.,\qquad\varsigma_{1\varphi}=\left\{
\begin{array}{cl}
\ell_{1\varphi} &\mbox{ for } f=x\\
\ell_{1\varphi}P_\circ &\mbox{ for } f=e \\
\end{array}\right. ,
\end{equation}
and $U_\circ$ and $P_\circ$ are as in Lemma \tref{lem_asympt_T_BEP}.
\end{lemma}
\begin{mproof}{\textit{Proof.}}
For $f=x$, by Lemma \tref{lem_asympt_T_BEP}(i),(ii),(iii), using $S(\ml)=\ml M_{xx}-M_{xd}M_{dd}^{-1}M_{dx}$; for $f=e$,
by Lemma \tref{lem_asympt_T_BEP}(ii),(x),(xi), using $S(\ml)=\ml M_{ee}-M_{ed}M_{dd}^{-1}M_{de}$.
\end{mproof}\smallskip

\begin{lemma}[Convergence to $\int B\partial W$]\label{lemma_ogawa}
Let $B$ and $W$ be two scalar standard Brownian motions, with $c_k=\int B\phi_k$ and $g^\circ_k=\int (\rd W)\phi_k$; then
as $K\toinf$ one has $\sum_{k=1}^{K}c_kg^\circ_k\overset{p}{\rightarrow }\int B\partial W$, a Stratonovich integral,
for any orthonormal basis $\{\phi_k\}_{k=1}^{\infty}$ of $L^2[0,1]$.
\end{lemma}
\begin{mproof}{\textit{Proof.}}
By linearity considerations, the claim follows from two special cases: (i) when $B=W$ is the same Brownian Motion, proved in \citet[][Example 3.1]{Og:17},
and (ii) when $B$ and $W$ are two independent standard Brownian Motions. Consider two steps to prove case (ii).

First, $\sum_{k=1}^{K}c_kg^\circ_k\overset{p}{\rightarrow }\int B\partial W$ for the \KLB\ by Wong-Zakai's theorem, see \citet{Twa:96}.
Second, it can be shown that the probability limit of $\sum_{k=1}^{K}c_kg^\circ_k$ is independent of the choice of an orthonormal basis.
Indeed, consider an orthonormal basis $\{\psi_k\}$ with $\gamma_k=\int B\psi_k$ and $\delta^\circ_k=\int \rd W\psi_k$.
Because of the independence of $B$ and $W$, the a.s. equality of $\sum_{k=1}^\infty c_kg^\circ_k$ and $\sum_{k=1}^\infty \gamma_k\delta^\circ_k$
can be established by conditioning on $B$ (effectively on $c_k,\gamma_k$).

Observe in fact that both $\{g^\circ_k\}$ and $\{\delta^\circ_k\}$ are i.i.d. $N(0,1)$, whereas $\sum_{k=1}^{K}c_k^2$ and
$\sum_{k=1}^{K}\gamma_k^2$ converge a.s.; hence it follows that $\sum_{k=1}^{K}c_kg^\circ_k$ and $\sum_{k=1}^{K}\gamma_k\delta^\circ_k$
converge in $L^2$-norm a.s. conditionally on $B$. Further,
\begin{align*}
\E_{B}\left( \sum_{k=1}^\infty c_kg^\circ_k-\sum_{k=1}^{%
\infty }\gamma_k\delta^\circ_k\right)^2
&=\underset{K\toinf}{\plim}\E_{B}\left( \sum_{k=1}^{K}c_k
g^\circ_k-\sum_{k=1}^{K}\gamma_k\delta^\circ_k\right)^2
\\
&=\underset{K\toinf}{\plim}\left\{ \sum_{k=1}^{K}(c_k^2+\gamma
_k^2)-2\sum_{k,l=1}^{K}c_k\gamma_{l}\int \phi_k\psi_{l}\right\}
\\
&= 2\int B^2-2\underset{K\toinf}{\plim}\int \left(
\sum_{k=1}^{K}c_k\phi_k\right) \left( \sum_{k=1}^{K}\gamma_k\psi
_k\right) \\
&=2\int B^2-2\int B^2=0,
\end{align*}
where $\E_{B}$ indicates the expectation operator conditional on $B$. This shows that $\sum_{k=1}^\infty c_kg^\circ_k=\sum_{k=1}^\infty \gamma_k\delta^\circ_k$ a.s. in conditional probability, and hence, also unconditionally.
\end{mproof}\smallskip

Lemma \tref{lem_asympt_K} below collects facts on large-$K$ asymptotic theory related to Parseval's theorem, the ergodic theorem and Ogawa's stochastic integration theory, see e.g. \citet{Ped:12} and \citet{Og:17}. Recall the $L^2$-expansion \eqref{eq_d_L2} of $W(u)$; by Parseval's theorem applied to almost all sample paths, it holds that
$$
\sum_{k=1}^\infty c_kc_k'=\int WW',\qquad c_k:=\int W\phi_k,
$$
where the series is a.s. convergent. If $W$ was continuously differentiable with gradient $\dot{W}$, again by Parseval's theorem it would hold that
\begin{equation}\label{dwy_dy}
\sum_{k=1}^\infty c_k^\circ c_k'=\int \dot{W}W',\qquad \sum_{k=1}^\infty c_k^\circ {c_k^\circ}'=\int\dot{W}\dot{W}'
\end{equation}
with $c_k^\circ:=\int \rd W\phi_k$. Although $W$ is not differentiable, $c_k^\circ$ are well-defined as stochastic integrals, and properties formally similar to the relations in \eqref{dwy_dy} do hold,
though with no connection to Parseval's theorem. On the one hand,
$$
\sum_{k=1}^\infty c_k^\circ c_k'=\int (\delta_{\phi}W)W',
$$
a noncausal stochastic integral with respect to the basis $\{\phi_k\}_{k=1}^\infty$ in the sense of \citet[][Chapter 3]{Og:17},
where the series on the l.h.s. is required to converge in probability. By Lemma \tref{lemma_ogawa} applied componentwise, it is seen
that $\int (\delta_{\phi}W)W'$ exists and equals the Stratonovich integral $\int \partial WW'$ independently of the choice of
$\{\phi_k\}_{k=1}^{\infty}$; this is the formal analogue of the first relation in \eqref{dwy_dy}. On the other hand,
the orthonormality of $\{\phi_k\}_{k=1}^\infty $ is sufficient for $\{c_k^\circ \}_{k=1}^\infty $ to be an i.i.d. $N(0,\Omega)$ sequence; thus by the ergodic theorem,
$$
K^{-1}\sum_{k=1}^{K}c_k^\circ {c_k^\circ }'\underset{K\toinf}{\overset{a.s.}{\longrightarrow }}\int \rd W\rd W'=\Omega ,
$$
where the integrand is understood as the quadratic variation $\rd W(u)\rd W(u)'=\Omega \rd u$; this is the formal analogue of the second relation
in \eqref{dwy_dy}. It is by virtue of this latter property that some interactions of stationary processes and the basis functions are not asymptotically negligible and can be used to cleanse limit distributions from dependence.

\begin{lemma}[Limits for $K\toinf$]\label{lem_asympt_K}
For any given orthonormal \cadlag\ basis $\{\phi_k\}_{k=1}^\infty$ of $L^2[0,1]$, consider
$$
\ell_{00}:=\int W W',\qquad\ell_{0\varphi}:=\int W\varphi_K',\qquad\ell^\circ_{\rd0\varphi}:=\int \rd W\varphi_K',
$$
where $\varphi_K:=(\phi_1,\dots,\phi_K)'$ and $W:=( W_1', W_2')'$, see \eqref{eq_x}, with $L^2$-representation in \eqref{eq_d_L2}; for $K\toinf$, one has that:
\begin{itemize}
\item[(i)] $\ell_{0\varphi}\ell_{\varphi0}\convas\ell_{00}$;

\item[(ii)] $K^{-1}\ell^\circ_{\rd0\varphi}\ell^\circ_{\varphi\rd0}\convas\Omega$.

\item[(iii)] $\ell^\circ_{\rd0\varphi}\ell_{\varphi0}\convp\int\partial W W'$, where $\int\partial W W'$ indicates a Stratonovich integral;

\item[(iv)] $\ell^\circ_{\rd2\varphi}\ell_{\varphi1}(\ell_{1\varphi}\ell_{\varphi1})^{-1}\convp \int \partial W_2 W_1'\left(\int W_1 W_1'\right)^{-1}$;

\item[(v)] $\ell^\circ_{\rd2\varphi}P_\circ\ell_{\varphi1}\left(\ell_{1\varphi}P_\circ\ell_{\varphi1}\right)^{-1}\convp\int \rd W_{2.1} W_1'\left(\int W_1 W_1'\right)^{-1}$, where $W_{2.1}(u)$ is as in \eqref{eq_def_b_q}.
\end{itemize}
\end{lemma}
\begin{mproof}{\textit{Proof.}}
(i) By definition, $\ell_{0\varphi}=(c_1,c_2,\dots,c_{K})$, so that $\ell_{0\varphi}\ell_{\varphi0}=\sum_{k=1}^K\nu_k^2\xi_k\xi_k'$;
by Parseval's theorem, $\ell_{00}=\sum_{k=1}^\infty \nu_k^2\xi_k\xi_k'$ a.s. and the statement follows from the a.s. convergence of the latter series.
(ii) By definition, $\ell^\circ_{\rd0\varphi}=(c^\circ_1,c^\circ_2,\dots,c^\circ_K)$, where $c^\circ_k:=\int\rd W\phi_k\sim N(0,\Omega)$ and
$\left\{c^\circ_k\right\}_{k=1}^{\infty}$ is i.i.d. because the basis $\{\phi_k\}_{k=1}^\infty$ is orthonormal. Hence
$\ell^\circ_{\rd0\varphi}\ell^\circ_{\varphi\rd0}=\sum_{k=1}^K c^\circ_k{c^\circ_k}'$ and the statement follows from the ergodic theorem.
(iii) This follows by Lemma \tref{lemma_ogawa} applied componentwise. (iv) By points (iii) and (i). (v) It is first shown that
$\ell^\circ_{\rd2\varphi}P_\circ\ell_{\varphi1}\convp\int \rd W_{2.1} W_1'$. Observe that
$\ell^\circ_{\rd2\varphi}P_\circ=\ell^\circ_{\rd2\varphi}-H_K\ell^\circ_{\rd1\varphi}$,
where $H_K:=\ell^\circ_{\rd2\varphi}\ell^\circ_{\varphi\rd1}(\ell^\circ_{\rd1\varphi}\ell^\circ_{\varphi\rd1})^{-1}$.
By point (ii), $H_K\convp\Omega_{21}\Omega_{11}^{-1}$ and hence, again by Lemma \tref{lemma_ogawa} applied componentwise,
\begin{align*}
\ell^\circ_{\rd2\varphi}P_\circ\ell_{\varphi1}&=\ell^\circ_{\rd2\varphi}\ell_{\varphi1}-H_{K}\ell^\circ_{\rd1\varphi}\ell_{\varphi 1}\\
&\convp\int \partial W_2 W_1'-\Omega_{21}\Omega_{11}^{-1}\int \partial W_1 W_1'=\int \partial[ W_2 -\Omega_{21}\Omega_{11}^{-1} W_1 ] W_1'\\
&=\int \partial W_{2.1} W_1'=\int \rd W_{2.1} W_1',
\end{align*}
where the last equality follows by the independence of the processes $W_{2.1}$ and $W_1$. Furthermore,
$$
\ell_{1\varphi}P_\circ\ell_{\varphi1} =\ell_{1\varphi}\ell_{\varphi1}-K^{-1}\ell_{1\varphi}\ell^\circ_{\varphi\rd1}(K^{-1}\ell^\circ_{\rd1\varphi}\ell^\circ_{\varphi\rd1})^{-1}\ell^\circ_{\rd1\varphi}\ell_{\varphi1}=\sum_{k=1}^{K}\nu_k^2\xi_{k1}\xi_{k1}'+O_{p}(K^{-1}),
$$
because $\ell_{1\varphi}\ell^\circ_{\varphi\rd1}= (\ell^\circ_{\rd1\varphi}\ell_{\varphi1})'\convp\int W_1 \partial W_1'$ and $K^{-1}\ell^\circ_{\rd1\varphi}\ell^\circ_{\varphi\rd1}\convp\Omega_{11}$ by points (iii) and (ii). Thus, $\ell_{1\varphi}P_\circ\ell_{\varphi1}\convp\sum_{k=1}^{\infty}\nu_k^2\xi_{k1}\xi_{k1}'=\int W_1 W_1'$ as $K\rightarrow\infty$.
\end{mproof}

\section{Proofs}\label{sec_app_proofs}

\begin{mproof}{\textit{Proof of Lemma \tref{lem_ident}.}}
Results (i) and (ii) are well known, see e.g. Exercise 3.7 and Section 13.2 in \citet{Joh:96}, while (iii) and (iv) are proved in eq. (B.1) and (B.2) in \citet{Par:97}, see also \citet{Par:06b}.
\end{mproof}\smallskip

\begin{mproof}{\textit{Proof of Theorem \tref{theorem_dim_coh}}.}
Consider $H$ of dimension $p\times m$ and full column rank. Assumption \tref{ass_DGP}.(i) is satisfied by $G(z):=H'C(z)$, of dimension $m\times n_\varepsilon$, because its disc of convergence is at least equal to the one of $C(z)$. Assumption \tref{ass_DGP}.(ii) is also satisfied by $G(z)$ because $m\leq p \leq n_\varepsilon$, so that $\rank H'C(z)<m$ only when $C(z)$ is of reduced rank; write $G(z)=G+G_1(z)(1-z)$, where $G_1:=G_1(1)=H'C_1$ and $G =H'C=H'\psi\kappa'$ of rank $q\leq s$. Here superscripts $\breve\,$ are omitted for brevity and because the results are invariant to the choice of bases of various spaces.
Let $\varrho$ be a basis of $\col G = \col (H'\psi)$ and write $H'\psi=\varrho \vartheta'$; then $G=\varrho \tau'$, where $\tau:=\kappa \vartheta$, is a rank decomposition of $G$ with $\varrho$ and $\tau$ full column rank matrices with $q$ columns.

Next it is shown that Assumption \tref{ass_DGP}.(iii) also holds for $G(z)$, i.e. that $\varrho_\bot'G_1\tau_\bot=\varrho_\bot'H'C_1\tau_\bot$ has full row rank $m-q$. First note that $H\varrho_\bot=\psi_\bot g$ where $g:=\bar\psi_\bot'H\varrho_\bot$; in fact,
$H\varrho_\bot=(\bar\psi\psi'H+\psi_\bot\bar\psi_\bot'H)\varrho_\bot = (\bar\psi\vartheta \varrho'+\psi_\bot\bar\psi_\bot'H)\varrho_\bot = \psi_\bot g$.
Next note that $\tau_\bot=(\kappa_\bot,\bar{\kappa}\vartheta_\bot)$, so that
$$
\varrho_\bot'G_1\tau_\bot=\varrho_\bot'H'C_1\tau_\bot=g'\psi_\bot'C_1(\kappa_\bot,\bar{\kappa}\vartheta_\bot)=g'(\psi_\bot'C_1\kappa_\bot,\psi_\bot'C_1\bar{\kappa}\vartheta_\bot),
$$
where $\rank (\psi_\bot'C_1\kappa_\bot) = p-s$ by Assumption \tref{ass_DGP}(iii) on $C(z)$ and
$g:=\bar\psi_\bot'H\varrho_\bot$ is $(p-s) \times (m-q)$ of full column rank $m-q$. This completes the proof.
\end{mproof}\smallskip

\begin{mproof}{\textit{Proof of Theorem \tref{theorem_lim_eig}.}}
(i) Consider $\CCA(x_t,d_t)$ and normalize $S(\ml):=\ml M_{xx}-M_{xd}M_{dd}^{-1}M_{dx}$ in \eqref{eq_BEP} as $A_T'S(\ml)A_T=\ml A_T'M_{xx}A_T-A_T'M_{xd}M_{dd}^{-1}M_{dx}A_T$, where $A_T=(T^{-\frac 12}\bar\psi,\beta)$; by Lemma \tref{lem_asympt_T_BEP_S} one finds
\begin{equation}\label{eq_pBEP}
A_T'S(\ml)A_T\convw\ml \left(\begin{array}{cc}
\ell_{11} & 0\\
0 & Q_{00}
\end{array}\right)-\left(\begin{array}{cc}
\ell_{1\varphi}\ell_{\varphi1} & 0\\
0 & 0
\end{array}\right),
\end{equation}
where $\ell_{\varphi1}=\ell_{1\varphi}'$. From \eqref{eq_pBEP} one finds that the $s$ largest eigenvalues in \eqref{eq_BEP} converge weakly to the ones of $\ell_{11}^{-1}\ell_{1\varphi}\ell_{\varphi1}$, which coincide with those of the symmetric and a.s. positive definite $s\times s$ matrix $\ell_{11}^{-1/2}\ell_{1\varphi}\ell_{\varphi1}\ell_{11}^{-1/2}$, and the $r=p-s$ smallest eigenvalues converge to 0 in probability. Observe that the eigenvalues of $\ell_{11}^{-1}\ell_{1\varphi}\ell_{\varphi1}$ are invariant with respect to the transformation of $W_1(u)$ into $A W_1(u)$ for any nonsingular $A\in \mathbb{R}^{s\times s}$; choosing $A=\Omega_{11}^{-1/2}$, where $\Omega_{11}$ is the variance of $W_1(1)$, see \eqref{eq_Xi}, one obtains \eqref{eq_large_eigs}. In order to prove (ii), set $\ml =\rho/T$ and consider
$$
B_T'S(\rho/T)B_T=\left(
\begin{array}{cc}
T^{-1}\bar{\psi}'S(\rho/T)\bar{\psi} &\bar{\psi}'S(\rho/T)\beta\\
\beta'S(\rho/T)\bar{\psi} & T\beta'S(\rho/T)\beta\\
\end{array}
\right),\qquad B_T:=(T^{-\frac 12}\bar\psi,T^{1/2}\beta);
$$
using Lemma \tref{lem_asympt_T}, one has
\begin{align*}
T^{-1}\bar{\psi}'S(\rho/T)\bar{\psi} & =\rho T^{-2}\bar{\psi}'M_{xx}\bar{\psi}-T^{-\frac 12}\bar{\psi}'M_{xd}M_{dd}^{-1}M_{dx}\bar{\psi}T^{-\frac 12}\convw -\ell_{1\varphi}\ell_{\varphi1},\\
\bar{\psi}'S(\rho/T)\beta & =\rho T^{-1}\bar{\psi}'M_{xx}\beta-T^{-\frac 12}\bar{\psi}'M_{xd}M_{dd}^{-1}M_{dx}\beta T^{1/2}\convw -\ell_{1\varphi}\ell^\circ_{\varphi\rd2},\\
T\beta'S(\rho/T)\beta & =\rho\beta'M_{xx}\beta-T^{1/2}\beta'M_{xd}M_{dd}^{-1}M_{dx}\beta T^{1/2}\convw\rho Q_{00}-\ell^\circ_{\rd2\varphi}\ell^\circ_{\varphi\rd2}
\end{align*}
jointly, where $\ell^\circ_{\varphi\rd2}=(\ell^\circ_{\rd2\varphi})'$, and hence
\begin{align*}
\left\vert B_T'S(\rho/T)B_T\right\vert & =\left\vert T^{-1}\bar{\psi}'S(\rho/T)\bar{\psi}\right\vert\left\vert T\beta'S(\rho/T)\beta-\beta'S(\rho/T)\bar{\psi}(T^{-1}\bar{\psi}'S(\rho/T)\bar{\psi})^{-1}\bar{\psi}'S(\rho/T)\beta\right\vert\\
&\convw\left\vert-\ell_{1\varphi}\ell_{\varphi1}\right\vert\left\vert\rho Q_{00}-\ell^\circ_{\rd2\varphi}\ell^\circ_{\varphi\rd2}+\ell^\circ_{\rd2\varphi}\ell_{\varphi1}(\ell_{1\varphi}\ell_{\varphi1})^{-1}\ell_{1\varphi}\ell^\circ_{\varphi\rd2}\right\vert.
\end{align*}
Because $\ell_{1\varphi}\ell_{\varphi1}$ is a.s. nonsingular, this shows that the $r$ smallest eigenvalues in \eqref{eq_BEP} multiplied by $T$ converge weakly to the ones of $Q_{00}^{-1}\ell^\circ_{\rd2\varphi}(I_K-\ell_{\varphi1}(\ell_{1\varphi}\ell_{\varphi1})^{-1}\ell_{1\varphi})\ell^\circ_{\varphi\rd2}$, which coincide with those of the symmetric and a.s. positive definite $r\times r$ matrix $Q_{00}^{-1/2}\ell^\circ_{\rd2\varphi}(I_K-\ell_{\varphi1}(\ell_{1\varphi}\ell_{\varphi1})^{-1}\ell_{1\varphi})\ell^\circ_{\varphi\rd2}Q_{00}^{-1/2}$. Observe that the eigenvalues of $\ell_{\varphi1}(\ell_{1\varphi}\ell_{\varphi1})^{-1}\ell_{1\varphi}$ are invariant with respect to the transformation of $W_1(u)$ into $A W_1(u)$ for any nonsingular $A\in \mathbb{R}^{s\times s}$; choosing $A=\Omega_{11}^{-1/2}$, one obtains \eqref{eq_small_eigs}.

(iii) It is next shown that $\Upsilon_K:=\left(\int B_1 B_1'\right)^{-1}\int B_1\varphi_K'\int \varphi_K B_1'\convas I_s$ as $K\toinf$.
Indeed, for the standard Brownian motion $B_1$ one has, see \eqref{eq_d_L2}, $
\int B_1 B_1'=\sum_{k=1}^\infty \nu_k^2\xi_k\xi_k'$ a.s. by Parseval Theorem and $\int B_1\varphi_K'=(\nu_1\xi_1,\nu_2\xi_2,\dots,\nu_K\xi_K)$ by definition; this implies
\begin{align}\label{eq_large_eigs_to_one}
\Upsilon_K=\left(\sum_{k=1}^\infty \nu_k^2\xi_k\xi_k'\right)^{-1}\sum_{k=1}^K\nu_k^2\xi_k\xi_k'\convas I_s
\end{align}
as $K\toinf$. This further implies that the eigenvalues $\mu_1\geq\mu_2\geq\dots\geq\mu_s$ of $\Upsilon_K$ all converge to one a.s. as $K\toinf$. Consider next \eqref{eq_small_eigs} and observe that
$$K^{-1}\Psi_K = K^{-1} Q_{00}^{-1}(\ell^\circ_{\rd2\varphi}
\ell^\circ_{\varphi\rd2}-
\ell^\circ_{\rd2\varphi}
\ell_{\varphi1}(\ell_{1\varphi}\ell_{\varphi1})^{-1}\ell_{1\varphi}
\ell^\circ_{\varphi\rd2});$$
from Lemma \ref{lem_asympt_K}.(ii) one has
$K^{-1} \ell^\circ_{\rd2\varphi}
\ell^\circ_{\varphi\rd2}\convas \Omega_{22}$ and
$K^{-1}
\ell^\circ_{\rd2\varphi}
\ell_{\varphi1}(\ell_{1\varphi}\ell_{\varphi1})^{-1}\ell_{1\varphi}
\ell^\circ_{\varphi\rd2} = O_{p}(K^{-1})$
by Lemma \ref{lem_asympt_K}.(i),(iii). This shows that $K^{-1}\Psi_K \convp Q_{00}^{-1} \Omega_{22}\aspos$.
(iv) 
By (iii),
the $s$ largest eigenvalues $\ml_1\geq\ml_2\geq\dots\geq\ml_s$ all converge weakly to one (and hence also in probability) as $(T,K)_{seq}\toinf$, while
$(\ml_{s+1}, \dots,\ml_p)$ converge to 0 in probability with $K^{-1}T(\ml_{s+1}, \dots,\ml_p)=O_p(1)$ as $(T,K)_{seq}\toinf$.
\end{mproof}\smallskip

\begin{mproof}{\textit{Proof of Corollary \tref{coro_cons_s}.}}
Direct consequence of Theorem \tref{theorem_lim_eig}.(iv), see \eqref{eq_order_eigs}.
\end{mproof}\smallskip

\begin{mproof}{\textit{Proof of Corollary \tref{coro_cons_other}.}}
Note that $K/T=o(1)$ as $(T,K)_{seq}\toinf$. Use Theorem \tref{theorem_lim_eig}.(i) and \eqref{eq_order_eigs2} to find $f_1(i)\asyp (K/T)^{|i-s|}$ for $i\in \mtI_1$, $f_2(i)\asyp 1_{i\neq s}+(T/K)1_{i=s}$ for $i\in \mtI_2$, $f_3(i)\asyp 1_{i\notin \{s-1,s\}}+1_{i=s-1,s>1}/\log(T/K)+\log(T/K)1_{i=s}$ for $i\in \mtI_3$. Hence asymptotically: $f_1(i)$ converges to 0 for all $i\neq s$ while it converges to $1/\det(Q_{00}^{-1}\Omega_{22})>0$ for $i=s$; hence it
has a unique maximum at $i=s$; $f_2(i)$ and $f_3(i)$ converge to finite values for $i\neq s$ and diverge for $i=s$, and hence have a unique maximum at $i=s$. This proves $\underset{i\in \mtI_j}{\argmax}\:f_j(i)\convp s$ for $j=1,2,3$ when $s\in \mtI_j$. When considering maximization over $\mtI_j^0$ for $j=2,3$, one still has $f_2(i)\asyp 1_{i\neq s}+(T/K)1_{i=s}$ for $i\in \mtI_2^0$, $f_3(i)\asyp 1_{i\notin \{s-1,s\}}+1_{i=s-1,s>0}/\log(T/K)+\log(T/K)1_{i=s}$ for $i\in \mtI_3^0$, and, as above, this implies $\underset{i\in \mtI_j^0}{\argmax}\:f_j(i)\convp s$ for $j=2,3$ when $s\in \mtI_j^0$. All of the above is still valid when $K$ is kept constant.
\end{mproof}\smallskip

\begin{mproof}{\textit{Proof of Theorem \tref{theorem_dec_rule}}}.
By dimensional coherence, see Theorem \tref{theorem_dim_coh}, Corollary \ref{coro_cons_s} applies to both $\wh s (x_t)$ and $\wh s ( b'x_t)$, the max-gap estimator based on $x_t$ and on $b'x_t$; hence $\myPr(\wh s ( b'x_t) = s=\wh s (x_t))\toone$ as $(T,K)_{seq}\toinf$. The case when $H_0$ is invalid is similar, with $\myPr(\wh s (x_t) = s)\toone$ and $\myPr(\wh s ( b'x_t) =\rank( b'\psi)<s)\toone$, and hence $\myPr(\wh s ( b'x_t) <\wh s (x_t))\toone$. This completes the proof.
\end{mproof}\smallskip

\begin{mproof}{\textit{Proof of Theorem \tref{theorem_asy_distr_s}.}}
Consider $\Upsilon_K$ in \eqref{eq_large_eigs}; from \eqref{eq_large_eigs_to_one} one has
$$
K\left(I_s-\Upsilon_K\right)=\left(\int B_1 B_1'\right)^{-1}K\sum_{k=K+1}^\infty \nu_k^2\xi_k\xi_k'.
$$
It is next shown that $K\sum_{k=K+1}^\infty \nu_k^2\xi_k\xi_k'\convp\frac1{\pi^2}I_s$ as $K\toinf$. Indeed,
$$
K\sum_{k=K+1}^\infty \nu_k^2\xi_k\xi_k'=I_sK\sum_{k=K+1}^\infty \nu_k^2+K\sum_{k=K+1}^\infty \nu_k^2\left(\xi_k\xi_k'-I_s\right)
$$
and for the {\KL} representation, see \eqref{eq_KLB}, $\nu_k=\frac1{(k-\frac 12)\pi}$; hence
\begin{align}\label{eq_pi2K}
\pi^2 K\sum_{k=K+1}^\infty \nu_k^2&=4K\sum_{k=K+1}^\infty \frac1{(2k-1)^2}=\frac1{K}\sum_{k=1}^\infty \frac1{(1+\frac{k}{K}-\frac1{2K})^2}=\frac1{K}\sum_{k=1}^\infty \frac1{(1+\frac{k}{K})^2}+o(1)\\ \nonumber
&\conv\int_0^\infty \frac{dx}{(1+x)^2}=1
\end{align}
by dominated convergence, whereas $\E\left(\sum_{k=K+1}^\infty \nu_k^2(\xi_k\xi_k'-I_s)\right) =0$ and, for every fixed $a,b\in \BR^s$,
\begin{align}\label{eq_sum4}
\E\left(\sum_{k=K+1}^\infty \nu_k^2(a'\xi_k\xi_k'b-a'b)\right)^2 =\sum_{k=K+1}^\infty \nu_k^4\E(a'\xi_k\xi_k'b-a'b)^2=[a'ab'b+2(a'b)^2]\sum_{k=K+1}^\infty \nu_k^4=O(K^{-3}).
\end{align}
Thus $\pi^2 K\sum_{k=K+1}^\infty \nu_k^2\xi_k\xi_k'\convp I_s$, which implies that $\pi^2 K\left(I_s-\Upsilon_K\right)\convp\left(\int B_1 B_1'
\right)^{-1}$. Because $\ml_i\convw\mu_i$, $i=1,\dots,s$, by Theorem \tref{theorem_lim_eig}.(i), where $\mu_1\geq\mu_2\geq\dots\geq\mu_s$ are the eigenvalues of $\Upsilon_K$, and the eigenvalues of $I_s-\Upsilon_K$ are $1-\mu_i$, $i=1,\dots,s$, this proves (i). In order to prove (ii), recall that using Theorem \tref{theorem_lim_eig}.(i), one has \eqref{eq_order_eigs} and hence $\ml_i\convp 0$, $i=s+1,\dots,p$, as $T\toinf$.
\end{mproof}\smallskip

\begin{mproof}{\textit{Proof of Corollary} \tref{theorem_sequence}.}
Same as in the proof of Theorem 12.3 in \cite{Joh:96}.
\end{mproof}\smallskip

\begin{lemma}[Super-consistency]\label{lem_cons_psi}
Let $d_t$ in \eqref{eq_d} be constructed using any orthonormal \cadlag\ basis of $L^2[0,1]$; then the one-step estimators
$\wh\psi^{(1)},\wh\beta^{(1)}$ and the ICC estimators $\wh\psi^{(2)}:=\wh\psi,\wh\beta^{(2)}:=\wh\beta$ of $\psi$ and $\beta$ are $T$-consistent, i.e.
\begin{equation}\label{eq_cons}
\wh\psi^{(j)}-\psi=o_p(T^{-\frac12}), \qquad \wh\beta^{(j)}-\beta=o_p(T^{-\frac12}), \qquad j=1,2,
\end{equation}
for any fixed $K\geq p$.
\end{lemma}
Thanks to \eqref{eq_dual_psi}, the results in \eqref{eq_cons} can be equivalently expressed as
$$
\wh\psi_*^{(j)}-\psi_*=-(\wh\beta_*^{(j)}-\beta_*)'=o_p(T^{-\frac12}), \qquad j=1,2.
$$
\begin{mproof}{\textit{Proof of Lemma \tref{lem_cons_psi}.}}
By the duality in \eqref{eq_dual_psi}, it is sufficient to prove the lemma for the estimators of $\psi$. Let $\Lambda=\diag(\Lambda_1,\Lambda_0)$ and $V=(V_1,V_0)$ in \eqref{eq_LV} satisfy
\begin{equation}\label{eq_LV2}
M_{ff}V\Lambda = M_{fd}M_{dd}^{-1}M_{df}V,\qquad V'M_{ff}V = I_p,\qquad V'M_{fd}M_{dd}^{-1}M_{df}V =\Lambda,
\end{equation}
where $f=x$ for $j=1$ and $f=e$ for $j=2$. Consider the first equation in \eqref{eq_LV2} for the $s$ largest eigenvalues,
\begin{equation}\label{eq_eig_prob_large_eigs_M}
M_{ff}V_1\Lambda_1=M_{fd}M_{dd}^{-1}M_{df}V_1,\qquad f=x,e,
\end{equation}
with associated normalization $V'_1M_{ff}V_1=I_s$. Then one has
$$
A_T'M_{ff}A_T(A_T^{-1}V_1)\Lambda_1 = A_T'M_{fd}M_{dd}^{-1}M_{df}A_T(A_T^{-1}V_1),
$$
where $A_T:=(T^{-\frac 12}\bar\psi,\beta)$ and $A_T^{-1}V_1=(T^{1/2}\psi,\bar\beta)'V_1=:(v_{1T}',v_{0T}')'$; hence, by Lemma \tref{lem_asympt_T_BEP_S} the limit eigenvalue problem is
\begin{equation}\label{eq_eig_lim}
\left(\begin{array}{cc}
\varsigma_{11} & 0\\
0 & Q_{00}
\end{array}\right)
\left(\begin{array}c
v_1^*\\
v_0^*
\end{array}\right)\Upsilon_K=\left(\begin{array}{cc}
\varsigma_{1\varphi}\varsigma_{\varphi 1} & 0\\
0 & 0
\end{array}\right)\left(\begin{array}c
v_1^*\\
v_0^*
\end{array}\right),
\end{equation}
where $\Lambda_1\convw\Upsilon_K$ and $\varsigma_{11},\varsigma_{1\varphi}\varsigma_{\varphi y}$ are a.s. nonsingular, with associated normalization of the eigenvectors $({v^*_1}',{v^*_0}') \diag(\varsigma_{11},Q_{00})({v^*_1}',{v^*_0}')'=I_s$.

The lower block of \eqref{eq_eig_lim} reads $Q_{00} v_0^*\Upsilon_K=0$, which implies $v_{0T}\convp v_0^*=0$ because $Q_{00}$ and $\Upsilon_K$ are a.s. nonsingular. The upper block of \eqref{eq_eig_lim} reads $\varsigma_{11} v_1^*\Upsilon_K=\varsigma_{1\varphi}\varsigma_{\varphi 1} v_1^*$;
hence it follows by the continuous mapping theorem that $v_{1T}\convw v_1^*$, where $v_1^*$ are eigenvectors of the limit eigenvalue problem $\det(\mu\varsigma_{11}-\varsigma_{1\varphi}\varsigma_{\varphi 1})=0$ such that ${v_1^*}'\varsigma_{11}v_1^*=I_s$. In particular, $v_1^*$ is a.s. nonsingular. Summarizing, $A_T^{-1}V_1$ converges in distribution to $({v_1^*}',0)'$, and
\begin{equation}\label{eqthetas}
\left(\begin{array}c
T^{-\frac12}\bar\psi'M_{ff}V_1\\
\beta'M_{ff}V_1
\end{array}\right)
=(A_T'M_{ff}A_T)(A_{T}^{-1}V_1)
\convw\left(\begin{array}{cc}
\varsigma_{11} & 0\\
0 & Q_{00}
\end{array}\right)
\left(\begin{array}c
v_1^*\\
0
\end{array}\right)=
\left(\begin{array}c
\vartheta_1\\
0
\end{array}\right),
\end{equation}
where $\vartheta_1:=\varsigma_{11}v_1^*$ is a.s. nonsingular, and thus
\begin{align}\label{eq_psi1}
T^{-\frac12}\bar{\psi}'M_{ff}V_1 &=O_{p}(1),\qquad (T^{-\frac12}\bar{\psi}'M_{ff}V_1)^{-1} =O_{p}(1), \qquad %
\beta'M_{ff}V_1=o_{p}(1) , \\
\label{eq_psi2}
T^{-\frac12}b'M_{ff}V_1 &=T^{-\frac12}b'\psi
\bar{\psi}'M_{ff}V_1+b'\beta o_{p}(T^{-\frac12%
})=T^{-\frac12}\bar{\psi}'M_{ff}V_1+o_{p}(T^{-\frac12}).
\end{align}%
Hence using \eqref{eq_psi1} and \eqref{eq_psi2}
\begin{align}\label{eq_part1}
\bar{\psi}'\widehat{\psi }^{(j)}& =T^{-\frac12}\bar{\psi}%
'M_{ff}V_1(T^{-\frac12}b^{\prime
}M_{ff}V_1)^{-1}=I_{s}+o_{p}(T^{-\frac12}),\\
\label{eq_part2}
\beta'\widehat{\psi }^{(j)}& =T^{-\frac12}\beta^{\prime
}M_{ff}V_1(T^{-\frac12}b'M_{ff}V_1)^{-1}=o_{p}(T^{-\frac{1}{%
2}})O_{p}(1)=o_{p}(T^{-\frac12}),
\end{align}%
and one finds $\widehat{\psi }^{(j)}=\psi \bar{\psi}'\widehat{\psi }%
^{(j)}+\beta \bar{\beta }'\widehat{\psi }^{(j)}=\psi
+o_{p}(T^{-\frac12})$.
This completes the proof.
\end{mproof}\smallskip

The following proofs employ a simplifying convention, without loss of generality.
By Lemma \tref{lem_cons_psi}, $\bar\psi'\wh\psi^{(1)}$ converges in probability to a positive-definite matrix
and $\beta'\wh\psi^{(1)}=o_p(T^{-\frac 12})$ as $T\conv\infty$, such that the statements in Lemmas \ref{lem_asympt_T_BEP}(x)-(xii)
and \ref{lem_asympt_T_BEP_S} remain valid with $e_t(\wh\psi^{(1)})$ in place of $e_t(\psi)$.
Thus, the ICC estimators $\wh\psi,\wh\beta$ are discussed as if computed using $e_t(\psi)$, abbreviated to $e_t$.

\begin{lemma}[Fixed $K$ asymptotic distribution of $\wh\psi^{(j)}$ and $\wh\beta^{(j)}$]\label{lem_psihat_distr}
Let $\wh\psi^{(1)}$, $\wh\beta^{(1)}$, $\wh\psi^{(2)}:=\wh\psi$ and $\wh\beta^{(2)}:=\wh\beta$ be as in Definition \tref{def_phi_hat};
then for fixed $K\geq p$ and $T\toinf$, one has
\begin{equation}\label{eq_psihat_distr}
T\bar c'(\wh\psi^{(j)}-\psi)=-T(\wh\beta^{(j)}-\beta)'\bar b\convw
\left\{
\begin{array}{cl}
\ell^\circ_{\rd2\varphi}\ell_{\varphi1}(\ell_{1\varphi}\ell_{\varphi1})^{-1} &\mbox{ for } j=1\\
\ell^\circ_{\rd2\varphi}P_\circ\ell_{\varphi1}(\ell_{1\varphi}P_\circ\ell_{\varphi1})^{-1} &\mbox{ for } j=2 \\
\end{array}\right.,
\end{equation}
where $P_\circ:=I_K-\ell^\circ_{\varphi\rd1}(\ell^\circ_{\rd1\varphi}\ell^\circ_{\varphi\rd1})^{-1}\ell^\circ_{\rd1\varphi}$.
\end{lemma}
\begin{mproof}{\textit{Proof of Lemma \tref{lem_psihat_distr}.}}
By the duality in \eqref{eq_dual_psi}, it is enough to prove the lemma for the estimators of $\psi$.
Consider the estimator $\wh\psi^{(j)} = M_{ff}V_1( b'M_{ff}V_1)^{-1}$, where $f=x$ for $j=1$ and $f=e$ for $j=2$, and drop the superscript$\,^{(j)}$ in the rest of the proof.
Write $\wh\psi =\wt\psi\theta^{-1}$ where
$\wt\psi := T^{-\frac 12}M_{ff} V_1$ and $\theta = T^{-\frac 12}b'M_{ff} V_1 = O_p(1)$, see \eqref{eq_psi2}. From \eqref{eq_eig_prob_large_eigs_M} one has $\wh\psi(\theta\Lambda_1\theta^{-1})=M_{fd}M_{dd}^{-1}M_{df}M_{ff}^{-1}\wh\psi$; pre-multiplying by $( b,\beta)'$ and using $A_T=T^{-\frac 12}(\bar\psi,T^{1/2}\beta)$ in the r.h.s., one has
$$
( b,\beta)'\wh\psi(\theta\Lambda_1\theta^{-1})=T^{-\frac 12}( b,\beta)'M_{fd}M_{dd}^{-1}(M_{df}A_T)(A_T'M_{ff}A_T)^{-1}(\bar\psi,T^{1/2}\beta)'\wh\psi
$$
and one finds
\begin{equation}\label{eq_goo}
\left(
\begin{array}c
I_s\\
\beta'\wh\psi
\end{array}
\right)
\theta\Lambda_1\theta^{-1}=T^{-\frac 12}( b,\beta)'M_{fd}Q_T,
\end{equation}
where by Lemma \tref{lem_asympt_T_BEP_S} and \eqref{eq_part1}, \eqref{eq_part2} one has
$$
Q_T:= M_{dd}^{-1}(M_{df}A_T)(A_T'M_{ff}A_T)^{-1}(\bar\psi,T^{1/2}\beta)'\wh\psi
\convw (\varsigma_{\varphi1},0)
\left(\begin{array}{cc}
\varsigma_{11} & 0\\
0 & Q_{00}
\end{array}\right)^{-1}\left(
\begin{array}c
I_s\\
0
\end{array}
\right)=\varsigma_{\varphi1}\varsigma_{11}^{-1}.
$$
Because $b=\bar\psi+\beta h$, $h:=\bar\beta'b$, and
$$
T^{-\frac 12} b'M_{fd}=T^{-\frac 12}\bar\psi'M_{fd}+h'T^{-\frac 12}\beta'M_{fd}\convw \varsigma_{1\varphi},
$$
from the first block of $s$ equalities in \eqref{eq_goo} one finds
$$
\theta\Lambda_1\theta^{-1}=T^{-\frac 12} b'M_{fd}Q_T\convw (\varsigma_{1\varphi}\varsigma_{\varphi1})\varsigma_{11}^{-1},
$$
which is a.s. nonsingular, and hence
$$
G_T:=Q_T(\theta\Lambda_1\theta^{-1})^{-1}\convw \varsigma_{\varphi1}(\varsigma_{1\varphi}\varsigma_{\varphi1})^{-1}=
\left\{
\begin{array}{cl}
\ell_{\varphi1}(\ell_{1\varphi}\ell_{\varphi1})^{-1} &\mbox{ for } j=1\medskip(f=x)\\
P_\circ\ell_{\varphi1}(\ell_{1\varphi}P_\circ\ell_{\varphi1})^{-1} &\mbox{ for } j=2\medskip(f=e) \\
\end{array}\right.,
$$
see \eqref{eq_def_varsigma}. From the second block of $r$ equalities in \eqref{eq_goo} one has
$$
T\beta'\wh\psi=(T^{\frac12}\beta'M_{fd})G_T,
$$
where $\beta'\wh\psi=\beta'(\wh\psi -\psi)$, and by Lemma \tref{lem_asympt_T}(vii), and Lemma \tref{lem_asympt_T_BEP}(xii),
$$
T^{\frac12}\beta'M_{fd}\convw\varsigma^\circ_{\rd2\varphi}:=
\left\{
\begin{array}{cl}
\ell^\circ_{\rd2\varphi} &\mbox{ for } j=1\medskip(f=x)\\
\ell^\circ_{\rd2\varphi}P_\circ &\mbox{ for } j=2\medskip(f=e)
\end{array}\right..
$$
This shows that
$$
T\beta'(\wh\psi^{(j)} -\psi)\convw\varsigma^\circ_{\rd2\varphi}\varsigma_{\varphi1}(\varsigma_{1\varphi}\varsigma_{\varphi1})^{-1}=
\left\{
\begin{array}{cl}
\ell^\circ_{\rd2\varphi}\ell_{\varphi1}(\ell_{1\varphi}\ell_{\varphi1})^{-1} &\mbox{ for } j=1\\
\ell^\circ_{\rd2\varphi}P_\circ\ell_{\varphi1}(\ell_{1\varphi}P_\circ\ell_{\varphi1})^{-1} &\mbox{ for } j=2 \\
\end{array}\right.;
$$
the statement follows from $\bar c'(\wh\psi^{(j)}-\psi)=\beta'(\wh\psi^{(j)}-\psi)$ and the duality in \eqref{eq_dual_psi}.
\end{mproof}\smallskip

\begin{mproof}{\textit{Proof of Theorem \tref{theorem_asy_distr_psi}.}}
Consider \eqref{eq_psihat_distr} in Lemma \tref{lem_psihat_distr}; by Lemma \tref{lem_asympt_K}(iv),(v)
$$
\begin{array}c
\ell^\circ_{\rd2\varphi}\ell_{\varphi1}(\ell_{1\varphi}\ell_{\varphi1})^{-1}\convp \int \partial W_2 W_1'\left(\int W_1 W_1'\right)^{-1}\\
\ell^\circ_{\rd2\varphi}P_\circ\ell_{\varphi1}(\ell_{1\varphi}P_\circ\ell_{\varphi1})^{-1}\convp\int \rd W_{2.1} W_1'\left(\int W_1 W_1'\right)^{-1}
\end{array}
$$
and thus the statement.
\end{mproof}\smallskip

\begin{mproof}{\textit{Proof of Theorem \tref{theorem_Xihat_cons}.}}
By Lemma \tref{lem_asympt_T}(viii),(ix), for fixed $K\geq p$ and $T\toinf$, one has that
$$
T^{1/2}\left(\begin{array}c
\bar\psi'M_{\Delta x d}\\
\beta'M_{xd}
\end{array}
\right)\convw\left(\begin{array}c
\int \rd W_1 \varphi_K'\\
\int \rd W_2 \varphi_K'
\end{array}
\right)=\int \rd W \varphi_K',
$$
by Lemma \tref{lem_asympt_T_BEP}(ii), for fixed $K\geq p$ and $T\toinf$, one has that $M_{dd}\conv I_K$ and by Lemma \tref{lem_asympt_K}(ii), for $K\toinf$,
$
K^{-1}\int \rd W\varphi_{K}'\int \varphi_{K}\rd W'\convas\Omega.
$
Combining these results, one finds that
$$
\frac TK\left(\begin{array}c
\bar\psi'M_{\Delta x d}\\
\beta'M_{xd}
\end{array}
\right)M_{dd}^{-1}\left(\begin{array}c
\bar\psi'M_{\Delta x d}\\
\beta'M_{xd}
\end{array}
\right)'\stackrel[(T,K)_{seq}\toinf]{w}{\conv}
\left(
\begin{array}{cc}
\Omega_{11}&\Omega_{12}\\
\Omega_{21} & \Omega_{22}\end{array}%
\right).
$$
Because $\Omega$ is non-stochastic, the last convergence is also in probability. By the super-consistency result in Theorem \tref{lem_cons_psi}, the same is achieved if $\wh\psi^{(j)}$ and $\wh\beta^{(j)}$
are used instead of $\psi$ and $\beta$, $j=1,2$.
\end{mproof}\smallskip

\begin{mproof}{\textit{Proof of Theorem \tref{theorem_inf_psi}.}}
Direct consequence of Theorem \tref{theorem_asy_distr_psi}, Remark \tref{rem_mixed_gauss} and Theorem \tref{theorem_Xihat_cons}.
\end{mproof}\smallskip

\begin{mproof}{\textit{Proof of Theorem \tref{theorem_MLE}.}}
Consider a VAR $A(L)X_t=\varepsilon_t$ under the $I(1)$ condition $|\alpha_\bot' A_1\beta_\bot|\neq 0$, where $A(z)=A+A_1(z)(1-z)$, $A=-\alpha \beta'$ and $A_1:=A_1(1)$; in this case, Granger's Representation Theorem, see \citet[][Theorem 4.2]{Joh:96}, $(1-z)A(z)^{-1}=C+C_1(z)(1-z)$ and hence $C(z)=C+C_1(z)(1-z)$ in \eqref{eq_DX} satisfies
\begin{equation}\label{eq_C}
C=\beta_\bot\Psi\alpha_\bot',\qquad\beta'C_1=\bar\alpha'(A_1C-I_p),\qquad
\begin{array}c
C_1:=C_1(1)\\
\Psi:=(\alpha_\bot'A_1\beta_\bot)^{-1}
\end{array}.
\end{equation}
Note that $\psi=\beta_\bot$ and $\kappa'=\Psi\alpha_\bot'$.

Using the present notation for the parameters, \citet[][Theorem 13.3]{Joh:96} states $T\bar b' (\wh\beta -\beta )\convw (\int GG')^{-1}\int G\rd V_\alpha' =: U$ where $G(u):=\bar{\beta}_\bot' CW_\varepsilon(u)$ and $V_\alpha(u):=(\alpha'\Omega_\varepsilon^{-1}\alpha)^{-1}\alpha'\Omega_\varepsilon^{-1}W_\varepsilon(u)$. It is next shown that $U=-Z'$, where $Z$ is as in \eqref{eq_asy_iter}. Note that $G(u)=W_1(u)$; in fact $\bar{\beta}_\bot' CW_\varepsilon(u)=\kappa'W_\varepsilon(u)=W_1(u)$, so that $U=(\int W_1W_1')^{-1}\int W_1\rd V_\alpha' $. Next it is shown that $V_\alpha=-W_{2.1}$; substituting $\psi=\beta_\bot$ and $\kappa'=\Psi\alpha_\bot'$ in \eqref{eq_x} one finds that
$$
W_1(u):=\kappa'W_\varepsilon(u)=\Psi\alpha_\bot'W_\varepsilon(u),\qquad W_2(u):=\beta'C_1W_\varepsilon(u)=\bar\alpha'(A_1C-I_p)W_\varepsilon(u)
$$
whose variance matrix at $u=1$ is
$$
\Omega:=
\left(
\begin{array}{cc}
\Omega_{11}&\Omega_{12}\\
\Omega_{21} &\Omega_{22}\end{array}
\right)=\left(
\begin{array}{cc}
\Psi\alpha_\bot'\Omega_\varepsilon \alpha_\bot\Psi'&\Psi\alpha_\bot'\Omega_\varepsilon (A_1 C-I_p)'\bar\alpha \\
\bar\alpha'(A_1C-I_p)\Omega_\varepsilon \alpha_\bot\Psi'&\bar\alpha'(A_1C-I_p)\Omega_\varepsilon (A_1 C-I_p)'\bar\alpha
\end{array}
\right).
$$
It is useful to recall the following non-orthogonal projection identity
\begin{equation}\label{eq_proj_id}
I_p=\Omega_\varepsilon \alpha_\bot(\alpha_\bot'\Omega_\varepsilon \alpha_\bot)^{-1}\alpha_\bot'+\alpha(\alpha'\Omega_\varepsilon^{-1}\alpha)^{-1}\alpha'\Omega_\varepsilon^{-1}=:P_1+P_2.
\end{equation}
One finds
\begin{align*}
\Omega_{21}\Omega_{11}^{-1}&=\bar\alpha'(A_1C-I_p)\Omega_\varepsilon \alpha_\bot\Psi'(\Psi\alpha_\bot'\Omega_\varepsilon \alpha_\bot\Psi')^{-1}=\bar\alpha'(I_p-\Omega_\varepsilon \alpha_\bot(\alpha_\bot'\Omega_\varepsilon \alpha_\bot)^{-1}\alpha_\bot')A_1\beta_\bot\\
&=(\alpha'\Omega_\varepsilon^{-1}\alpha)^{-1}\alpha'\Omega_\varepsilon^{-1}A_1\beta_\bot =\bar{\alpha}'P_2A_1\beta_\bot,
\end{align*}
where the second equality follows by substituting $C=\beta_\bot\Psi\alpha_\bot' $ and $\Psi$ from \eqref{eq_C} and the third one by the non-orthogonal projection identity \eqref{eq_proj_id}. Substituting the above expressions into the definition of $W_{2.1}(u)$, see \eqref{eq_def_b_q}, one finds, using \eqref{eq_proj_id} again,
\begin{align*}
 W_{2.1}(u)&:= W_2(u)-\Omega_{21}\Omega_{11}^{-1} W_1(u)=\bar\alpha'(A_1C-I_p)W_\varepsilon(u)-\bar{\alpha}'P_2A_1CW_\varepsilon(u)\\
&=\bar\alpha'(-I_p+(I_p-P_2)A_1C)W_\varepsilon(u)=\bar\alpha'(-P_2+P_1(A_1C-I_p))W_\varepsilon(u),
\end{align*}
where $P_1(A_1C-I_p)=0$ follows from $\alpha_\bot'(A_1 C-I_p)=\alpha_\bot'A_1\beta_\bot(\alpha_\bot'A_1\beta_\bot)^{-1}\alpha_\bot'-\alpha_\bot'=0$; hence
$$
 W_{2.1}(u)=-\bar\alpha'P_2W_\varepsilon(u)=-(\alpha'\Omega_\varepsilon^{-1}\alpha)^{-1}\alpha'\Omega_\varepsilon^{-1}W_\varepsilon(u)=-V_\alpha(u).
$$
This completes the proof that $U=-Z'$.
\end{mproof}\smallskip

\begin{mproof}{\textit{Proof of Theorem \tref{theorem_IV}.}}
If $C(z)$ in \eqref{eq_DX} satisfies Assumption \tref{ass_DGP}, then $Q(z)$ satisfies the summability condition (\textbf{L}) in \citet{Phi:14}, as e.g. in the discussion in \citet[][page 809]{FP:20}.
$Q(1)$ nonsingular and it is normalized to $I_p$ without loss of generality.
From \eqref{eq_PCB} one has
\begin{align*}
\left(\begin{array}c
\Delta y_t\\
\Delta z_t
\end{array}
\right)&=\left(\begin{array}c
A u_{1t}+\Delta u_{0t}\\
u_{1t}
\end{array}
\right)=\left(\begin{array}{cc}
0 & A\\
0 & I_s
\end{array}
\right)\left(\begin{array}c
u_{rt}\\
u_{st}
\end{array}
\right)+\left(\begin{array}{cc}
I_r & 0\\
0 & 0
\end{array}
\right)\left(\begin{array}c
\Delta u_{0t}\\
\Delta u_{1t}
\end{array}
\right)\\
&=\left(\left(\begin{array}{cc}
0 & A\\
0 & I_s
\end{array}
\right)+\left(\begin{array}{cc}
I_r & 0\\
0 & 0
\end{array}
\right)\Delta \right)Q(L)\varepsilon_t,
\end{align*}
where $Q(z)=I_p+Q_1(z)(1-z)$ and $Q_1:=Q_1(1)$; hence $C(z)=C+C_1(z)(1-z)$ in \eqref{eq_DX} satisfies
$$
C=\left(\begin{array}{cc}
0 & A\\
0 & I_s
\end{array}
\right),\qquad C_1=\left(\begin{array}{cc}
0 & A\\
0 & I_s
\end{array}
\right)Q_1+\left(\begin{array}{cc}
I_r & 0\\
0 & 0
\end{array}
\right).
$$
This shows that one can choose $\psi=(A',I_s)'$ with $\kappa'=(0, I_s)$ and $\beta'=(I_r,-A)$. Note also that $\beta'C_1\kappa_\bot=(I_r,-A)C_1(I_r,0)'=I_r$ of full row rank so that Assumption \tref{ass_DGP} holds. Substituting $\psi=(A',I_s)'$ with $\kappa'=(0, I_s)$ and $\beta'=(I_r,-A)$ in \eqref{eq_X}, from
\eqref{eq_x} one finds that
$$
 W_1(u):=\kappa'W_\varepsilon(u)=(0, I_s)W_\varepsilon(u),\qquad
 W_2(u):=\beta'C_1W_\varepsilon(u)=(I_r,0)W_\varepsilon(u).
$$
Using the present notation for the parameters, \citet[][point (a) in Section 3]{Phi:14} reads
$ T (\wh\beta -\beta )'\bar b\convw -\int \rd B_{0.x}B_x'(\int B_x B_x')^{-1} =: U$, where $B_x(u):=(0, I_s)W_\varepsilon(u)= W_1(u)$,
$B_0(u):=(I_r,0)W_\varepsilon(u)= W_2(u)$ and $B_{0.x}(u):=B_0(u)-\Omega_{0x}\Omega_{xx}^{-1}B_x(u) = W_{2.1}(u)$.

It is then shown that $U=-Z$, where $Z$ is as in \eqref{eq_asy_iter}. Note that the variance matrix of $B_x(u)= W_1(u)$ and $B_0(u)= W_2(u)$ is
$$
\left(
\begin{array}{cc}
\Omega_{xx}&\Omega_{x0}\\
\Omega_{0x} &\Omega_{00}\end{array}%
\right)
=
\left(
\begin{array}{cc}
\Omega_{11}&\Omega_{12}\\
\Omega_{21} &\Omega_{22}
\end{array}
\right);
$$
because $W_{2.1}(u)=B_{0.x}(u)$ and $W_1(u)=B_x(u)$, one can write $Z$ in \eqref{eq_asy_iter} as $Z=\int \rd B_{0.x} B_{x}'\left(\int B_{x}B_{x}'\right)^{-1}$, which shows that $U=-Z$.
\end{mproof}

\section{Simulations}\label{sec_app_sim}

Results for the Monte Carlo simulations are split between inference on $s$ and $\psi$.

\subsection{Estimation and inference on $s$}\label{sec_app_sim_s}

\begin{figure}[htbp]
\begin{center}
\includegraphics[width=\linewidth]{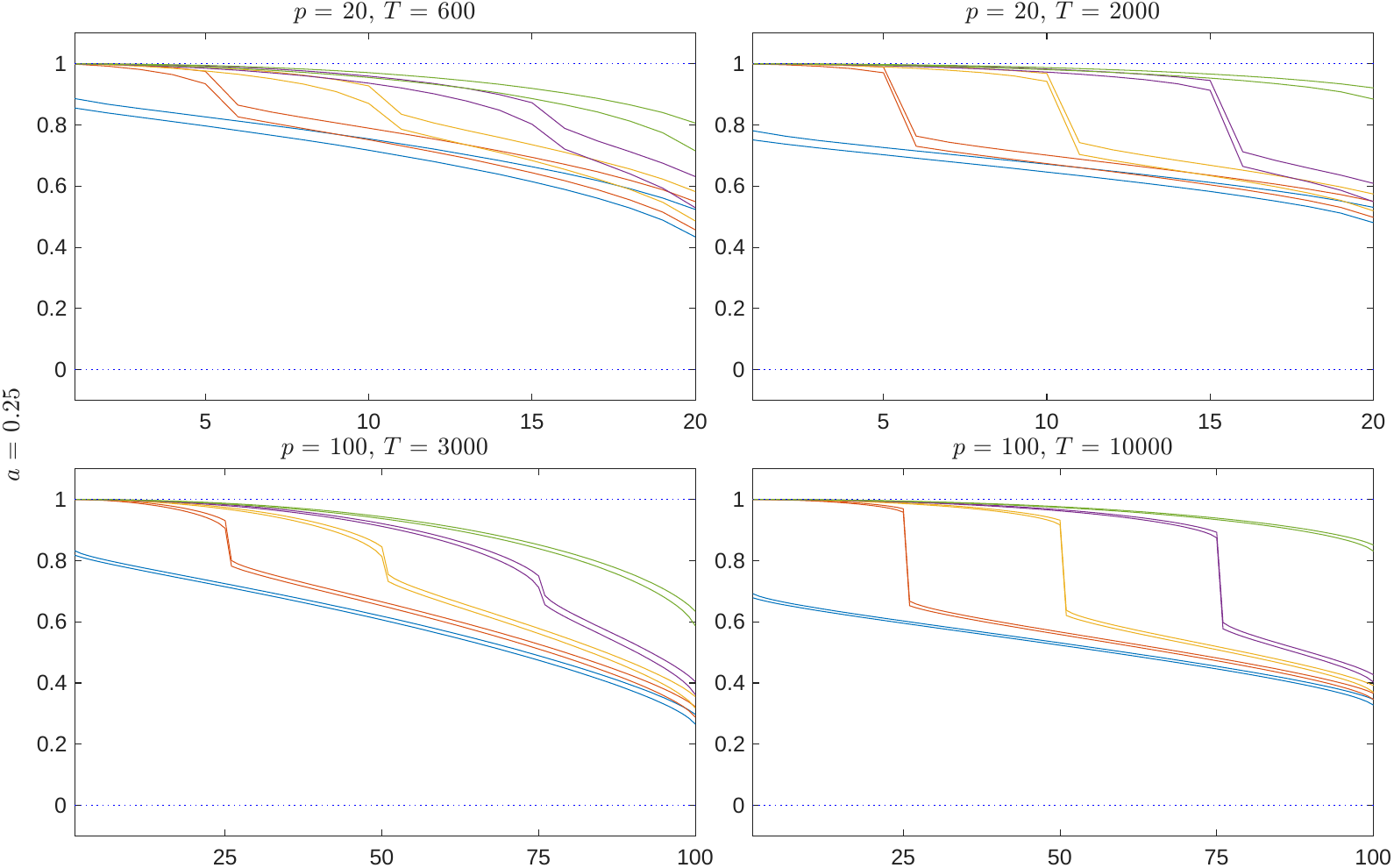}
\includegraphics[width=\linewidth]{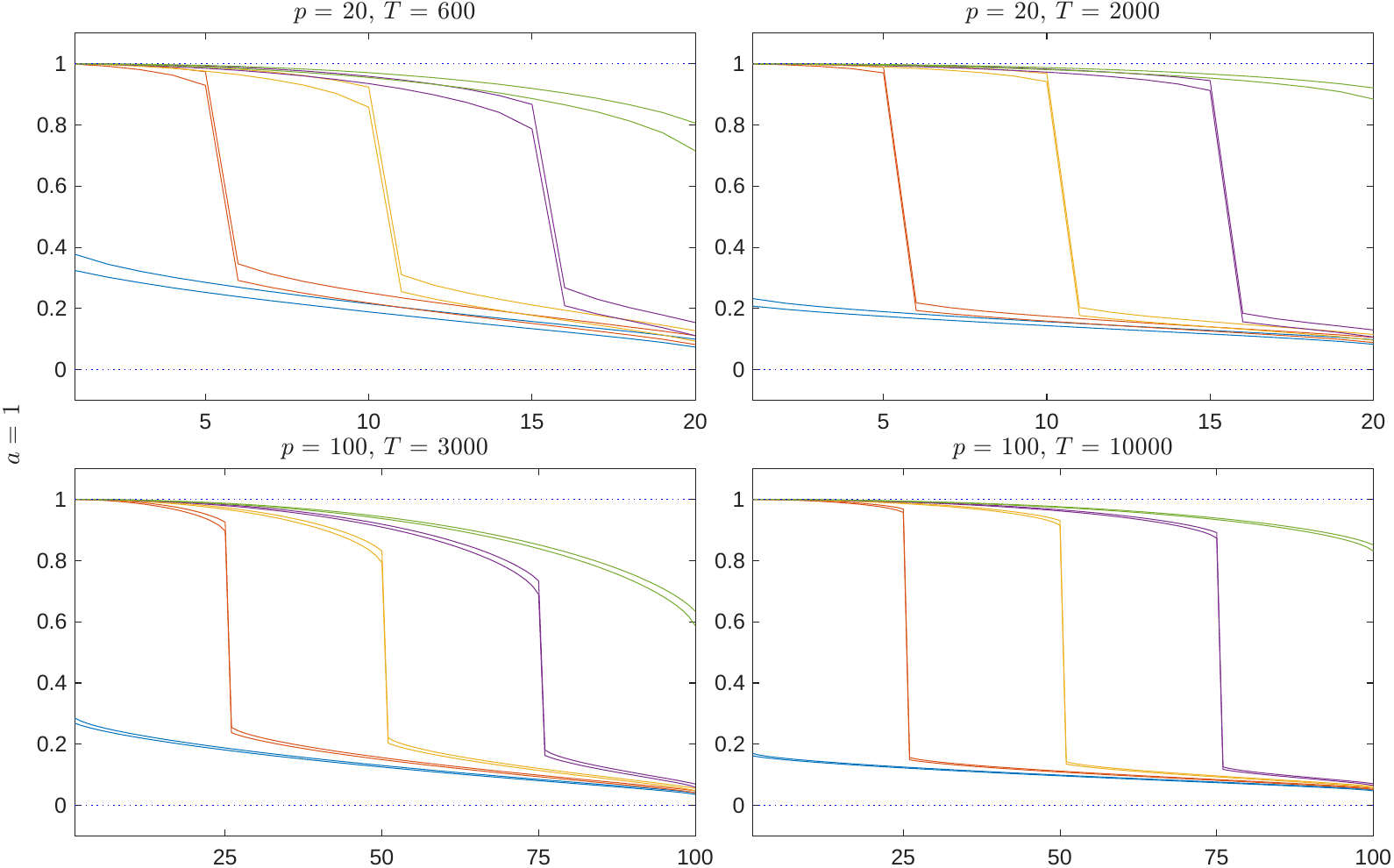}
\end{center}
\caption{\footnotesize Quantiles $0.05$ and $0.95$ of $\lambda_i$, $i=1,\dots,p$. $p=20,100$, $T/p=30,100$ and $a=0.25$ (first and second panels from top) and $a=1$ (bottom two panels) in DGP \eqref{eq_DGP_OW}. Color codes for $s=\lceil jp/4 \rceil$, $j=0,1,2,3,4$: blue $j=0$, orange $j=1$, yellow $j=2$, purple $j=3$, green $j=4$. {\Ntext} replications of DGP \eqref{eq_DGP_OW}.}
\label{fig_OW_eigs}
\end{figure}

First, the \SCC\ profile $\{\lambda_i\}_{i=1,\dots,p}$ was simulated, to illustrate Theorem \tref{theorem_lim_eig}.(iv). Fig. \tref{fig_OW_eigs} reports quantiles 0.05 and 0.95 ($(l,u)$, lower-upper) of the \SCC\ profile $\lambda_i$ for $a=0.25,1$, $p=20,100$ and $T=30p, 100p$. Lines with the same colour refer to $(l,u)$, colors (blue, red, orange, purple, green) correspond to $s=0,\lceil p/4 \rceil,\lceil p/2 \rceil,\lceil 3p/4 \rceil,p$, the first and second (third and fourth) row of panels correspond to $a=0.25$ ($a=1$), the first and third (second and fourth) row of panels correspond to $p=20$ ($p=100$), the first (second) column of panels correspond to $T=30p$ ($T=100p$). The eight panels illustrate how the curve $(\lambda_i)_{i=1,\dots,p}$ converges to $1_{i\leq s}$, see Theorem \tref{theorem_lim_eig}. Note that $(l,u)$ lines get closer passing from the first column to the second one.


For fixed $(p,T,s)$, the MC behavior of the max-gap estimator worsens as $a$ decreases	
because of slower convergence to 0 of the last part of the \SCC\ profile $\{\lambda_{i}\}_{i=s+1}^p$
for lower values of $a$. 
This can be seen by comparing the top 2 rows with the bottom 2 rows in Fig. \tref{fig_OW_eigs}.

Next consider the finite sample properties of the max-gap estimator and of the sequential testing procedures $\{F_{j,n}\}_{j=p,\dots,1}$, $n=1,\infty$.
Table \tref{table_max_gap} reports MC statistics on the max-gap $\wh s$. The top panel reports the MC frequency of correct selection of $s$,
$\frac1N\sum_{j=1}^N1_{\wh s_j =s}$, where $\wh s_j $ indicates the value of $\wh s$ in MC replication $j$.
The bottom panel reports the Mean Absolute Error (MAE), $\frac1N\sum_{j=1}^N|\wh s_j - s|$. The number of replications of DGP \eqref{eq_DGP_OW} is {\Ntext}.

When $a=1$ or $s=p$, it can be seen that the max-gap estimator selects $s$ with MC frequency 1 except for $T=10p$ at $s=\lceil 3p/4\rceil$, $p=10,20,50$, and $T=10p$, $p=10$ at $s=\lceil p/2\rceil$. For $a=0.25,0.5,0.75$, the cases of less-than-perfect selection of the max-gap estimator increase, with worse performance for low values of $a$, and high values of $s$ (excluding $s=p$). The MAE indicates that estimation errors in $s$ are moderate except for $a=0.25$ and large values of $p$.

Table \tref{table_mx} reports MC statistics for the estimator $\wt s $ of $s$ based on the test sequence $\{F_{j,\infty}\}_{j=p,\dots,1}$. The top two panels are of the same type as the ones reported in Table \tref{table_max_gap}
for the max-gap estimator. The lower panel reports the MC size of test $F_{s,\infty}$ for the true value of $s$.

It can be seen that the test $F_{s,\infty}$ appears to be undersized in finite samples. The MC frequency of reaching the test $F_{s,\infty}$ in the test sequence is 1 for $T=30p$ except for $a=0.25$. At $a=0.25$ this frequency is also 1 for $s=\lceil p/4\rceil$ but deteriorates for higher values of $s$. Unlike for the max-gap, the predicted asymptotic behavior for $\{F_{j,\infty}\}_{j=p,\dots,1}$ is in line with MC frequencies at $T=30p$ for all values of $a$ except for $a=0.25$ and $s=\lceil jp/4\rceil$, $j=2,3$.

Table \tref{table_tr} reports MC statistics for estimator of $s$ employing the test sequence $\{F_{j,1}\}_{j=p,\dots,1}$, also indicated as $\wt s $ in the caption. The structure of the table is the same as for Table \tref{table_mx}.
It can be seen that also the test $F_{s,1}$ is undersized in finite samples. Similarly to $\{F_{j,\infty}\}_{j=p,\dots,1}$, the predicted asymptotic behavior for $\{F_{j,1}\}_{j=p,\dots,1}$ is in line with MC frequencies at $T=30p$ for all value of $a$ except for $a=0.25$ and $s=\lceil jp/4\rceil$, $j=2,3$.

The max-gap criterion and the tests $F_{p,n}$ can be combined into a hybrid estimator of $s$, that first tests the hypothesis $H_0:s=p$ using the $F_{p,n}$ statistic, and, in case of rejection, employs the restricted max-gap estimator that replaces the domain $\{0,1 , \dots p\}$ with $\{0,1 , \dots p-1\}$; formally the hybrid estimator is defined as
	\begin{align}\label{eq_hybrid}
	\wh s_{n}:=  1_{\{F_{p,n}\leq c_{n,\eta}\}}
	\cdot
	p
	+ 1_ {\{F_{p,n}> c_{n,\eta}\}}\cdot
	\argmax_{j \in \{0,1 , \dots p-1\}}(\lambda_
	{j}-\lambda_
	{j + 1}), \qquad n=1, \infty.
	\end{align}
	Note that $\wh s_{n}$ does not involve $\ml_{p+1}:=0$.
Under the conditions of Theorem \ref{theorem_sequence} one finds that  the hybrid estimator $\wh s_{n}$  satisfies $\Pr(\wh s_{n}=s)\conv 1-\eta$ if $s=p$ and $\Pr(\wh s_{n}=s)\conv 1$ if $s<p$, where $\eta$ is the asymptotic significance level of the test $F_{p,n}$.

Monte Carlo results for the hybrid estimators $\wh s_{n}$, $n=1, \infty$, are reported in Tables \ref{table_hyb_mx} and \ref{table_hyb_tr}. The performance of the hybrid estimator based on $F_{p,n}$ compares favorably with the one of the max-gap estimator for all values of $a$.

\cp

\cp

\begin{table}[htbp]\tiny
\renewcommand{\arraystretch}{0.8}
\addtolength{\tabcolsep}{-1pt}
\begin{center}
\begin{tabular}{|cc||cccc||cccc||cccc||cccc||c|}
\hline
\multicolumn{19}{|c|}{Max-gap}\\
\hline
\hline
\multicolumn{19}{|c|}{Frequency of correct selection}\\
\hline
\hline
\multicolumn{2}{|c||}{} & \multicolumn{4}{c||}{$s=0$, $a=$} & \multicolumn{4}{c||}{$s=\lceil p/4\rceil$, $a=$} & \multicolumn{4}{c||}{$s=\lceil p/2\rceil$, $a=$} & \multicolumn{4}{c||}{$s=\lceil 3p/4\rceil$, $a=$} & \multicolumn{1}{c|}{$s=p$} \\
  $p$ & $T/p$ & 0.25 & 0.5 & 0.75 & 1 & 0.25 & 0.5 & 0.75 & 1 & 0.25 & 0.5 & 0.75 & 1 & 0.25 & 0.5 & 0.75 & 1 & \\
\hline
10 & 10 & 0 & 0 & 0.96 & 1 & 0 & 0 & 0.80 & 1 & 0 & 0 & 0.59 & 0.99 & 0 & 0.01 & 0.34 & 0.93 & 1 \\
 & 20 & 0 & 0.01 & 1 & 1 & 0 & 0.01 & 1 & 1 & 0 & 0 & 0.98 & 1 & 0 & 0.01 & 0.85 & 1 & 1 \\
 & 30 & 0 & 0.06 & 1 & 1 & 0 & 0.03 & 1 & 1 & 0 & 0.02 & 1 & 1 & 0 & 0.02 & 0.99 & 1 & 1 \\
\hline
20 & 10 & 0 & 0.17 & 1 & 1 & 0 & 0.02 & 1 & 1 & 0 & 0 & 0.90 & 1 & 0 & 0 & 0.55 & 0.98 & 1 \\
 & 20 & 0 & 0.63 & 1 & 1 & 0 & 0.21 & 1 & 1 & 0 & 0.03 & 1 & 1 & 0 & 0 & 1 & 1 & 1 \\
 & 30 & 0 & 0.98 & 1 & 1 & 0 & 0.75 & 1 & 1 & 0 & 0.28 & 1 & 1 & 0 & 0.07 & 1 & 1 & 1 \\
\hline
50 & 10 & 0 & 1 & 1 & 1 & 0 & 0.58 & 1 & 1 & 0 & 0.01 & 0.99 & 1 & 0 & 0 & 0.78 & 0.99 & 1 \\
 & 20 & 0 & 1 & 1 & 1 & 0 & 1 & 1 & 1 & 0 & 0.72 & 1 & 1 & 0 & 0.04 & 1 & 1 & 1 \\
 & 30 & 0 & 1 & 1 & 1 & 0 & 1 & 1 & 1 & 0 & 1 & 1 & 1 & 0 & 0.87 & 1 & 1 & 1 \\
\hline
100 & 10 & 0 & 1 & 1 & 1 & 0 & 0.99 & 1 & 1 & 0 & 0.04 & 1 & 1 & 0 & 0 & 0.86 & 1 & 1 \\
 & 20 & 0 & 1 & 1 & 1 & 0 & 1 & 1 & 1 & 0 & 1 & 1 & 1 & 0 & 0.39 & 1 & 1 & 1 \\
 & 30 & 0 & 1 & 1 & 1 & 0 & 1 & 1 & 1 & 0 & 1 & 1 & 1 & 0 & 1 & 1 & 1 & 1 \\
\hline
300 & 10 & 1 & 1 & 1 & 1 & 0 & 1 & 1 & 1 & 0 & 0.27 & 1 & 1 & 0 & 0 & 0.59 & 1 & 1 \\
 & 20 & 1 & 1 & 1 & 1 & 0 & 1 & 1 & 1 & 0 & 1 & 1 & 1 & 0 & 1 & 1 & 1 & 1 \\
 & 30 & 1 & 1 & 1 & 1 & 0 & 1 & 1 & 1 & 0 & 1 & 1 & 1 & 0 & 1 & 1 & 1 & 1
\\
\hline
\hline
\multicolumn{19}{|c|}{MAE}\\
\hline
\hline
\multicolumn{2}{|c||}{} & \multicolumn{4}{c||}{$s=0$, $a=$} & \multicolumn{4}{c||}{$s=\lceil p/4\rceil$, $a=$} & \multicolumn{4}{c||}{$s=\lceil p/2\rceil$, $a=$} & \multicolumn{4}{c||}{$s=\lceil 3p/4\rceil$, $a=$} & \multicolumn{1}{c|}{$s=p$} \\
  $p$ & $T/p$ & 0.25 & 0.5 & 0.75 & 1 & 0.25 & 0.5 & 0.75 & 1 & 0.25 & 0.5 & 0.75 & 1 & 0.25 & 0.5 & 0.75 & 1 & \\
\hline
10 & 10 & 9.99 & 9.93 & 0.35 & 0 & 6.99 & 6.96 & 1.30 & 0 & 4.99 & 4.96 & 1.92 & 0.01 & 2 & 1.96 & 1.26 & 0.10 & 0 \\
 & 20 & 10 & 9.88 & 0 & 0 & 7 & 6.95 & 0.01 & 0 & 5 & 4.98 & 0.09 & 0 & 2 & 1.99 & 0.30 & 0 & 0 \\
 & 30 & 10 & 9.42 & 0 & 0 & 7 & 6.78 & 0 & 0 & 5 & 4.92 & 0 & 0 & 2 & 1.96 & 0.01 & 0 & 0 \\
\hline
20 & 10 & 20 & 16.64 & 0 & 0 & 15 & 14.69 & 0.03 & 0 & 10 & 9.98 & 0.81 & 0 & 5 & 4.99 & 2.13 & 0.02 & 0 \\
 & 20 & 20 & 7.35 & 0 & 0 & 15 & 11.91 & 0 & 0 & 10 & 9.69 & 0 & 0 & 5 & 4.98 & 0.01 & 0 & 0 \\
 & 30 & 20 & 0.41 & 0 & 0 & 15 & 3.70 & 0 & 0 & 10 & 7.23 & 0 & 0 & 5 & 4.66 & 0 & 0 & 0 \\
\hline
50 & 10 & 50 & 0 & 0 & 0 & 37 & 15.51 & 0 & 0 & 25 & 24.75 & 0.01 & 0 & 12 & 12 & 2.05 & 0.01 & 0 \\
 & 20 & 50 & 0 & 0 & 0 & 37 & 0.01 & 0 & 0 & 25 & 7.11 & 0 & 0 & 12 & 11.51 & 0 & 0 & 0 \\
 & 30 & 50 & 0 & 0 & 0 & 37 & 0 & 0 & 0 & 25 & 0 & 0 & 0 & 12 & 1.56 & 0 & 0 & 0 \\
\hline
100 & 10 & 99.99 & 0 & 0 & 0 & 75 & 0.18 & 0 & 0 & 50 & 47.63 & 0 & 0 & 25 & 25 & 1.39 & 0 & 0 \\
 & 20 & 100 & 0 & 0 & 0 & 75 & 0 & 0 & 0 & 50 & 0.03 & 0 & 0 & 25 & 15.14 & 0 & 0 & 0 \\
 & 30 & 100 & 0 & 0 & 0 & 75 & 0 & 0 & 0 & 50 & 0 & 0 & 0 & 25 & 0 & 0 & 0 & 0 \\
\hline
300 & 10 & 0 & 0 & 0 & 0 & 225 & 0 & 0 & 0 & 150 & 45.35 & 0 & 0 & 75 & 74.98 & 1.70 & 0 & 0 \\
 & 20 & 0 & 0 & 0 & 0 & 225 & 0 & 0 & 0 & 150 & 0 & 0 & 0 & 75 & 0.14 & 0 & 0 & 0 \\
 & 30 & 0 & 0 & 0 & 0 & 225 & 0 & 0 & 0 & 150 & 0 & 0 & 0 & 75 & 0 & 0 & 0 & 0
\\
\hline
\end{tabular}
\end{center}
\caption{\footnotesize Max-gap. Top panel: frequency of correct selection of $s$, $\frac1N\sum_{n=1}^N1_{\wh s_n =s}$; bottom panel: Mean Absolute Error (MAE), $\frac1N\sum_{n=1}^N|\wh s_n - s|$. {\Ntext} replications of DGP \eqref{eq_DGP_OW}.} \label{table_max_gap}
\end{table}

\cp

\begin{table}[htbp]\tiny
\renewcommand{\arraystretch}{0.6}
\addtolength{\tabcolsep}{-1pt}
\begin{center}
\begin{tabular}{|cc||cccc||cccc||cccc||cccc||c|}
\hline
\multicolumn{19}{|c|}{$F_{j,\infty}$}\\
\hline
\hline
\multicolumn{19}{|c|}{Frequency of correct selection}\\
\hline
\hline
\multicolumn{2}{|c||}{} & \multicolumn{4}{c||}{$s=0$, $a=$} & \multicolumn{4}{c||}{$s=\lceil p/4\rceil$, $a=$} & \multicolumn{4}{c||}{$s=\lceil p/2\rceil$, $a=$} & \multicolumn{4}{c||}{$s=\lceil 3p/4\rceil$, $a=$} & \multicolumn{1}{c|}{$s=p$} \\
  $p$ & $T/p$ & 0.25 & 0.5 & 0.75 & 1 & 0.25 & 0.5 & 0.75 & 1 & 0.25 & 0.5 & 0.75 & 1 & 0.25 & 0.5 & 0.75 & 1 & \\
\hline
10 & 10 & 0 & 0.81 & 1 & 1 & 0 & 0.05 & 0.98 & 0.98 & 0 & 0.01 & 0.88 & 0.98 & 0 & 0.07 & 0.76 & 0.98 & 0.98 \\
 & 20 & 0.72 & 1 & 1 & 1 & 0.04 & 0.98 & 0.97 & 0.97 & 0.01 & 0.98 & 0.97 & 0.97 & 0.08 & 0.97 & 0.97 & 0.97 & 0.96 \\
 & 30 & 1 & 1 & 1 & 1 & 0.97 & 0.97 & 0.96 & 0.96 & 0.77 & 0.97 & 0.97 & 0.97 & 0.68 & 0.97 & 0.97 & 0.97 & 0.96 \\
\hline
20 & 10 & 0 & 1 & 1 & 1 & 0 & 0.34 & 0.99 & 0.98 & 0 & 0 & 0.98 & 0.98 & 0 & 0 & 0.77 & 0.99 & 0.99 \\
 & 20 & 1 & 1 & 1 & 1 & 0.27 & 0.98 & 0.98 & 0.97 & 0 & 0.98 & 0.98 & 0.97 & 0 & 0.98 & 0.97 & 0.97 & 0.97 \\
 & 30 & 1 & 1 & 1 & 1 & 0.99 & 0.98 & 0.97 & 0.97 & 0.96 & 0.97 & 0.97 & 0.97 & 0.72 & 0.97 & 0.97 & 0.97 & 0.96 \\
\hline
50 & 10 & 0 & 1 & 1 & 1 & 0 & 0.84 & 1 & 0.99 & 0 & 0 & 1 & 1 & 0 & 0 & 0.22 & 1 & 1 \\
 & 20 & 1 & 1 & 1 & 1 & 0.91 & 0.99 & 0.98 & 0.98 & 0 & 0.99 & 0.99 & 0.98 & 0 & 0.99 & 0.99 & 0.98 & 0.99 \\
 & 30 & 1 & 1 & 1 & 1 & 0.99 & 0.98 & 0.98 & 0.97 & 1 & 0.99 & 0.98 & 0.98 & 0.76 & 0.98 & 0.98 & 0.97 & 0.98 \\
\hline
100 & 10 & 0 & 0 & 0 & 0 & 0 & 0 & 0 & 0 & 0 & 0 & 0 & 0 & 0 & 0 & 0 & 0 & 1 \\
 & 20 & 1 & 1 & 1 & 1 & 1 & 1 & 0.99 & 0.99 & 0 & 1 & 1 & 0.99 & 0 & 1 & 1 & 1 & 1 \\
 & 30 & 1 & 1 & 1 & 1 & 1 & 0.99 & 0.98 & 0.98 & 1 & 1 & 0.99 & 0.99 & 0.35 & 1 & 1 & 1 & 1 \\
\hline
300 & 10 & 0 & 0 & 0 & 0 & 0 & 0 & 0 & 0 & 0 & 0 & 0 & 0 & 0 & 0 & 0 & 0 & 1 \\
 & 20 & 0 & 0.55 & 1 & 1 & 0 & 0 & 1 & 1 & 0 & 0 & 1 & 1 & 0 & 0 & 0.12 & 1 & 1 \\
 & 30 & 1 & 1 & 1 & 1 & 1 & 1 & 1 & 1 & 0.98 & 1 & 1 & 1 & 0 & 1 & 1 & 1 & 1
\\
\hline
\hline
\multicolumn{19}{|c|}{MAE}\\
\hline
\hline
\multicolumn{2}{|c||}{} & \multicolumn{4}{c||}{$s=0$, $a=$} & \multicolumn{4}{c||}{$s=\lceil p/4\rceil$, $a=$} & \multicolumn{4}{c||}{$s=\lceil p/2\rceil$, $a=$} & \multicolumn{4}{c||}{$s=\lceil 3p/4\rceil$, $a=$} & \multicolumn{1}{c|}{$s=p$} \\
  $p$ & $T/p$ & 0.25 & 0.5 & 0.75 & 1 & 0.25 & 0.5 & 0.75 & 1 & 0.25 & 0.5 & 0.75 & 1 & 0.25 & 0.5 & 0.75 & 1 & \\
\hline
10 & 10 & 9.61 & 1.20 & 0 & 0 & 6.76 & 3.68 & 0.02 & 0.02 & 4.84 & 3.18 & 0.14 & 0.02 & 1.93 & 1.46 & 0.26 & 0.02 & 0.02 \\
 & 20 & 1.87 & 0 & 0 & 0 & 3.91 & 0.02 & 0.03 & 0.03 & 3.20 & 0.02 & 0.03 & 0.03 & 1.41 & 0.03 & 0.03 & 0.03 & 0.04 \\
 & 30 & 0 & 0 & 0 & 0 & 0.03 & 0.03 & 0.04 & 0.04 & 0.29 & 0.03 & 0.03 & 0.03 & 0.35 & 0.03 & 0.03 & 0.03 & 0.04 \\
\hline
20 & 10 & 19.39 & 0.01 & 0 & 0 & 14.64 & 3.31 & 0.01 & 0.02 & 9.82 & 5.86 & 0.02 & 0.02 & 4.93 & 3.59 & 0.25 & 0.01 & 0.01 \\
 & 20 & 0.01 & 0 & 0 & 0 & 4.08 & 0.02 & 0.02 & 0.03 & 5.62 & 0.02 & 0.02 & 0.03 & 3.20 & 0.02 & 0.03 & 0.03 & 0.03 \\
 & 30 & 0 & 0 & 0 & 0 & 0.01 & 0.02 & 0.03 & 0.03 & 0.04 & 0.03 & 0.03 & 0.04 & 0.31 & 0.03 & 0.03 & 0.03 & 0.04 \\
\hline
50 & 10 & 49.92 & 0 & 0 & 0 & 36.98 & 1.14 & 0 & 0.01 & 25 & 19.84 & 0 & 0 & 12 & 11.47 & 1.41 & 0 & 0 \\
 & 20 & 0 & 0 & 0 & 0 & 0.39 & 0.01 & 0.02 & 0.02 & 12.98 & 0.01 & 0.01 & 0.02 & 7.80 & 0.01 & 0.01 & 0.02 & 0.01 \\
 & 30 & 0 & 0 & 0 & 0 & 0.01 & 0.02 & 0.02 & 0.03 & 0 & 0.01 & 0.02 & 0.02 & 0.25 & 0.02 & 0.02 & 0.03 & 0.02 \\
\hline
100 & 10 & 100 & 100 & 100 & 100 & 75 & 75 & 75 & 75 & 50 & 50 & 50 & 50 & 25 & 25 & 25 & 25 & 0 \\
 & 20 & 0 & 0 & 0 & 0 & 0 & 0 & 0.01 & 0.01 & 31.38 & 0 & 0 & 0.01 & 19.63 & 0 & 0 & 0 & 0 \\
 & 30 & 0 & 0 & 0 & 0 & 0 & 0.01 & 0.02 & 0.02 & 0 & 0 & 0.01 & 0.01 & 0.85 & 0 & 0 & 0 & 0 \\
\hline
300 & 10 & 300 & 300 & 300 & 300 & 225 & 225 & 225 & 225 & 150 & 150 & 150 & 150 & 75 & 75 & 75 & 75 & 0 \\
 & 20 & 300 & 134.84 & 0 & 0 & 225 & 225 & 0 & 0 & 150 & 150 & 0 & 0 & 75 & 75 & 65.67 & 0 & 0 \\
 & 30 & 0 & 0 & 0 & 0 & 0 & 0 & 0 & 0 & 0.03 & 0 & 0 & 0 & 59.35 & 0 & 0 & 0 & 0
\\
\hline
\multicolumn{19}{|c|}{Empirical size}\\
\hline
\hline
\multicolumn{2}{|c||}{} & \multicolumn{4}{c||}{$s=0$, $a=$} & \multicolumn{4}{c||}{$s=\lceil p/4\rceil$, $a=$} & \multicolumn{4}{c||}{$s=\lceil p/2\rceil$, $a=$} & \multicolumn{4}{c||}{$s=\lceil 3p/4\rceil$, $a=$} & \multicolumn{1}{c|}{$s=p$} \\
  $p$ & $T/p$ & 0.25 & 0.5 & 0.75 & 1 & 0.25 & 0.5 & 0.75 & 1 & 0.25 & 0.5 & 0.75 & 1 & 0.25 & 0.5 & 0.75 & 1 & \\
\hline
10 & 10 &  &  &  &  & 0 & 0 & 0.01 & 0.02 & 0 & 0 & 0.01 & 0.02 & 0 & 0 & 0.01 & 0.02 & 0.02 \\
 & 20 &  &  &  &  & 0.01 & 0.02 & 0.03 & 0.03 & 0 & 0.02 & 0.03 & 0.03 & 0 & 0.02 & 0.03 & 0.03 & 0.04 \\
 & 30 &  &  &  &  & 0.02 & 0.03 & 0.04 & 0.04 & 0.01 & 0.03 & 0.03 & 0.03 & 0.01 & 0.03 & 0.03 & 0.03 & 0.04 \\
\hline
20 & 10 &  &  &  &  & 0 & 0 & 0.01 & 0.02 & 0 & 0 & 0.01 & 0.02 & 0 & 0 & 0 & 0.01 & 0.01 \\
 & 20 &  &  &  &  & 0 & 0.02 & 0.02 & 0.03 & 0 & 0.02 & 0.02 & 0.03 & 0 & 0.02 & 0.03 & 0.03 & 0.03 \\
 & 30 &  &  &  &  & 0.01 & 0.02 & 0.03 & 0.03 & 0.01 & 0.03 & 0.03 & 0.03 & 0.01 & 0.03 & 0.03 & 0.03 & 0.04 \\
\hline
50 & 10 &  &  &  &  & 0 & 0 & 0 & 0.01 & 0 & 0 & 0 & 0 & 0 & 0 & 0 & 0 & 0 \\
 & 20 &  &  &  &  & 0 & 0.01 & 0.02 & 0.02 & 0 & 0.01 & 0.01 & 0.02 & 0 & 0.01 & 0.01 & 0.02 & 0.01 \\
 & 30 &  &  &  &  & 0.01 & 0.02 & 0.02 & 0.03 & 0 & 0.01 & 0.02 & 0.02 & 0 & 0.02 & 0.02 & 0.03 & 0.02 \\
\hline
100 & 10 &  &  &  &  & 0 & 0 & 0 & 0 & 0 & 0 & 0 & 0 & 0 & 0 & 0 & 0 & 0 \\
 & 20 &  &  &  &  & 0 & 0 & 0.01 & 0.01 & 0 & 0 & 0 & 0.01 & 0 & 0 & 0 & 0 & 0 \\
 & 30 &  &  &  &  & 0 & 0.01 & 0.02 & 0.02 & 0 & 0 & 0.01 & 0.01 & 0 & 0 & 0 & 0 & 0 \\
\hline
300 & 10 &  &  &  &  & 0 & 0 & 0 & 0 & 0 & 0 & 0 & 0 & 0 & 0 & 0 & 0 & 0 \\
 & 20 &  &  &  &  & 0 & 0 & 0 & 0 & 0 & 0 & 0 & 0 & 0 & 0 & 0 & 0 & 0 \\
 & 30 &  &  &  &  & 0 & 0 & 0 & 0 & 0 & 0 & 0 & 0 & 0 & 0 & 0 & 0 & 0
\\
\hline
\end{tabular}
\end{center}
\caption{\footnotesize $F_{j,\infty}$. Top panel: frequency of correct selection of $s$, $\frac1N\sum_{j=1}^N1_{\wt s_j =s}$, with test sequence at $5\%$ significance level; second from top panel: Mean Absolute Error (MAE), $\frac1N\sum_{j=1}^N|\wt s_j - s|$; third from top panel: MC size of test $F_{s,\infty}$. Note that $F_{0,\infty}$ is undefined.
{\Ntext} replications of DGP \eqref{eq_DGP_OW}.} \label{table_mx}
\end{table}

\cp

\begin{table}[htbp]\tiny
\renewcommand{\arraystretch}{0.6}
\addtolength{\tabcolsep}{-1pt}
\begin{center}
\begin{tabular}{|cc||cccc||cccc||cccc||cccc||c|}
\hline
\multicolumn{19}{|c|}{$F_{j,1}$}\\
\hline
\hline
\multicolumn{19}{|c|}{Frequency of correct selection}\\
\hline
\hline
\multicolumn{2}{|c||}{} & \multicolumn{4}{c||}{$s=0$, $a=$} & \multicolumn{4}{c||}{$s=\lceil p/4\rceil$, $a=$} & \multicolumn{4}{c||}{$s=\lceil p/2\rceil$, $a=$} & \multicolumn{4}{c||}{$s=\lceil 3p/4\rceil$, $a=$} & \multicolumn{1}{c|}{$s=p$} \\
  $p$ & $T/p$ & 0.25 & 0.5 & 0.75 & 1 & 0.25 & 0.5 & 0.75 & 1 & 0.25 & 0.5 & 0.75 & 1 & 0.25 & 0.5 & 0.75 & 1 & \\
\hline
10 & 10 & 0.31 & 1 & 1 & 1 & 0 & 0.19 & 0.95 & 0.98 & 0 & 0.05 & 0.58 & 0.97 & 0 & 0.07 & 0.36 & 0.76 & 0.99 \\
 & 20 & 1 & 1 & 1 & 1 & 0.22 & 0.98 & 0.97 & 0.97 & 0.06 & 0.96 & 0.98 & 0.98 & 0.09 & 0.82 & 0.98 & 0.98 & 0.98 \\
 & 30 & 1 & 1 & 1 & 1 & 0.94 & 0.97 & 0.97 & 0.97 & 0.58 & 0.98 & 0.98 & 0.97 & 0.45 & 0.98 & 0.98 & 0.98 & 0.98 \\
\hline
20 & 10 & 1 & 1 & 1 & 1 & 0 & 0.27 & 0.99 & 0.99 & 0 & 0.01 & 0.38 & 0.94 & 0 & 0 & 0.12 & 0.48 & 1 \\
 & 20 & 1 & 1 & 1 & 1 & 0.36 & 0.99 & 0.98 & 0.98 & 0.01 & 0.96 & 0.99 & 0.99 & 0.01 & 0.68 & 0.99 & 0.99 & 0.99 \\
 & 30 & 1 & 1 & 1 & 1 & 0.99 & 0.98 & 0.98 & 0.98 & 0.50 & 0.99 & 0.98 & 0.98 & 0.26 & 0.99 & 0.99 & 0.99 & 0.99 \\
\hline
50 & 10 & 1 & 1 & 1 & 1 & 0 & 0.02 & 0.94 & 1 & 0 & 0 & 0.01 & 0.34 & 0 & 0 & 0 & 0.01 & 1 \\
 & 20 & 1 & 1 & 1 & 1 & 0.08 & 1 & 0.99 & 0.99 & 0 & 0.62 & 1 & 1 & 0 & 0.09 & 0.72 & 0.97 & 1 \\
 & 30 & 1 & 1 & 1 & 1 & 0.99 & 0.99 & 0.99 & 0.99 & 0.10 & 1 & 1 & 1 & 0.01 & 0.90 & 1 & 1 & 1 \\
\hline
100 & 10 & 1 & 1 & 1 & 1 & 0 & 0 & 0.40 & 1 & 0 & 0 & 0 & 0 & 0 & 0 & 0 & 0 & 1 \\
 & 20 & 1 & 1 & 1 & 1 & 0 & 1 & 1 & 1 & 0 & 0.01 & 0.75 & 1 & 0 & 0 & 0 & 0.07 & 1 \\
 & 30 & 1 & 1 & 1 & 1 & 0.88 & 1 & 1 & 1 & 0 & 0.94 & 1 & 1 & 0 & 0.04 & 0.66 & 0.96 & 1 \\
\hline
300 & 10 & 1 & 1 & 1 & 1 & 0 & 0 & 0 & 0 & 0 & 0 & 0 & 0 & 0 & 0 & 0 & 0 & 1 \\
 & 20 & 1 & 1 & 1 & 1 & 0 & 0 & 0.96 & 1 & 0 & 0 & 0 & 0 & 0 & 0 & 0 & 0 & 1 \\
 & 30 & 1 & 1 & 1 & 1 & 0 & 1 & 1 & 1 & 0 & 0 & 0 & 0 & 0 & 0 & 0 & 0 & 1
\\
\hline
\hline
\multicolumn{19}{|c|}{MAE}\\
\hline
\hline
\multicolumn{2}{|c||}{} & \multicolumn{4}{c||}{$s=0$, $a=$} & \multicolumn{4}{c||}{$s=\lceil p/4\rceil$, $a=$} & \multicolumn{4}{c||}{$s=\lceil p/2\rceil$, $a=$} & \multicolumn{4}{c||}{$s=\lceil 3p/4\rceil$, $a=$} & \multicolumn{1}{c|}{$s=p$} \\
  $p$ & $T/p$ & 0.25 & 0.5 & 0.75 & 1 & 0.25 & 0.5 & 0.75 & 1 & 0.25 & 0.5 & 0.75 & 1 & 0.25 & 0.5 & 0.75 & 1 & \\
\hline
10 & 10 & 2.57 & 0 & 0 & 0 & 5.06 & 0.97 & 0.05 & 0.02 & 4.18 & 1.48 & 0.42 & 0.03 & 1.91 & 1.28 & 0.65 & 0.24 & 0.01 \\
 & 20 & 0 & 0 & 0 & 0 & 0.91 & 0.02 & 0.03 & 0.03 & 1.43 & 0.04 & 0.02 & 0.02 & 1.19 & 0.18 & 0.02 & 0.02 & 0.02 \\
 & 30 & 0 & 0 & 0 & 0 & 0.06 & 0.03 & 0.03 & 0.03 & 0.42 & 0.02 & 0.02 & 0.03 & 0.55 & 0.02 & 0.02 & 0.02 & 0.02 \\
\hline
20 & 10 & 0 & 0 & 0 & 0 & 7.90 & 0.74 & 0.01 & 0.01 & 7.28 & 1.97 & 0.62 & 0.06 & 4.54 & 2.16 & 0.98 & 0.52 & 0 \\
 & 20 & 0 & 0 & 0 & 0 & 0.65 & 0.01 & 0.02 & 0.02 & 1.85 & 0.04 & 0.01 & 0.01 & 1.96 & 0.32 & 0.01 & 0.01 & 0.01 \\
 & 30 & 0 & 0 & 0 & 0 & 0.01 & 0.02 & 0.02 & 0.02 & 0.50 & 0.01 & 0.01 & 0.02 & 0.78 & 0.01 & 0.01 & 0.01 & 0.01 \\
\hline
50 & 10 & 0 & 0 & 0 & 0 & 14.58 & 1.18 & 0.06 & 0 & 16.08 & 3.72 & 1.27 & 0.66 & 10.48 & 4.63 & 2.31 & 1.45 & 0 \\
 & 20 & 0 & 0 & 0 & 0 & 0.99 & 0 & 0.01 & 0.01 & 3.32 & 0.38 & 0 & 0 & 3.95 & 0.94 & 0.28 & 0.03 & 0 \\
 & 30 & 0 & 0 & 0 & 0 & 0.01 & 0.01 & 0.01 & 0.01 & 0.96 & 0 & 0 & 0 & 1.56 & 0.10 & 0 & 0 & 0 \\
\hline
100 & 10 & 0 & 0 & 0 & 0 & 25.23 & 2.01 & 0.60 & 0 & 32.03 & 7.28 & 2.83 & 1.58 & 22.49 & 10.11 & 5.40 & 3.66 & 0 \\
 & 20 & 0 & 0 & 0 & 0 & 1.62 & 0 & 0 & 0 & 6.29 & 1.04 & 0.25 & 0 & 8.13 & 2.22 & 1.11 & 0.93 & 0 \\
 & 30 & 0 & 0 & 0 & 0 & 0.12 & 0 & 0 & 0 & 1.98 & 0.06 & 0 & 0 & 3.32 & 0.96 & 0.34 & 0.04 & 0 \\
\hline
300 & 10 & 0 & 0 & 0 & 0 & 90.99 & 6.92 & 2.64 & 1.21 & 126.50 & 31.28 & 13.61 & 8.54 & 75 & 61.70 & 38.74 & 29.26 & 0 \\
 & 20 & 0 & 0 & 0 & 0 & 5.48 & 1 & 0.04 & 0 & 24.94 & 5.90 & 3.09 & 2.12 & 39.06 & 14.63 & 9.70 & 7.88 & 0 \\
 & 30 & 0 & 0 & 0 & 0 & 1.79 & 0 & 0 & 0 & 8.72 & 2.67 & 1.57 & 1 & 18.11 & 7.20 & 5.05 & 4.17 & 0
\\
\hline
\multicolumn{19}{|c|}{Empirical size}\\
\hline
\hline
\multicolumn{2}{|c||}{} & \multicolumn{4}{c||}{$s=0$, $a=$} & \multicolumn{4}{c||}{$s=\lceil p/4\rceil$, $a=$} & \multicolumn{4}{c||}{$s=\lceil p/2\rceil$, $a=$} & \multicolumn{4}{c||}{$s=\lceil 3p/4\rceil$, $a=$} & \multicolumn{1}{c|}{$s=p$} \\
  $p$ & $T/p$ & 0.25 & 0.5 & 0.75 & 1 & 0.25 & 0.5 & 0.75 & 1 & 0.25 & 0.5 & 0.75 & 1 & 0.25 & 0.5 & 0.75 & 1 & \\
\hline
10 & 10 &  &  &  &  & 0 & 0 & 0.01 & 0.02 & 0 & 0 & 0.01 & 0.01 & 0 & 0 & 0.01 & 0.01 & 0.01 \\
 & 20 &  &  &  &  & 0.01 & 0.02 & 0.03 & 0.03 & 0 & 0.01 & 0.02 & 0.02 & 0 & 0.01 & 0.02 & 0.02 & 0.02 \\
 & 30 &  &  &  &  & 0.02 & 0.03 & 0.03 & 0.03 & 0.01 & 0.02 & 0.02 & 0.03 & 0.01 & 0.02 & 0.02 & 0.02 & 0.02 \\
\hline
20 & 10 &  &  &  &  & 0 & 0 & 0.01 & 0.01 & 0 & 0 & 0 & 0 & 0 & 0 & 0 & 0 & 0 \\
 & 20 &  &  &  &  & 0 & 0.01 & 0.02 & 0.02 & 0 & 0.01 & 0.01 & 0.01 & 0 & 0 & 0.01 & 0.01 & 0.01 \\
 & 30 &  &  &  &  & 0.01 & 0.02 & 0.02 & 0.02 & 0 & 0.01 & 0.01 & 0.02 & 0 & 0.01 & 0.01 & 0.01 & 0.01 \\
\hline
50 & 10 &  &  &  &  & 0 & 0 & 0 & 0 & 0 & 0 & 0 & 0 & 0 & 0 & 0 & 0 & 0 \\
 & 20 &  &  &  &  & 0 & 0 & 0.01 & 0.01 & 0 & 0 & 0 & 0 & 0 & 0 & 0 & 0 & 0 \\
 & 30 &  &  &  &  & 0 & 0.01 & 0.01 & 0.01 & 0 & 0 & 0 & 0 & 0 & 0 & 0 & 0 & 0 \\
\hline
100 & 10 &  &  &  &  & 0 & 0 & 0 & 0 & 0 & 0 & 0 & 0 & 0 & 0 & 0 & 0 & 0 \\
 & 20 &  &  &  &  & 0 & 0 & 0 & 0 & 0 & 0 & 0 & 0 & 0 & 0 & 0 & 0 & 0 \\
 & 30 &  &  &  &  & 0 & 0 & 0 & 0 & 0 & 0 & 0 & 0 & 0 & 0 & 0 & 0 & 0 \\
\hline
300 & 10 &  &  &  &  & 0 & 0 & 0 & 0 & 0 & 0 & 0 & 0 & 0 & 0 & 0 & 0 & 0 \\
 & 20 &  &  &  &  & 0 & 0 & 0 & 0 & 0 & 0 & 0 & 0 & 0 & 0 & 0 & 0 & 0 \\
 & 30 &  &  &  &  & 0 & 0 & 0 & 0 & 0 & 0 & 0 & 0 & 0 & 0 & 0 & 0 & 0
\\
\hline
\end{tabular}
\end{center}
\caption{\footnotesize $F_{j,1}$. Top panel: frequency of correct selection of $s$, $\frac1N\sum_{n=1}^N1_{\wt s_n =s}$, with test sequence at $5\%$ significance level; second from top panel: Mean Absolute Error (MAE), $\frac1N\sum_{j=1}^N|\wt s_j - s|$; third from top panel: MC size of test $F_{s,1}$.
Note that $F_{0,1}$ is undefined.
{\Ntext} replications of DGP \eqref{eq_DGP_OW}.} \label{table_tr}
\end{table}

\cp

\begin{table}[htbp]\tiny
	\renewcommand{\arraystretch}{0.8}
	\addtolength{\tabcolsep}{-1pt}
	\begin{center}
\begin{tabular}{|cc||cccc||cccc||cccc||cccc||c|}
\hline
\multicolumn{19}{|c|}{Hybrid with $F_{p,\infty}$}\\
\hline
\hline
\multicolumn{19}{|c|}{Frequency of correct selection}\\
\hline
\hline
\multicolumn{2}{|c||}{} & \multicolumn{4}{c||}{$s=0$, $a=$} & \multicolumn{4}{c||}{$s=\lceil p/4\rceil$, $a=$} & \multicolumn{4}{c||}{$s=\lceil p/2\rceil$, $a=$} & \multicolumn{4}{c||}{$s=\lceil 3p/4\rceil$, $a=$} & \multicolumn{1}{c|}{$s=p$} \\
  $p$ & $T/p$ & 0.25 & 0.5 & 0.75 & 1 & 0.25 & 0.5 & 0.75 & 1 & 0.25 & 0.5 & 0.75 & 1 & 0.25 & 0.5 & 0.75 & 1 & \\
\hline
10 & 10 & 0 & 0.65 & 1 & 1 & 0 & 0.40 & 0.96 & 1 & 0 & 0.26 & 0.87 & 0.99 & 0.01 & 0.18 & 0.75 & 0.96 & 0.98 \\
 & 20 & 0.46 & 1 & 1 & 1 & 0.26 & 0.99 & 1 & 1 & 0.18 & 0.95 & 1 & 1 & 0.17 & 0.89 & 1 & 1 & 0.96 \\
 & 30 & 0.88 & 1 & 1 & 1 & 0.76 & 1 & 1 & 1 & 0.63 & 1 & 1 & 1 & 0.62 & 0.99 & 1 & 1 & 0.96 \\
\hline
20 & 10 & 0 & 1 & 1 & 1 & 0 & 0.83 & 1 & 1 & 0 & 0.40 & 0.95 & 1 & 0 & 0.20 & 0.84 & 0.98 & 0.99 \\
 & 20 & 0.88 & 1 & 1 & 1 & 0.52 & 1 & 1 & 1 & 0.19 & 1 & 1 & 1 & 0.14 & 0.98 & 1 & 1 & 0.97 \\
 & 30 & 1 & 1 & 1 & 1 & 0.96 & 1 & 1 & 1 & 0.82 & 1 & 1 & 1 & 0.64 & 1 & 1 & 1 & 0.96 \\
\hline
50 & 10 & 0.03 & 1 & 1 & 1 & 0 & 0.98 & 1 & 1 & 0 & 0.55 & 0.99 & 1 & 0 & 0.06 & 0.88 & 0.99 & 1 \\
 & 20 & 1 & 1 & 1 & 1 & 0.89 & 1 & 1 & 1 & 0.24 & 1 & 1 & 1 & 0.07 & 1 & 1 & 1 & 0.99 \\
 & 30 & 1 & 1 & 1 & 1 & 1 & 1 & 1 & 1 & 0.98 & 1 & 1 & 1 & 0.83 & 1 & 1 & 1 & 0.98 \\
\hline
100 & 10 & 0 & 0 & 0 & 0 & 0 & 0 & 0 & 0 & 0 & 0 & 0 & 0 & 0 & 0 & 0 & 0 & 1 \\
 & 20 & 1 & 1 & 1 & 1 & 1 & 1 & 1 & 1 & 0.33 & 1 & 1 & 1 & 0.03 & 1 & 1 & 1 & 1 \\
 & 30 & 1 & 1 & 1 & 1 & 1 & 1 & 1 & 1 & 1 & 1 & 1 & 1 & 0.94 & 1 & 1 & 1 & 1 \\
\hline
300 & 10 & 0 & 0 & 0 & 0 & 0 & 0 & 0 & 0 & 0 & 0 & 0 & 0 & 0 & 0 & 0 & 0 & 1 \\
 & 20 & 0 & 0.62 & 1 & 1 & 0 & 0 & 1 & 1 & 0 & 0 & 1 & 1 & 0 & 0 & 0.25 & 1 & 1 \\
 & 30 & 1 & 1 & 1 & 1 & 1 & 1 & 1 & 1 & 1 & 1 & 1 & 1 & 0.99 & 1 & 1 & 1 & 1
\\
\hline
\hline
\multicolumn{19}{|c|}{MAE}\\
\hline
\hline
\multicolumn{2}{|c||}{} & \multicolumn{4}{c||}{$s=0$, $a=$} & \multicolumn{4}{c||}{$s=\lceil p/4\rceil$, $a=$} & \multicolumn{4}{c||}{$s=\lceil p/2\rceil$, $a=$} & \multicolumn{4}{c||}{$s=\lceil 3p/4\rceil$, $a=$} & \multicolumn{1}{c|}{$s=p$} \\
  $p$ & $T/p$ & 0.25 & 0.5 & 0.75 & 1 & 0.25 & 0.5 & 0.75 & 1 & 0.25 & 0.5 & 0.75 & 1 & 0.25 & 0.5 & 0.75 & 1 & \\
\hline
10 & 10 & 9.52 & 2.74 & 0.01 & 0 & 6.71 & 2.82 & 0.09 & 0 & 4.81 & 2.39 & 0.22 & 0.01 & 1.92 & 1.36 & 0.27 & 0.05 & 0.03 \\
 & 20 & 4.55 & 0 & 0 & 0 & 3.80 & 0.05 & 0 & 0 & 2.80 & 0.11 & 0 & 0 & 1.33 & 0.11 & 0 & 0 & 0.04 \\
 & 30 & 1.03 & 0 & 0 & 0 & 1.27 & 0 & 0 & 0 & 1.15 & 0 & 0 & 0 & 0.41 & 0.01 & 0 & 0 & 0.04 \\
\hline
20 & 10 & 19.07 & 0.04 & 0 & 0 & 14.46 & 1.54 & 0 & 0 & 9.74 & 3.19 & 0.05 & 0 & 4.92 & 2.40 & 0.21 & 0.02 & 0.01 \\
 & 20 & 2.23 & 0 & 0 & 0 & 6.12 & 0 & 0 & 0 & 5.92 & 0.01 & 0 & 0 & 2.63 & 0.04 & 0 & 0 & 0.04 \\
 & 30 & 0.07 & 0 & 0 & 0 & 0.49 & 0 & 0 & 0 & 1.27 & 0 & 0 & 0 & 0.97 & 0 & 0 & 0 & 0.04 \\
\hline
50 & 10 & 48.22 & 0 & 0 & 0 & 36.96 & 0.04 & 0 & 0 & 25 & 2.48 & 0.01 & 0 & 12 & 8.96 & 0.15 & 0.01 & 0 \\
 & 20 & 0 & 0 & 0 & 0 & 3.31 & 0 & 0 & 0 & 14.95 & 0 & 0 & 0 & 7.59 & 0 & 0 & 0 & 0.01 \\
 & 30 & 0 & 0 & 0 & 0 & 0 & 0 & 0 & 0 & 0.29 & 0 & 0 & 0 & 0.99 & 0 & 0 & 0 & 0.02 \\
\hline
100 & 10 & 100 & 100 & 100 & 100 & 75 & 75 & 75 & 75 & 50 & 50 & 50 & 50 & 25 & 25 & 25 & 25 & 0 \\
 & 20 & 0 & 0 & 0 & 0 & 0.16 & 0 & 0 & 0 & 24.55 & 0 & 0 & 0 & 17.94 & 0 & 0 & 0 & 0 \\
 & 30 & 0 & 0 & 0 & 0 & 0 & 0 & 0 & 0 & 0.01 & 0 & 0 & 0 & 0.44 & 0 & 0 & 0 & 0 \\
\hline
300 & 10 & 300 & 300 & 300 & 300 & 225 & 225 & 225 & 225 & 150 & 150 & 150 & 150 & 75 & 75 & 75 & 75 & 0 \\
 & 20 & 300 & 113.73 & 0 & 0 & 225 & 224.98 & 0 & 0 & 150 & 150 & 0 & 0 & 75 & 75 & 55.94 & 0 & 0 \\
 & 30 & 0 & 0 & 0 & 0 & 0 & 0 & 0 & 0 & 0 & 0 & 0 & 0 & 0.02 & 0 & 0 & 0 & 0
\\
\hline
\end{tabular}
	\end{center}
	\caption{\footnotesize Hybrid with $F_{p,\infty}$. Top panel: frequency of correct selection of $s$, $\frac1N\sum_{n=1}^N1_{\wh s_n =s}$; bottom panel: Mean Absolute Error (MAE), $\frac1N\sum_{n=1}^N|\wh s_n - s|$. {\Ntext} replications of DGP \eqref{eq_DGP_OW}.} \label{table_hyb_mx}
\end{table}

\cp

\begin{table}[htbp]\tiny
	\renewcommand{\arraystretch}{0.8}
	\addtolength{\tabcolsep}{-1pt}
	\begin{center}
\begin{tabular}{|cc||cccc||cccc||cccc||cccc||c|}
\hline
\multicolumn{19}{|c|}{Hybrid with $F_{p,1}$}\\
\hline
\hline
\multicolumn{19}{|c|}{Frequency of correct selection}\\
\hline
\hline
\multicolumn{2}{|c||}{} & \multicolumn{4}{c||}{$s=0$, $a=$} & \multicolumn{4}{c||}{$s=\lceil p/4\rceil$, $a=$} & \multicolumn{4}{c||}{$s=\lceil p/2\rceil$, $a=$} & \multicolumn{4}{c||}{$s=\lceil 3p/4\rceil$, $a=$} & \multicolumn{1}{c|}{$s=p$} \\
  $p$ & $T/p$ & 0.25 & 0.5 & 0.75 & 1 & 0.25 & 0.5 & 0.75 & 1 & 0.25 & 0.5 & 0.75 & 1 & 0.25 & 0.5 & 0.75 & 1 & \\
\hline
10 & 10 & 0 & 0.65 & 1 & 1 & 0.01 & 0.42 & 0.96 & 1 & 0.02 & 0.32 & 0.87 & 0.99 & 0.03 & 0.27 & 0.77 & 0.96 & 0.99 \\
 & 20 & 0.46 & 1 & 1 & 1 & 0.28 & 0.99 & 1 & 1 & 0.22 & 0.95 & 1 & 1 & 0.27 & 0.89 & 1 & 1 & 0.98 \\
 & 30 & 0.88 & 1 & 1 & 1 & 0.76 & 1 & 1 & 1 & 0.63 & 1 & 1 & 1 & 0.64 & 0.99 & 1 & 1 & 0.98 \\
\hline
20 & 10 & 0.02 & 1 & 1 & 1 & 0 & 0.83 & 1 & 1 & 0 & 0.40 & 0.95 & 1 & 0.02 & 0.25 & 0.84 & 0.98 & 1 \\
 & 20 & 0.88 & 1 & 1 & 1 & 0.52 & 1 & 1 & 1 & 0.19 & 1 & 1 & 1 & 0.15 & 0.98 & 1 & 1 & 0.99 \\
 & 30 & 1 & 1 & 1 & 1 & 0.96 & 1 & 1 & 1 & 0.82 & 1 & 1 & 1 & 0.64 & 1 & 1 & 1 & 0.99 \\
\hline
50 & 10 & 0.83 & 1 & 1 & 1 & 0 & 0.98 & 1 & 1 & 0 & 0.55 & 0.99 & 1 & 0.01 & 0.17 & 0.88 & 0.99 & 1 \\
 & 20 & 1 & 1 & 1 & 1 & 0.89 & 1 & 1 & 1 & 0.24 & 1 & 1 & 1 & 0.07 & 1 & 1 & 1 & 1 \\
 & 30 & 1 & 1 & 1 & 1 & 1 & 1 & 1 & 1 & 0.98 & 1 & 1 & 1 & 0.83 & 1 & 1 & 1 & 1 \\
\hline
100 & 10 & 1 & 1 & 1 & 1 & 0 & 1 & 1 & 1 & 0 & 0.55 & 1 & 1 & 0.01 & 0.11 & 0.89 & 1 & 1 \\
 & 20 & 1 & 1 & 1 & 1 & 1 & 1 & 1 & 1 & 0.33 & 1 & 1 & 1 & 0.03 & 1 & 1 & 1 & 1 \\
 & 30 & 1 & 1 & 1 & 1 & 1 & 1 & 1 & 1 & 1 & 1 & 1 & 1 & 0.94 & 1 & 1 & 1 & 1 \\
\hline
300 & 10 & 1 & 1 & 1 & 1 & 0.01 & 1 & 1 & 1 & 0 & 0.33 & 1 & 1 & 0 & 0.02 & 0.59 & 1 & 1 \\
 & 20 & 1 & 1 & 1 & 1 & 1 & 1 & 1 & 1 & 0.47 & 1 & 1 & 1 & 0.01 & 1 & 1 & 1 & 1 \\
 & 30 & 1 & 1 & 1 & 1 & 1 & 1 & 1 & 1 & 1 & 1 & 1 & 1 & 0.99 & 1 & 1 & 1 & 1
\\
\hline
\hline
\multicolumn{19}{|c|}{MAE}\\
\hline
\hline
\multicolumn{2}{|c||}{} & \multicolumn{4}{c||}{$s=0$, $a=$} & \multicolumn{4}{c||}{$s=\lceil p/4\rceil$, $a=$} & \multicolumn{4}{c||}{$s=\lceil p/2\rceil$, $a=$} & \multicolumn{4}{c||}{$s=\lceil 3p/4\rceil$, $a=$} & \multicolumn{1}{c|}{$s=p$} \\
  $p$ & $T/p$ & 0.25 & 0.5 & 0.75 & 1 & 0.25 & 0.5 & 0.75 & 1 & 0.25 & 0.5 & 0.75 & 1 & 0.25 & 0.5 & 0.75 & 1 & \\
\hline
10 & 10 & 8.15 & 2.73 & 0.01 & 0 & 5.27 & 2.65 & 0.09 & 0 & 3.85 & 1.92 & 0.22 & 0.01 & 1.90 & 1.11 & 0.25 & 0.05 & 0.01 \\
 & 20 & 4.52 & 0 & 0 & 0 & 3.65 & 0.05 & 0 & 0 & 2.42 & 0.11 & 0 & 0 & 1.03 & 0.11 & 0 & 0 & 0.03 \\
 & 30 & 1.03 & 0 & 0 & 0 & 1.27 & 0 & 0 & 0 & 1.15 & 0 & 0 & 0 & 0.37 & 0.01 & 0 & 0 & 0.03 \\
\hline
20 & 10 & 17.27 & 0.04 & 0 & 0 & 12.65 & 1.54 & 0 & 0 & 7.69 & 3.16 & 0.05 & 0 & 4.12 & 1.83 & 0.21 & 0.02 & 0 \\
 & 20 & 2.23 & 0 & 0 & 0 & 6.12 & 0 & 0 & 0 & 5.91 & 0.01 & 0 & 0 & 2.42 & 0.04 & 0 & 0 & 0.01 \\
 & 30 & 0.07 & 0 & 0 & 0 & 0.49 & 0 & 0 & 0 & 1.27 & 0 & 0 & 0 & 0.97 & 0 & 0 & 0 & 0.02 \\
\hline
50 & 10 & 7.94 & 0 & 0 & 0 & 33 & 0.04 & 0 & 0 & 21.31 & 2.14 & 0.01 & 0 & 8.89 & 3.75 & 0.15 & 0.01 & 0 \\
 & 20 & 0 & 0 & 0 & 0 & 3.31 & 0 & 0 & 0 & 14.95 & 0 & 0 & 0 & 7.59 & 0 & 0 & 0 & 0 \\
 & 30 & 0 & 0 & 0 & 0 & 0 & 0 & 0 & 0 & 0.29 & 0 & 0 & 0 & 0.99 & 0 & 0 & 0 & 0 \\
\hline
100 & 10 & 0 & 0 & 0 & 0 & 66.74 & 0 & 0 & 0 & 42.71 & 1 & 0 & 0 & 18.86 & 5.36 & 0.16 & 0 & 0 \\
 & 20 & 0 & 0 & 0 & 0 & 0.16 & 0 & 0 & 0 & 24.55 & 0 & 0 & 0 & 17.94 & 0 & 0 & 0 & 0 \\
 & 30 & 0 & 0 & 0 & 0 & 0 & 0 & 0 & 0 & 0.01 & 0 & 0 & 0 & 0.44 & 0 & 0 & 0 & 0 \\
\hline
300 & 10 & 0 & 0 & 0 & 0 & 139.04 & 0 & 0 & 0 & 85.73 & 2.27 & 0 & 0 & 75 & 16.71 & 1.55 & 0 & 0 \\
 & 20 & 0 & 0 & 0 & 0 & 0 & 0 & 0 & 0 & 30.16 & 0 & 0 & 0 & 56.60 & 0 & 0 & 0 & 0 \\
 & 30 & 0 & 0 & 0 & 0 & 0 & 0 & 0 & 0 & 0 & 0 & 0 & 0 & 0.02 & 0 & 0 & 0 & 0
\\
\hline
\end{tabular}
	\end{center}
	\caption{\footnotesize Hybrid with $F_{p,1}$. Top panel: frequency of correct selection of $s$, $\frac1N\sum_{n=1}^N1_{\wh s_n =s}$; bottom panel: Mean Absolute Error (MAE), $\frac1N\sum_{n=1}^N|\wh s_n - s|$. {\Ntext} replications of DGP \eqref{eq_DGP_OW}.} \label{table_hyb_tr}
\end{table}

\cp

\subsection{Inference on $\psi$}\label{sec_app_sim_psi}

For hypotheses tests on $\psi$, the estimation of the long-run variance matrix $\Omega_{22.1}$ was performed using \eqref{eq_Xihat} (LRV-P),
as well as the estimator in \cite{An:91} and \cite{AM:92} with Parzen's kernel (LRV-A). The latter calculations were performed using the
\texttt{hac.m} function in \textsc{matlab}, see \cite{Mat:25}; the bandwidth was calculated for the first MC replication and it was kept fixed for the remaining ones. The unfeasible solution LRV-U was computed substituting $\Omega_{22.1}$ with $a^{-2}I_r$, the value in the DGP.

Table \tref{table_W1_Wr} reports the MC frequency of rejection under the two null hypotheses on $\psi$ introduced in the main text of the paper, a scalar hypothesis
tested by means of a $t$-test and an $r$-vector hypothesis assessed using a Wald $Q$ test,
over {\Ntext} replications of DGP \eqref{eq_DGP_OW} at the $5\%$ nominal significance level. The top panel reports results for the $t$-test of the scalar hypothesis
and the bottom panel for the $Q$ test of the vector hypothesis.

The $t$-test is oversized in finite samples for all LRV estimators considered. The best performance is observed
for the unfeasible LRV-U variant. The MC test size approaches the nominal level as $a$ and $T$ increase. Results for $p=10$ and $p=20$ are very similar.
Tests employing the LRV-P and LRV-A variants are more oversized; their size also becomes closer to the nominal level for large $T$, especially so when $a$ is large.

The Wald $Q$ test is more pronouncedly oversized in finite samples than the corresponding $t$-test for all LRV choices.
The MC sizes tend to be closer to the nominal level as $a$ and $T$ increase, though convergence is slower than for the $t$-test.
Comparing values for $p=10,20$, the MC sizes deteriorate when passing from $p=10$ to $20$, corresponding to an increase in
the number of scalar restrictions under test (i.e., the degrees of freedom of the test).

\begin{table}[htbp]\tiny
\renewcommand{\arraystretch}{0.58}
\begin{center}
\begin{tabular}{|cc||cccc||cccc||cccc|}
\hline
\hline
\multicolumn{14}{|c|}{Empirical size of $t$-test}\\
\hline
\hline
\multicolumn{2}{|c||}{LRV-U} & \multicolumn{4}{c||}{$s=\lceil p/4\rceil$, $a=$} & \multicolumn{4}{c||}{$s=\lceil p/2\rceil$, $a=$} & \multicolumn{4}{c|}{$s=\lceil 3p/4\rceil$, $a=$} \\
  $p$ & $T/p$ & 0.25 & 0.5 & 0.75 & 1 & 0.25 & 0.5 & 0.75 & 1 & 0.25 & 0.5 & 0.75 & 1\\
\hline
10 & 10 & 0.13 & 0.11 & 0.09 & 0.09 & 0.19 & 0.17 & 0.13 & 0.12 & 0.30 & 0.31 & 0.24 & 0.21 \\
 & 20 & 0.09 & 0.07 & 0.07 & 0.06 & 0.15 & 0.10 & 0.09 & 0.09 & 0.26 & 0.14 & 0.12 & 0.11 \\
 & 30 & 0.08 & 0.06 & 0.06 & 0.06 & 0.10 & 0.08 & 0.07 & 0.07 & 0.17 & 0.11 & 0.10 & 0.09 \\
 & 60 & 0.06 & 0.06 & 0.06 & 0.05 & 0.07 & 0.07 & 0.07 & 0.07 & 0.09 & 0.08 & 0.07 & 0.07 \\
 & 90 & 0.06 & 0.05 & 0.05 & 0.05 & 0.06 & 0.05 & 0.05 & 0.05 & 0.08 & 0.07 & 0.07 & 0.07 \\
\hline
20 & 10 & 0.13 & 0.10 & 0.09 & 0.08 & 0.22 & 0.19 & 0.14 & 0.13 & 0.32 & 0.35 & 0.26 & 0.22 \\
 & 20 & 0.08 & 0.07 & 0.06 & 0.06 & 0.16 & 0.11 & 0.10 & 0.09 & 0.27 & 0.14 & 0.13 & 0.12 \\
 & 30 & 0.07 & 0.07 & 0.06 & 0.06 & 0.10 & 0.08 & 0.08 & 0.08 & 0.16 & 0.11 & 0.10 & 0.09 \\
 & 60 & 0.06 & 0.06 & 0.06 & 0.06 & 0.08 & 0.07 & 0.07 & 0.07 & 0.09 & 0.07 & 0.07 & 0.07 \\
 & 90 & 0.06 & 0.06 & 0.06 & 0.06 & 0.07 & 0.06 & 0.06 & 0.06 & 0.08 & 0.07 & 0.06 & 0.06
\\
\hline
\hline
\multicolumn{2}{|c||}{LRV-P} & \multicolumn{4}{c||}{$s=\lceil p/4\rceil$, $a=$} & \multicolumn{4}{c||}{$s=\lceil p/2\rceil$, $a=$} & \multicolumn{4}{c|}{$s=\lceil 3p/4\rceil$, $a=$} \\
  $p$ & $T/p$ & 0.25 & 0.5 & 0.75 & 1 & 0.25 & 0.5 & 0.75 & 1 & 0.25 & 0.5 & 0.75 & 1\\
\hline
10 & 10 & 0.38 & 0.24 & 0.16 & 0.13 & 0.40 & 0.34 & 0.25 & 0.21 & 0.49 & 0.51 & 0.44 & 0.38 \\
 & 20 & 0.31 & 0.15 & 0.10 & 0.08 & 0.39 & 0.21 & 0.15 & 0.12 & 0.50 & 0.30 & 0.22 & 0.18 \\
 & 30 & 0.26 & 0.12 & 0.09 & 0.07 & 0.32 & 0.16 & 0.11 & 0.10 & 0.43 & 0.23 & 0.16 & 0.14 \\
 & 60 & 0.20 & 0.10 & 0.07 & 0.06 & 0.23 & 0.12 & 0.09 & 0.08 & 0.28 & 0.14 & 0.11 & 0.10 \\
 & 90 & 0.17 & 0.08 & 0.06 & 0.06 & 0.18 & 0.09 & 0.07 & 0.06 & 0.22 & 0.11 & 0.09 & 0.08 \\
\hline
20 & 10 & 0.35 & 0.21 & 0.15 & 0.13 & 0.34 & 0.35 & 0.27 & 0.23 & 0.37 & 0.51 & 0.46 & 0.42 \\
 & 20 & 0.27 & 0.13 & 0.10 & 0.09 & 0.39 & 0.21 & 0.16 & 0.14 & 0.48 & 0.30 & 0.23 & 0.21 \\
 & 30 & 0.23 & 0.12 & 0.09 & 0.08 & 0.30 & 0.16 & 0.12 & 0.10 & 0.40 & 0.22 & 0.17 & 0.15 \\
 & 60 & 0.17 & 0.09 & 0.07 & 0.06 & 0.21 & 0.11 & 0.09 & 0.08 & 0.24 & 0.13 & 0.10 & 0.09 \\
 & 90 & 0.15 & 0.08 & 0.07 & 0.06 & 0.17 & 0.09 & 0.08 & 0.07 & 0.20 & 0.11 & 0.09 & 0.08
\\
\hline
\hline
\multicolumn{2}{|c||}{LRV-A} & \multicolumn{4}{c||}{$s=\lceil p/4\rceil$, $a=$} & \multicolumn{4}{c||}{$s=\lceil p/2\rceil$, $a=$} & \multicolumn{4}{c|}{$s=\lceil 3p/4\rceil$, $a=$} \\
  $p$ & $T/p$ & 0.25 & 0.5 & 0.75 & 1 & 0.25 & 0.5 & 0.75 & 1 & 0.25 & 0.5 & 0.75 & 1\\
\hline
10 & 10 & 0.40 & 0.26 & 0.17 & 0.12 & 0.51 & 0.39 & 0.26 & 0.17 & 0.60 & 0.53 & 0.37 & 0.28 \\
 & 20 & 0.27 & 0.15 & 0.11 & 0.08 & 0.39 & 0.22 & 0.15 & 0.11 & 0.62 & 0.35 & 0.23 & 0.15 \\
 & 30 & 0.20 & 0.12 & 0.09 & 0.07 & 0.29 & 0.16 & 0.12 & 0.08 & 0.46 & 0.24 & 0.16 & 0.11 \\
 & 60 & 0.13 & 0.09 & 0.08 & 0.06 & 0.18 & 0.11 & 0.09 & 0.07 & 0.25 & 0.14 & 0.11 & 0.08 \\
 & 90 & 0.10 & 0.08 & 0.06 & 0.05 & 0.12 & 0.08 & 0.07 & 0.06 & 0.18 & 0.11 & 0.09 & 0.07 \\
\hline
20 & 10 & 0.39 & 0.23 & 0.15 & 0.10 & 0.47 & 0.42 & 0.27 & 0.18 & 0.53 & 0.62 & 0.47 & 0.30 \\
 & 20 & 0.24 & 0.13 & 0.10 & 0.07 & 0.43 & 0.24 & 0.17 & 0.11 & 0.58 & 0.35 & 0.23 & 0.15 \\
 & 30 & 0.17 & 0.11 & 0.09 & 0.07 & 0.31 & 0.17 & 0.12 & 0.09 & 0.48 & 0.25 & 0.17 & 0.12 \\
 & 60 & 0.11 & 0.08 & 0.07 & 0.06 & 0.18 & 0.12 & 0.09 & 0.07 & 0.24 & 0.14 & 0.10 & 0.08 \\
 & 90 & 0.09 & 0.07 & 0.07 & 0.06 & 0.14 & 0.09 & 0.08 & 0.06 & 0.18 & 0.11 & 0.09 & 0.07
\\
\hline
\hline
\multicolumn{14}{|c|}{Empirical size of $Q$-test}\\
\hline
\hline
\multicolumn{2}{|c||}{LRV-U} & \multicolumn{4}{c||}{$s=\lceil p/4\rceil$, $a=$} & \multicolumn{4}{c||}{$s=\lceil p/2\rceil$, $a=$} & \multicolumn{4}{c|}{$s=\lceil 3p/4\rceil$, $a=$} \\
  $p$ & $T/p$ & 0.25 & 0.5 & 0.75 & 1 & 0.25 & 0.5 & 0.75 & 1 & 0.25 & 0.5 & 0.75 & 1\\
\hline
10 & 10 & 0.32 & 0.20 & 0.16 & 0.15 & 0.45 & 0.34 & 0.26 & 0.23 & 0.48 & 0.45 & 0.34 & 0.29 \\
 & 20 & 0.17 & 0.11 & 0.10 & 0.09 & 0.29 & 0.16 & 0.13 & 0.12 & 0.37 & 0.19 & 0.15 & 0.14 \\
 & 30 & 0.12 & 0.08 & 0.08 & 0.07 & 0.18 & 0.12 & 0.11 & 0.10 & 0.22 & 0.13 & 0.12 & 0.11 \\
 & 60 & 0.08 & 0.07 & 0.06 & 0.06 & 0.10 & 0.08 & 0.07 & 0.07 & 0.12 & 0.09 & 0.09 & 0.08 \\
 & 90 & 0.06 & 0.06 & 0.06 & 0.06 & 0.08 & 0.07 & 0.07 & 0.07 & 0.09 & 0.08 & 0.07 & 0.07 \\
\hline
20 & 10 & 0.46 & 0.26 & 0.21 & 0.19 & 0.74 & 0.55 & 0.43 & 0.39 & 0.79 & 0.72 & 0.58 & 0.52 \\
 & 20 & 0.21 & 0.13 & 0.12 & 0.11 & 0.45 & 0.24 & 0.20 & 0.19 & 0.60 & 0.30 & 0.24 & 0.22 \\
 & 30 & 0.15 & 0.10 & 0.10 & 0.09 & 0.25 & 0.16 & 0.14 & 0.13 & 0.33 & 0.19 & 0.16 & 0.15 \\
 & 60 & 0.08 & 0.07 & 0.07 & 0.07 & 0.13 & 0.10 & 0.09 & 0.09 & 0.14 & 0.11 & 0.10 & 0.10 \\
 & 90 & 0.07 & 0.06 & 0.06 & 0.06 & 0.10 & 0.08 & 0.08 & 0.07 & 0.11 & 0.09 & 0.08 & 0.08
\\
\hline
\hline
\multicolumn{2}{|c||}{LRV-P} & \multicolumn{4}{c||}{$s=\lceil p/4\rceil$, $a=$} & \multicolumn{4}{c||}{$s=\lceil p/2\rceil$, $a=$} & \multicolumn{4}{c|}{$s=\lceil 3p/4\rceil$, $a=$} \\
  $p$ & $T/p$ & 0.25 & 0.5 & 0.75 & 1 & 0.25 & 0.5 & 0.75 & 1 & 0.25 & 0.5 & 0.75 & 1\\
\hline
10 & 10 & 0.85 & 0.74 & 0.56 & 0.47 & 0.73 & 0.76 & 0.64 & 0.55 & 0.67 & 0.70 & 0.62 & 0.55 \\
 & 20 & 0.84 & 0.48 & 0.30 & 0.24 & 0.84 & 0.51 & 0.34 & 0.28 & 0.70 & 0.44 & 0.32 & 0.27 \\
 & 30 & 0.74 & 0.35 & 0.21 & 0.17 & 0.74 & 0.37 & 0.25 & 0.20 & 0.60 & 0.32 & 0.22 & 0.19 \\
 & 60 & 0.56 & 0.22 & 0.14 & 0.11 & 0.53 & 0.22 & 0.15 & 0.12 & 0.40 & 0.19 & 0.14 & 0.12 \\
 & 90 & 0.45 & 0.17 & 0.11 & 0.09 & 0.43 & 0.17 & 0.12 & 0.10 & 0.31 & 0.15 & 0.11 & 0.10 \\
\hline
20 & 10 & 0.91 & 0.91 & 0.79 & 0.72 & 0.71 & 0.94 & 0.90 & 0.86 & 0.70 & 0.89 & 0.92 & 0.88 \\
 & 20 & 0.96 & 0.65 & 0.47 & 0.39 & 0.95 & 0.74 & 0.58 & 0.51 & 0.88 & 0.70 & 0.56 & 0.50 \\
 & 30 & 0.89 & 0.49 & 0.32 & 0.27 & 0.90 & 0.54 & 0.39 & 0.33 & 0.83 & 0.51 & 0.38 & 0.33 \\
 & 60 & 0.71 & 0.29 & 0.18 & 0.15 & 0.70 & 0.32 & 0.21 & 0.18 & 0.57 & 0.27 & 0.20 & 0.17 \\
 & 90 & 0.59 & 0.22 & 0.14 & 0.12 & 0.56 & 0.23 & 0.15 & 0.13 & 0.45 & 0.20 & 0.15 & 0.13
\\
\hline
\hline
\multicolumn{2}{|c||}{LRV-A} & \multicolumn{4}{c||}{$s=\lceil p/4\rceil$, $a=$} & \multicolumn{4}{c||}{$s=\lceil p/2\rceil$, $a=$} & \multicolumn{4}{c|}{$s=\lceil 3p/4\rceil$, $a=$} \\
  $p$ & $T/p$ & 0.25 & 0.5 & 0.75 & 1 & 0.25 & 0.5 & 0.75 & 1 & 0.25 & 0.5 & 0.75 & 1\\
\hline
10 & 10 & 0.95 & 0.85 & 0.59 & 0.31 & 0.88 & 0.86 & 0.64 & 0.39 & 0.78 & 0.73 & 0.54 & 0.40 \\
 & 20 & 0.90 & 0.56 & 0.32 & 0.17 & 0.89 & 0.58 & 0.35 & 0.21 & 0.82 & 0.52 & 0.33 & 0.21 \\
 & 30 & 0.77 & 0.40 & 0.23 & 0.11 & 0.76 & 0.41 & 0.25 & 0.14 & 0.66 & 0.35 & 0.22 & 0.14 \\
 & 60 & 0.45 & 0.22 & 0.14 & 0.08 & 0.47 & 0.23 & 0.15 & 0.10 & 0.37 & 0.20 & 0.14 & 0.10 \\
 & 90 & 0.30 & 0.16 & 0.11 & 0.07 & 0.32 & 0.17 & 0.12 & 0.08 & 0.26 & 0.15 & 0.11 & 0.08 \\
\hline
20 & 10 & 0.98 & 0.99 & 0.81 & 0.42 & 0.88 & 0.98 & 0.90 & 0.61 & 0.83 & 0.94 & 0.92 & 0.69 \\
 & 20 & 0.99 & 0.79 & 0.49 & 0.21 & 0.98 & 0.84 & 0.57 & 0.31 & 0.95 & 0.80 & 0.54 & 0.33 \\
 & 30 & 0.94 & 0.60 & 0.33 & 0.15 & 0.96 & 0.66 & 0.40 & 0.19 & 0.93 & 0.61 & 0.37 & 0.21 \\
 & 60 & 0.65 & 0.32 & 0.18 & 0.09 & 0.70 & 0.37 & 0.21 & 0.12 & 0.62 & 0.32 & 0.20 & 0.12 \\
 & 90 & 0.46 & 0.22 & 0.14 & 0.08 & 0.50 & 0.25 & 0.16 & 0.09 & 0.43 & 0.22 & 0.15 & 0.09
\\
\hline
\hline
\end{tabular}
\end{center}
\caption{\footnotesize Frequency of rejection under the null of $t$ and $Q$ test on $\psi$ over {\Ntext} replications of DGP \eqref{eq_DGP_OW} at $5\%$ significance level. LRV-U: $\wh\Omega$ equal to its true value, LRV-P: $\wh\Omega$ as in \eqref{eq_Xihat}, LRV-A: $\wh\Omega$ as in \cite{An:91} and \cite{AM:92}.} \label{table_W1_Wr}
\end{table}

\cp

\section{Empirical application}\label{sec_app_emp}

Data were downloaded from the Federal Reserve Economic Data (FRED) website: \\
\url{https://fred.stlouisfed.org/}.

Developed Markets (DM) and Emerging Markets (EM) are defined as in the MSCI 2024 Market Classification, available at:\\
\url{https://www.msci.com/our-solutions/indexes/market-classification}.

EM, Emerging Markets: Brazil (BZ), China (CH), India (IN), Malaysia (MA), Mexico (MX), South Africa (SA), South Korea (SK), Taiwan (TA), Thailand (TH); DM, Developed Markets are composed by EU, European countries: UK and SZ: United Kindom (UK) and Switzerland (SZ); Nc and EZ, Nc and EZ and Eurozone: Denmark (DK), Eurozone (euro), Norway (NK), Sweden (SK) and Non-EU, non European countries: Australia (AU), Canada (CA), Hong Kong (HK), Japan (JP), Singapore (SG).

The empirical analysis was performed for different partitions of the data in groups of countries, following the tree structure in Figure \tref{fig_tree}, which reports the dimension $p$ of the system and the estimated number of common trends $\wh s$ via the max-gap criterion, see Table \tref{table_other} for results obtained also via the $\{F_{j,n}\}_{j=p,\dots,1}$ test sequences, $n=1,\infty$.

\begin{table}[htbp]\small
\begin{center}
\begin{tabular}{lc|ccc|c}
	\hline\hline
 & & \multicolumn{3}{c|}{estimate of $s$} & \\
	Group & $p$ & max-gap & $%
	F_{j,\infty }$ & $F_{j,1}$ & countries\\ \hline\hline
	WM & 20 & 19 & 19 & 17 & all \\ \hline
	DM & 11 & 10 & 10 & 10 & AU, CA, DK, EU, HK, JP, NO, SG, SW, SZ\\
	EM & 9 & 9 & 8 & 8 & BZ, CH, IN, MA, MX, SA, SK, TA,TH\\ \hline
	EU & 6 & 5 & 5 & 5 & DK, EU, NO, SW, SZ, UK \\
	Non-EU & 5 & 5 & 5 & 5 & AU, CA, HK, JP, SG\\ \hline
	Nc and EZ & 4 & 3 & 3 & 3 & DK, EU, NK, SK\\
	UK and SZ & 2 & 2 & 2 & 2 & SZ, UK\\ \hline\hline
\end{tabular}
\vspace{1em}
\caption{\footnotesize Estimates of $s$: max-gap estimator and $F_{j,\infty}$ and $F_{j,1}$ sequential tests.}
\label{table_other}
\end{center}
\end{table}

The semiparametric results are compared with the high dimensional VAR analysis of \citet{OW:18} and \citet{BG:24} for inference on $s$, and with the likelihood analysis of \citet{Joh:96} for inference on $s$ and $\psi$.

\begin{figure}[htbp]
 \includegraphics[width=\textwidth]{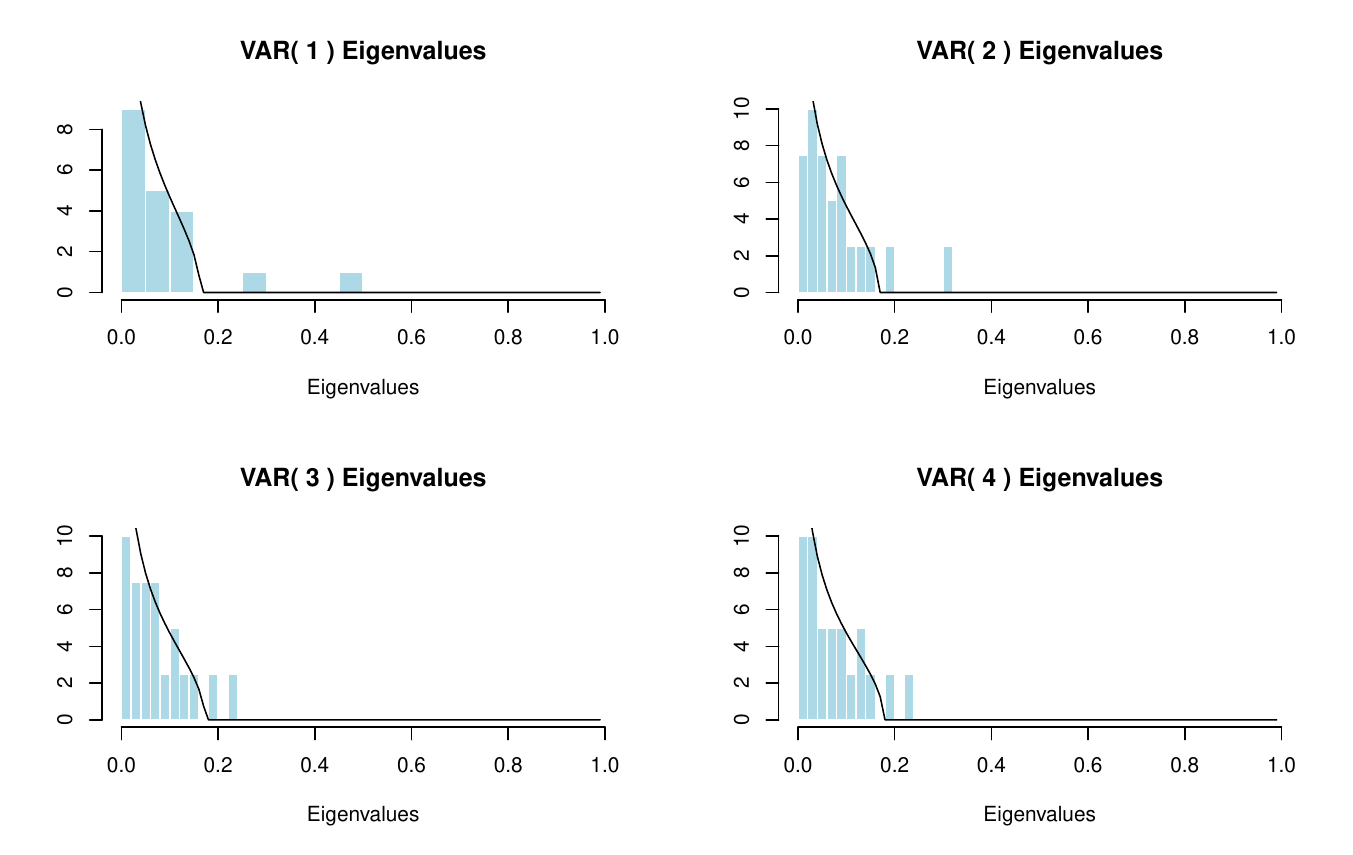}
 \vspace{-3em}
\caption{\footnotesize \citet{BG:24} analysis of the 20 WM currencies. Eigenvalues (histogram) and Wachter distribution (line) based on VAR$(k)$, $k=1,2,3,4$.}
\label{fig_BG}
\end{figure}

The result on the presence of cointegration ($s<p$) among the of the 20 WM currencies is further investigated using Wachter plots as in \citet{OW:18} and \citet{BG:24}, and the test of no cointegration of \citet{BG:24}. For a VAR$(k)$, $k=1,2,3,4$, Figure \tref{fig_BG} displays eigenvalues and the Wachter distribution under $s=p$ and Table \tref{table_BG} reports $p$-values of the \citet{BG:24} test of no cointegration. Computations were performed in R with the package \texttt{Largevars} documented in \citet{BGK:24}. The plots in Figure \tref{fig_BG} display eigenvalues to the right of the support of the Wachter distribution, which is an indication of the presence of cointegration, see \citet{OW:18} and \citet{BG:24}. This is in line with the results of the test of no cointegration in Table \tref{table_BG}, which reports $p$-values below 0.01.

\cp

\begin{table}[htbp]\small
\begin{center}
\begin{tabular}{c|cccc}
\hline\hline
$k$ & $1$ & $2$ & $3$ & $4$ \\
\hline
$p$-value & $<0.01$ & $<0.01$ & $<0.01$ & $<0.01$ \\
\hline\hline
\end{tabular}
\vspace{1em}
\caption{\footnotesize \citet{BG:24} test for the null of no cointegration among the 20 WM currencies based on VAR$(k)$, $k=1,2,3,4$.}
\label{table_BG}
\end{center}
\end{table}

For the group of Nc and EZ currencies, the semiparametric results are compared with those based on the likelihood analysis of the Error Correction Model $\Delta X_t = \alpha \beta' X_{t-1} + \Gamma \Delta X_{t-1} +\varepsilon_t$, with $\alpha,\beta\in\BR^{p\times r}$, see \citet{Joh:96}. Table \tref{table_VAR} reports the VAR likelihood ratio test on $r$ and the QML estimate of $\beta$ in \citet{Joh:96} (with no deterministic components).

The trace test strongly rejects $r=0$, i.e. $s=4$, and does not reject $r\leq1$, i.e. $s\geq 3$, hence implying the same conclusion as the semiparametric analysis of $s=3$ stochastic trends and $r=1$ cointegrating relation in the Nc and EZ system. Moreover, the QML estimate of $\beta$ reported in the table is very similar to the semiparametric estimate in \eqref{eq_psi_hat}.

\begin{table}[htbp]\small
\begin{center}
\begin{tabular}{ccrc|crr}
\hline\hline
\multicolumn{1}{c}{$r$} & \multicolumn{1}{c}{$s$} & \multicolumn{1}{c}{Trace test} & \multicolumn{1}{c|}{5\% critical value} &
\multicolumn{1}{c}{currency} & \multicolumn{1}{c}{$\wh \beta$} & \multicolumn{1}{c}{$p$-values} \\
\hline
0   & 4 & 128.63 & 39.71 & DK & $-1.015$ & $< 10^{-4}$ \\
$\leq 1$ & $\geq 3$ &12.75 & 24.08 & Euro & 1   &      \\
$\leq 2$ & $\geq 2$ &3.63  & 12.21 & NK & 0.015 & $< 10^{-4}$ \\
$\leq 3$ & $\geq 1$ &0.22  & 4.14 & SK & 0.001 & 0.741  \\
\hline\hline
\end{tabular}
\vspace{1em}
 \caption{\footnotesize Likelihood based analysis of the 4 Nc and EZ currencies. Trace test and corresponding $5\%$ critical values and ML estimate of $\beta$ with $p$-values, see \citet{Joh:96}.}
\label{table_VAR}
\end{center}
\end{table}

\bibliographystyle{Chicago}

\end{document}